\documentclass[pra,twocolumn]{revtex4}
\usepackage{epsfig}
\usepackage{graphicx}
\usepackage{amsmath}
\usepackage{natbib}
\newcommand{\bn}{\begin{eqnarray}}
\newcommand{\en}{\end{eqnarray}}
\newcommand{\eml}{\end{multline}}
\newcommand{\bml}{\begin{multline}}

\begin{document}

\title {Euler top with a rotor: classical analogies of spin squeezing and quantum phase transitions in a generalized Lipkin-Meshkov-Glick model}
 \author{Tom\'{a}\v{s} Opatrn\'{y},  Luk\'{a}\v{s} Richterek, and Martin  Opatrn\'{y}}
 \affiliation{Palack\'{y} University, Faculty of Science, 17. Listopadu 12,
 77146 Olomouc, Czech Republic}

\begin{abstract}
We show that the classical model of Euler top (freely rotating, generally asymmetric rigid body), possibly  supplemented with a rotor, corresponds to a generalized  Lipkin-Meshkov-Glick (LMG) model describing phenomena of various branches of quantum physics.
Classical effects such as free precession of a symmetric top, Feynman's wobbling plate, tennis-racket instability and the Dzhanibekov effect, attitude control of satellites by momentum wheels, or twisting somersault dynamics, have their counterparts in quantum effects that include spin squeezing by one-axis twisting and two-axis countertwisting, transitions between the Josephson and  Rabi regimes of a Bose-Einstein condensate in a double-well potential, and other quantum critical phenomena. The parallels enable us to expand the range of explored quantum phase transitions in the generalized LMG model, as well as to present a classical analogy of the recently proposed LMG Floquet time crystal.
\end{abstract}

\date{\today }

 \maketitle

\section{Introduction}

{\it ``The same equations have the same solutions''}  is a well known Feynman's quote from his lecture on electrostatic analogs \cite{Feynman-lecture}. Taking advantage of known solutions of Maxwell's equations, Feynman shows how to apply them for solving problems of heat transport, neutron diffusion, fluid dynamics, and photometry. The message is that analogs are powerful tools that allow the exchange of know-how between different branches of physics. Here we follow this approach and focus on quantum analogs of the  Euler dynamical equations, initially introduced to study rotations of rigid bodies \cite{Euler}. We show that already the simplest version of Euler equations describing a free spinning top is relevant to the quantum mechanical problem of spin squeezing \cite{Kitagawa,Wineland1994}, i.e., noise suppression important for improving precision of atomic clocks and measuring devices. 
If a freely spinning rotor with its axis fixed with respect to the top is added,
plethora of new phenomena occur with analogies across diverse fields. Equations of motion that were originally purely quadratic get additional linear terms. As a consequence, one can observe features of the Lipkin-Meshkov-Glick (LMG) model of nuclear physics \cite{Lipkin}, excited state quantum phase transitions \cite{Sachdev,Caprio,Cejnar},  self-trapping of Bose-Einstein condensates in potential wells \cite{Smerzi,Micheli2003}, or twist-and-turn scenario of spin squeezing \cite{Muessel}. 
These quantum phenomena correspond to purely classical effects such as satellite stabilization by momentum wheels \cite{Krishnaprasad,Bloch-1992,Casu} or  motion of an athlete executing a twisted somersault \cite{Bharadwaj,Dullin}.

The aim of this paper is to identify mutually equivalent relations of various fields and help the reader to use the intuition accumulated in one branch of physics for solving problems of another. We  apply here this approach to show new types of excited state quantum phase transitions in a generalized LMG model that correspond to different kinds of motion in rigid body dynamics. We also propose a classical analogue of the recently introduced LMG Floquet time crystal \cite{Russomanno-2017}.

The paper is organized in such a way that more complex models follow the simple ones.
In Sec. \ref{Sec-Eqs} the basic equations are introduced. In Sec. \ref{Sec-symmetrictop} analogies between motion of a symmetric top  and spin squeezing by one-axis twisting are studied.  In Sec. \ref{Sec-asymmetrictop} we deal with the asymmetric top, tennis-racket instability, and two-axis countertwisting scenario of spin squeezing.
In Sec. \ref{SecCoaxial} the dynamics of a symmetric top with a coaxial rotor and their quantum analogies are studied.  Sec. \ref{SecPerpendicularRotor} is focused on a symmetric top with a perpendicular rotor and the corresponding quantum model with effects such as the twist-and-turn spin squeezing scenario, or transitions between the Rabi and Josephson regimes in trapped Bose-Einstein condensates. 
Sec. \ref{AsymmetricPrincipal} deals with the main features of an asymmetric top with a rotor along one of the principal axes and their analogies in the LMG model. In Sec. \ref{SecPhase} 
we present a general treatment of stationary angular momenta and their stability, relevant to the excited-state quantum phase transitions in a generalized LMG model.
In Sec. \ref{SecTimeCrystal} we introduce a classical analogue of the LMG time crystal, and we conclude with Sec. \ref{SecConclusion}. 
Some lengthy formulas and derivations are presented in Appendixes.

\section{Equations of motion}
\label{Sec-Eqs}
\subsection{Classical motion}
\begin{figure}
\centerline{\epsfig{file=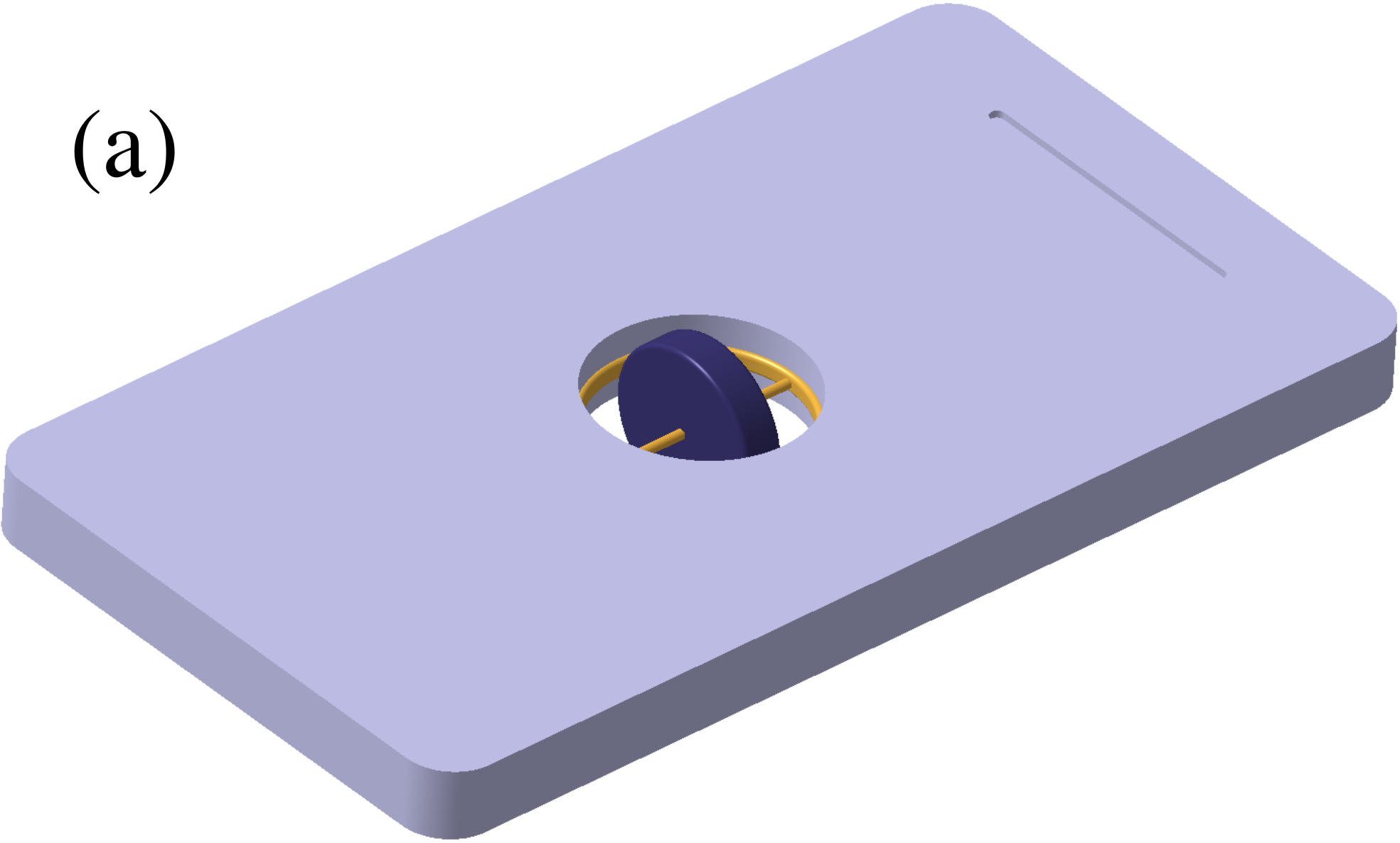,width=0.5\linewidth}
\epsfig{file=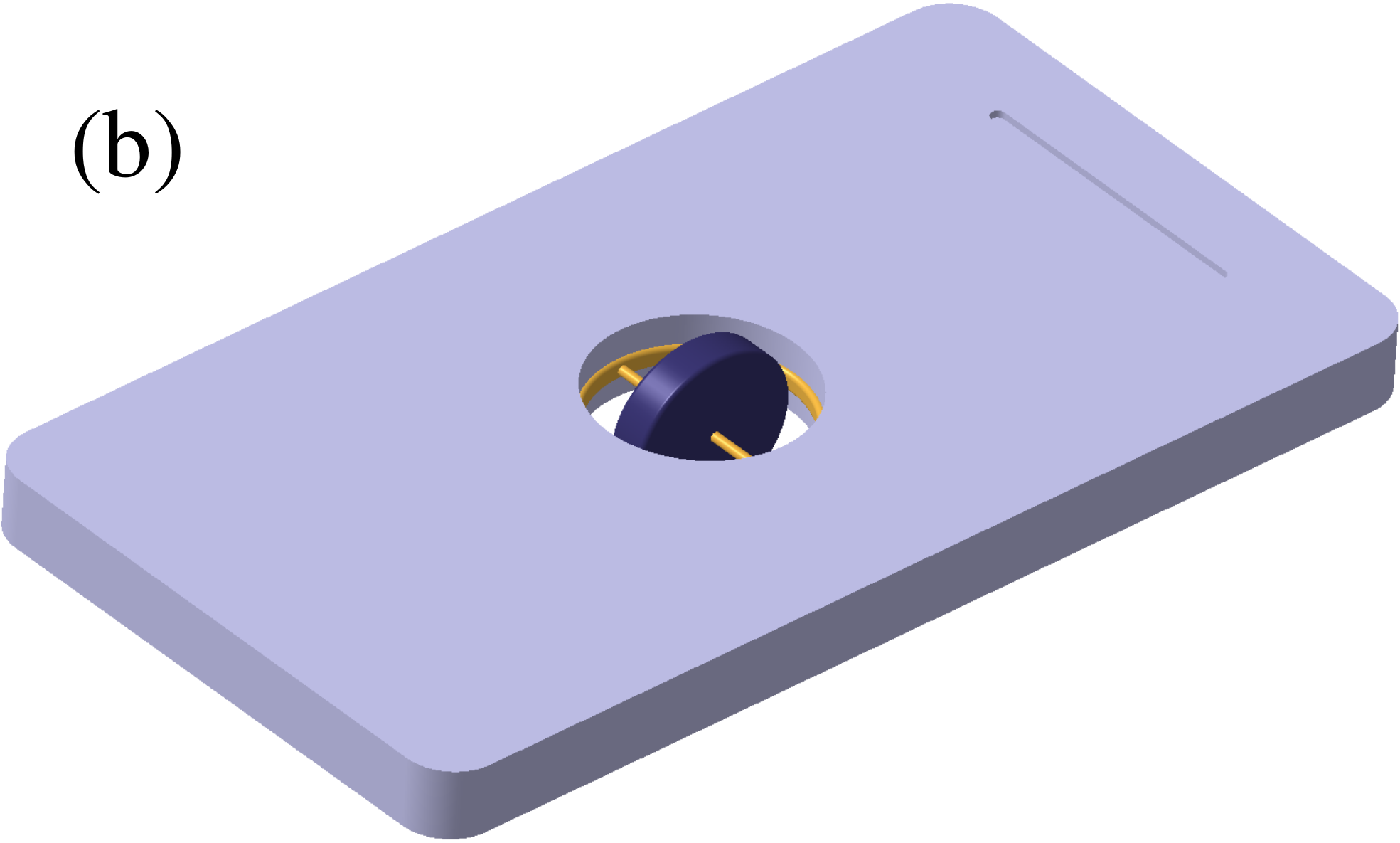,width=0.5\linewidth}}
\centerline{\epsfig{file=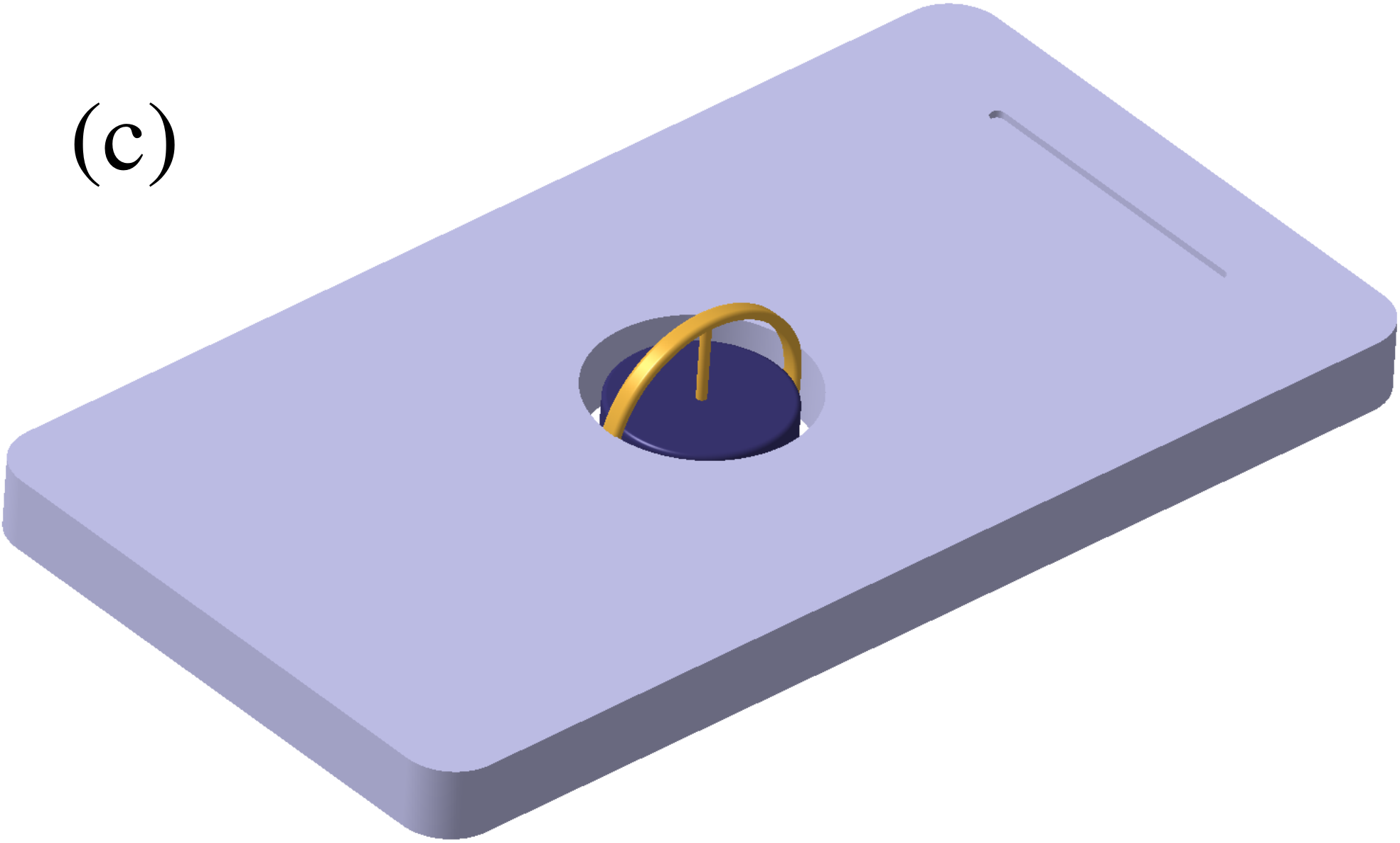,width=0.5\linewidth}
\epsfig{file=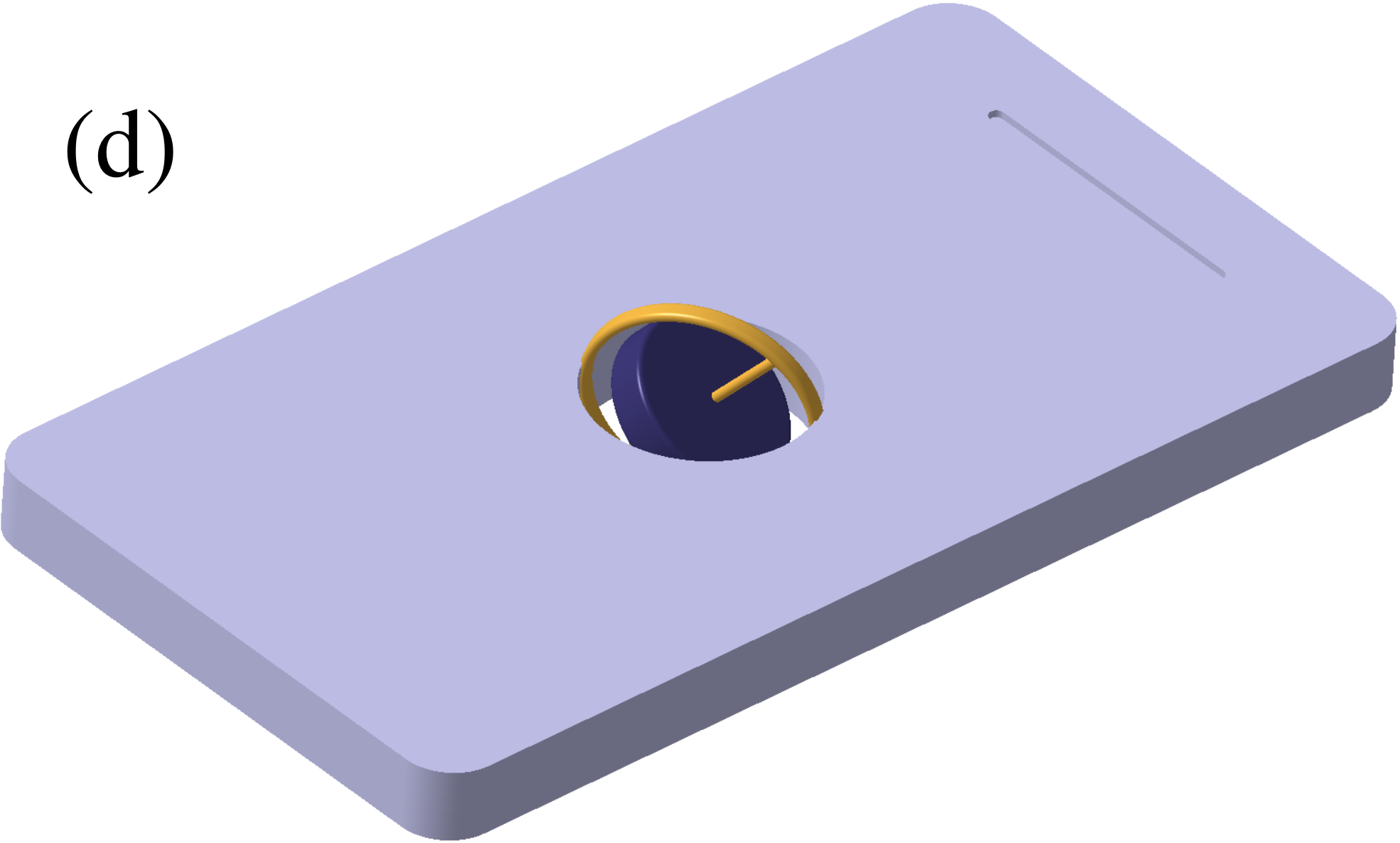,width=0.5\linewidth}}
\caption{\label{f-setrvacniky1}
Rigid bodies supplemented with a rotor. In panels (a)---(c) the rotor axis lies along one of the principal axes of the body, in panel (d) the rotor axis has general orientation.
}
\end{figure}

Evolution of the angular momentum  $\vec L$ of a rigid body is governed by the equation
\begin{eqnarray}
\frac{d\vec{L}}{dt} = \vec{M},
\label{dLdt}
\end{eqnarray}
where $\vec{M}$ is the torque. We use the relation between the time derivative $d\vec{A}/dt$ of a vector $\vec A$ in an inertial coordinate system and the time derivative $d'\vec{A}/dt$ in a coordinate system that rotates with angular velocity $\vec{\omega}$ with respect to the inertial system
\begin{eqnarray}
\frac{d'\vec{A}}{dt} = \frac{d\vec{A}}{dt} - \vec{\omega}\times \vec{A}.
\end{eqnarray}
Applying that on Eq. (\ref{dLdt}), we get
\begin{eqnarray}
\frac{d'\vec{L}}{dt} = \vec{M} -\vec{\omega}\times \vec{L}.
\label{dLdt2}
\end{eqnarray}
Assume that the torque stems from a rotor whose axis is fixed with respect to the rigid body as in Fig. \ref{f-setrvacniky1}. We have
\begin{eqnarray}
\vec{M} = -\vec{M}_{\rm rotor} = -\frac{d\vec{K}}{dt} ,
\end{eqnarray}
where $\vec{M}_{\rm rotor}$ is the torque with which the rigid body acts on the rotor with angular momentum $\vec{K}$.
Using the expression for the time derivative in the rotating system, we have
\begin{eqnarray}
\vec{M} &=& -\frac{d'\vec{K}}{dt} - \vec{\omega}\times \vec{K}  \nonumber \\
&=&  - \vec{\omega}\times \vec{K},
\end{eqnarray}
since $d'\vec{K}/dt=0$ as the rotor changes neither the  magnitude of rotation nor the axis orientation with respect to the rigid body. Using this in Eq. (\ref{dLdt2}) we have
\begin{eqnarray}
\frac{d'\vec{L}}{dt} = -\vec{\omega} \times \left(\vec{L} +  \vec{K} \right).
\label{dLdt3}
\end{eqnarray}
If the axes of the rotating coordinate system are the principal axes of the tensor of inertia of the body, we have
\begin{eqnarray}
L_k = I_k \omega_k , \qquad k=1,2,3,
\end{eqnarray}
where $I_{1,2,3}$ are the principal moments of inertia. This allows us to write
\begin{eqnarray}
\dot{\omega}_1 &=& \frac{I_2-I_3}{I_1}\omega_2 \omega_3 +\frac{{K}_2\omega_3 - {K}_3\omega_2}{I_1} , \nonumber  \\
\dot{\omega}_2 &=& \frac{I_3-I_1}{I_2}\omega_3 \omega_1 +\frac{{K}_3\omega_1 - {K}_1\omega_3}{I_2} , \nonumber \\
\dot{\omega}_3 &=& \frac{I_1-I_2}{I_3}\omega_1 \omega_2 +\frac{{K}_1\omega_2 - {K}_2\omega_1}{I_3} ,
\label{Eq-omegadot}
\end{eqnarray}
where the dot denotes time derivative in the rotating system.
These are the well known Euler dynamical equations which for $\vec{K}=0$ correspond to a free top, and here the special case corresponds to the torque coming from the rotor.
These equations can be expressed in terms of the angular momentum,
\begin{eqnarray}
\dot{L}_1 &=& \left( \frac{1}{I_3} - \frac{1}{I_2}\right) L_2 L_3 + \frac{{K}_2}{I_3}L_3 -  \frac{{K}_3}{I_2}L_2 , 
\nonumber \\
\dot{L}_2 &=& \left( \frac{1}{I_1} - \frac{1}{I_3}\right) L_3 L_1 + \frac{{K}_3}{I_1}L_1 -  \frac{{K}_1}{I_3}L_3 , \nonumber \\ 
\dot{L}_3 &=& \left( \frac{1}{I_2} - \frac{1}{I_1}\right) L_1 L_2 + \frac{{K}_1}{I_2}L_2 -  \frac{{K}_2}{I_1}L_1 . 
\label{dotLs}
\end{eqnarray}
It is suitable to work with the total angular momentum $\vec{J}\equiv \vec{L}+\vec{K}$, for which one finds
\begin{eqnarray}
\dot{J}_1 &=& \left( \frac{1}{I_3} - \frac{1}{I_2}\right) J_2 J_3 + \frac{{K}_2}{I_2}J_3 -  \frac{{K}_3}{I_3}J_2 , 
\nonumber \\
\dot{J}_2 &=& \left( \frac{1}{I_1} - \frac{1}{I_3}\right) J_3 J_1 + \frac{{K}_3}{I_3}J_1 -  \frac{{K}_1}{I_1}J_3 , \nonumber \\ 
\dot{J}_3 &=& \left( \frac{1}{I_2} - \frac{1}{I_1}\right) J_1 J_2 + \frac{{K}_1}{I_1}J_2 -  \frac{{K}_2}{I_2}J_1 . 
\label{dotJs}
\end{eqnarray}
As can be checked, the evolution equations conserve the kinetic energy of the rigid body and the magnitude of the total angular momentum, i.e.,
\begin{eqnarray}
\dot{E}_{\rm body}= 0, \qquad \dot{J^2} = 0,
\end{eqnarray}
where
\begin{eqnarray}
\label{eqEnergie}
E_{\rm body} &=& \frac{L_1^2}{2I_1} +  \frac{L_2^2}{2I_2} + \frac{L_3^2}{2I_3}, \\
J^2 &=& J_1^2 + J_2^2 + J_3^2.
\label{eqLsquare}
\end{eqnarray}
Thus, the trajectories in the angular momentum space are intersections of the energy ellipsoid $E_{\rm body}=$ const and the total angular momentum sphere ${J}=$ const, their centers being displaced by $\vec{K}$.
This geometric interpretation is especially helpful for finding stationary angular momenta and determining their stability.
%

\subsection{Quantum motion}
\label{SubsecQuantumMotion}
Assume two bosonic modes described by annihilation operators $\hat a$ and $\hat b$ with total number of particles $N$.
These operators commute as $[\hat a,\hat a^{\dag}] = [\hat b,\hat b^{\dag}] =1$ and the remaining commutators vanish. One can introduce operator $\hat{\vec{J}}$ with components defined as
\begin{eqnarray}
\hat J_x &=& \frac{1}{2}(\hat a^{\dag}\hat b+\hat a\hat b^{\dag}), \\
\hat J_y &=& \frac{1}{2i}(\hat a^{\dag}\hat b-\hat a\hat b^{\dag}), \\
\hat J_z &=& \frac{1}{2}(\hat a^{\dag}\hat a-\hat b^{\dag}\hat b),
\end{eqnarray}
with
$N = \hat a^{\dag}\hat a+\hat b^{\dag}\hat b$. These operators satisfy the angular momentum  commutation relations $[\hat J_x,\hat J_y]=i\hat J_z$,  $[\hat J_y,\hat J_z]=i\hat J_x$, and $[\hat J_z,\hat J_x]=i\hat J_y$.
Assume a general quadratic Hamiltonian in the form
\begin{eqnarray}
\hat H &=& \sum_{k,l}\chi_{kl}\hat J_k\hat J_l + \sum_{k}\Omega_k \hat J_k,
\label{Ham-nlin}
\end{eqnarray}
where the indexes $k,l$ run through $x,y,z$, the quantities $\chi_{kl}=\chi_{lk}$ are components of a twisting tensor $\chi$ \cite{Opatrny2015}, and we put $\hbar =1$.
As discussed below, this Hamiltonian describes dynamics of symmetrical samples of interacting two-level systems in the Lipkin-Meshkov-Glick model \cite{Lipkin}, as well as various regimes of spin squeezing \cite{Kitagawa}.

By a suitable rotation of the coordinate system, the twisting tensor can be cast into diagonal form such that the Hamiltonian is
\begin{eqnarray}
\hat H &=&   \sum_{k=1}^3 \left( \chi_k \hat J_k^2 +  \Omega_k \hat J_k \right)  ,
\label{Ham123}
\end{eqnarray}
where $\chi_k$ are the eigenvalues of the twisting tensor and 
the commutation rules are $[\hat J_1,\hat J_2]=i\hat J_3$ with cyclic permutations. The Heisenberg evolution equations $i d \hat A/dt = [\hat A, \hat H]$ then yield
\begin{eqnarray}
\frac{d\hat J_1}{dt} &=& (\chi_2-\chi_3) (\hat J_2 \hat J_3 + \hat J_3 \hat J_2) + \Omega_2 \hat J_3- \Omega_3 \hat J_2,
\nonumber \\
\frac{d\hat J_2}{dt} &=& (\chi_3-\chi_1) (\hat J_3 \hat J_1 + \hat J_1 \hat J_3) + \Omega_3 \hat J_1- \Omega_1 \hat J_3,
\nonumber \\
\frac{d\hat J_3}{dt} &=& (\chi_1-\chi_2) (\hat J_1 \hat J_2 + \hat J_2 \hat J_1) + \Omega_1 \hat J_2- \Omega_2 \hat J_1.
\label{dotJ1to3}
\end{eqnarray}
Note that Hamiltonian (\ref{Ham123}) commutes with the total angular momentum so that $\hat J^2$ is a conserved quantity with $\hat J^2  \equiv \hat J_1^2 + \hat J_2^2 + \hat J_3^2 = \frac{N}{2}(\frac{N}{2}+1)$.

\subsection{Correspondence of the models}
Equations  (\ref{dotJ1to3}) and (\ref{dotJs}) have the same structure, except for (\ref{dotJs})  being classical equations  whereas (\ref{dotJ1to3}) are operator equations with symmetrized  products of operators. Thus, both models yield analogous predictions.
This happens for short times provided the quantum system was initialized in a classical-like spin coherent state. For longer times, interference phenomena occur and the predictions of the two models deviate.

The two sets of equations correspond to each other provided we make the change
\begin{eqnarray}
\chi_{k} \leftrightarrow -\frac{1}{2I_k}, \qquad \Omega_k  \leftrightarrow \frac{{K}_k}{I_k},
\end{eqnarray}
or 
\begin{eqnarray}
I_{k} \leftrightarrow -\frac{1}{2\chi_k}, \qquad {K}_k  \leftrightarrow -\frac{\Omega_k}{2\chi_k}.
\label{Ikchi}
\end{eqnarray}
Note that the dimension of the quantities is set by our choice $\hbar =1$; to have the usual dimensionality, the relation between $\chi_k$ and $I_k$ would be changed to $\chi_{k} \leftrightarrow -\hbar /(2I_k)$. 

Note also that, whereas there is a straightforward correspondence between the quantum and classical angular momenta $\hat{\vec{J}} \leftrightarrow \vec{J}$, the relation between the energy of the body (\ref{eqEnergie}) and the quantum Hamiltonian 
(\ref{Ham123}) is rather
\begin{eqnarray}
\hat H  \leftrightarrow - E_{\rm body} + \sum_{k=1}^{3} \frac{K_k^2}{2I_k}.
\end{eqnarray}
The last term is a constant that can be considered trivial.
On the other hand, the difference of signs of $\hat H$ and $E_{\rm body}$ is interesting: as a result, 
the quantum vector $\hat{\vec{J}}$ moves on the sphere of $\hat J^2 =$ const along a constant energy contour such that the higher energy area is on the left, the classical vector  $\vec{J}$ moves on the  sphere of $J^2 =$ const with the higher energy area on the right.

\subsection{Invariance with respect to transformation of $\chi_k$ and $I_k$}
Since in Eq. (\ref{dotJ1to3}) only the differences between the twisting tensor eigenvalues occur, the dynamics are not changed  if a constant is added to all eigenvalues of $\chi$, i.e., $\chi_k\to \chi_k+\chi_0$. This transformation just shifts the Hamiltonian by a constant $\chi_0 \hat J^2$. This means one has a freedom in choosing a reference value of $\chi$.
The same holds in the classical dynamics if the moments of inertia are modified as 
\begin{eqnarray}
\frac{1}{I_k} \to  \frac{1}{I_k}+ \frac{1}{I_0} 
\label{1/Ik}
\end{eqnarray}
and the angular momentum of the rotor as
\begin{eqnarray}
K_k \to  \frac{K_k}{1+ \frac{I_k}{I_0}} 
\label{K/Ik}
\end{eqnarray}
with $I_0$ independent of $k$. With $J$ fixed, the energy of the body is then shifted by 
\begin{eqnarray}
\Delta E_{\rm body} = \frac{J^2}{2I_0} - \sum_{k=1}^3 \frac{K_k^2}{2(I_0+I_k)}. 
\end{eqnarray}

As a consequence, for any quantum system described by twisting tensor $\chi$ and frequency vector $\vec{\Omega}$,  one can find a classical rigid body characterized by tensor of inertia $I$ supplemented with a  rotor with angular momentum $\vec{K}$ such that the combined system has the same dynamics. To show that, recall that mass can be assembled such as to have arbitrary principal moments of inertia $I_k$, provided these values are positive and satisfy the triangle inequality $I_j \leq I_k+I_l$ for any permutation of indexes $j,k,l$. 
The first condition is satisfied by a suitable choice of the additive constant $\chi_0$  making all values $\chi_k$ negative such that all values $I_k$ resulting from Eq. (\ref{Ikchi}) are positive.  If the resulting values $I_k$ violate the triangle inequality, one can fix it as follows. Assume that, say,  $I_1>I_2+I_3$. Then, by choosing $I_0$ satisfying
\begin{eqnarray}
0 < I_0 < \frac{I_2 I_3 + \sqrt{I_2^2 I_3^2 + I_1 I_2 I_3 (I_1-I_2-I_3)}}{I_1 -I_2 -I_3}
\end{eqnarray} 
and applying Eq. (\ref{1/Ik}), one finds a realistic tensor of inertia corresponding to the given twisting tensor $\chi$. The linear part of the equations of motion can then easily be adjusted by a suitable choice of  $\vec{K}$.

Let us note that the invariance with respect to transformations (\ref{1/Ik}) is valid only for the dynamics of momenta, Eqs. (\ref{dotLs}) or (\ref{dotJs}), but not for the evolution of the angular velocity, Eq. (\ref{Eq-omegadot}). This means that one can mutually map the quantum and classical evolution only with respect to where the angular momentum points, but not with respect to how the rigid body is oriented itself. The latter would follow from the kinematic Euler equations which are not included in our study. We anticipate that with expanding the analogy, the classical body orientation would be related to the global phase of the quantum system. Even though connections to other interesting phenomena might be found, such as, e.g., the Montgomery phase \cite{Montgomery,Natario} referring to the change of body orientation after $\vec{J}$ returns to its initial value, these are beyond the scope of this paper.

\subsection{Lipkin-Meshkov-Glick model}

In 1965 Lipkin, Meshkov and Glick formulated a toy model of multiparticle interaction that can be, under certain conditions, solved exactly, and thus serve as a basis for testing various approximation methods \cite{Lipkin}. Although the original motivation was in modeling energy spectra of atomic nuclei, the scheme turned out to be useful for studying interesting phenomena in more general systems. These include  quantum criticality and phase transitions \cite{Kwok-Lin,Zibold2010, Vidal2004,Castanos,Engelhardt,Gallemi,Campbell-2016,Leyvraz,Ribeiro2007,Ribeiro, Hooley}, 
multi-particle entanglement \cite{Vidal2004,Dusuel2005,Vidal2006,Vidal-entanglement,Carrasco}, spin squeezing \cite{Kitagawa,Wineland1994,Micheli2003,Opatrny2015,Kajtoch,Bennett-2013,Duan-2001,Sorensen-Molmer-2002,Takeuchi-2005,Schleier-SmithPRA2010,Orzel2001,Esteve2008,Gross2010,Riedel,Leroux-2010,Liu2011,Shen-Duan-2013,Zhang2014,Huang-2015,Wu-2015,Helmerson,OKD2014,Korkmaz,Yu-2016,Zhang-2017,Borregaard-2017, Bhattacharya,Bhattacharya-2015,Rudner-2011,Jin-2009,Muessel,Kolar-2015,Opatrny-classqueez,TO-Counterdiabatic,Chaudhury-2007,Fernholz-2008,Pan2017,Pezze,Kruse-2016,Opatrny-2017},
molecular magnetism \cite{Chudnovsky2001}, or circuit quantum electrodynamics \cite{Larsen}.

The LMG Hamiltonian can be written in the form
\begin{eqnarray}
\hat H=\epsilon \hat J_3 + V(\hat J_1^2-\hat J_2^2) + W(\hat J_1^2+\hat J_2^2),
\label{Ham-LMG2}
\end{eqnarray}
where $\epsilon, V$ and $W$ are real parameters. In the original paper  \cite{Lipkin} the LMG model describes $N$ fermions in two degenerate levels whose energies differ by $\epsilon$. The term proportional to $V$ scatters pairs of particles of the same level to the other level, and the term proportional to $W$ scatters one particle up while another particle is scattered down. 

\begin{figure}
\centerline{\epsfig{file=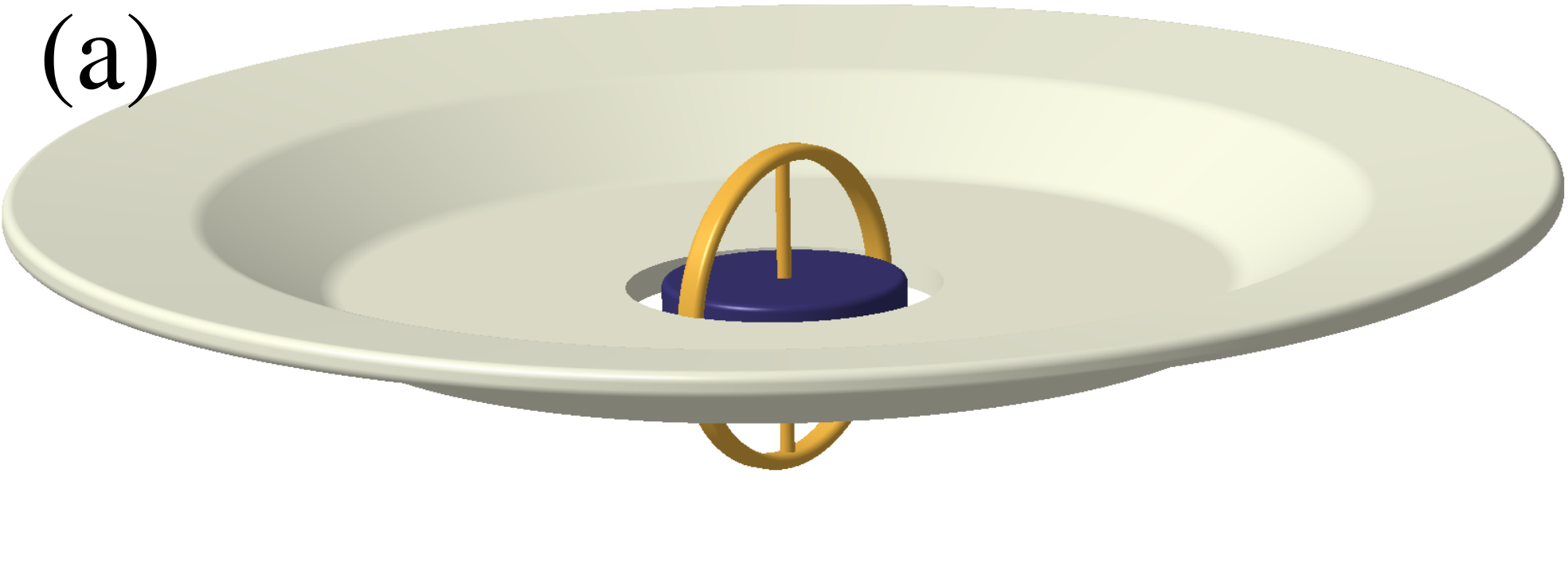,width=0.7\linewidth}}
\centerline{\epsfig{file=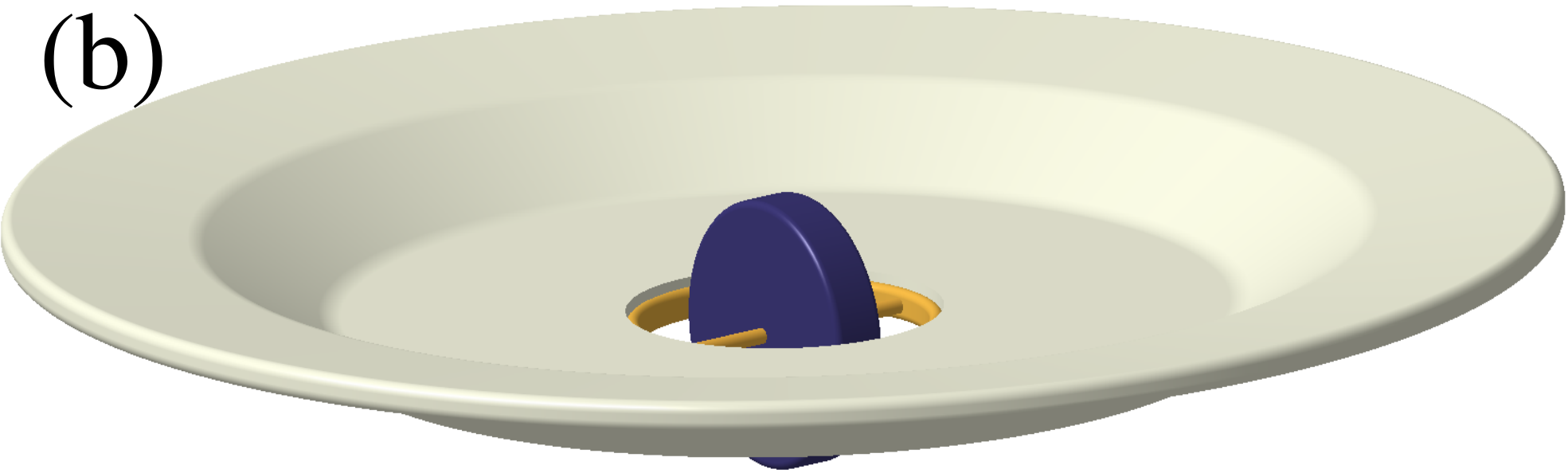,width=0.7\linewidth}}
\caption{\label{f-talir}
Plate as a symmetric top  with a coaxial rotor (a), and with a perpendicular axis rotor (b).
}
\end{figure}

As discussed in Sec. \ref{SubsecQuantumMotion}, the quadratic part of the Hamiltonian corresponds to the diagonal twisting tensor $\chi$,
\begin{eqnarray}
\chi = \left( 
\begin{array}{ccc}
W+V & 0 & 0 \\
0 & W-V & 0 \\
0 & 0 & 0
\end{array}
\right).
\label{chi-LMG}
\end{eqnarray}
Since any multiple of a unit tensor can be added to $\chi$ without changing the dynamics, any diagonal $\chi$ can be expressed in a form equivalent to (\ref{chi-LMG}). In particular, for a diagonal $\chi$ with terms $\chi_{1},\chi_2,\chi_{3}$, by subtracting $\chi_3$ from all the diagonal terms, one gets the LMG parameters $W=(\chi_1+\chi_2)/2 -\chi_3$ and $V=(\chi_1-\chi_2)/2$.
Since for general quadratic Hamiltonians the labeling of principal axes 1,2,3 is arbitrary, any quadratic Hamiltonian with the linear part parallel to one of the principal axes is equivalent to the LMG Hamiltonian (\ref{Ham-LMG2}). Thus,  in the classical analogy, the LMG model corresponds to a freely rotating rigid body supplemented with a rotor with its rotational axis fixed along one of the principal axes of the body as in Fig. \ref{f-setrvacniky1} (a)---(c). 
The special case of $V=0$ corresponds to a symmetrical top with an coaxial rotor as, e.g., in Fig. \ref{f-talir}(a). For $W>0$ the top is a prolate and for $W<0$ oblate. For $V=W$ the LMG model corresponds to a symmetric top with a perpendicular axis rotor as, e.g., in Fig. \ref{f-talir}(b), with $W>0$ referring to oblate and $W<0$ to prolate tops.


\section{Free symmetric top, Feynman's plate, and spin squeezing by one-axis twisting}
\label{Sec-symmetrictop}
As the simplest model, consider a symmetric top with $I_1=I_2 \neq I_3$ with no rotor, i.e., ${K}_k=0$. Equations 
(\ref{Eq-omegadot}) then simplify to
\begin{eqnarray}
\dot{\omega}_1 &=& -\tilde\Omega \omega_2  , \nonumber  \\
\dot{\omega}_2 &=& \tilde\Omega \omega_1  , \nonumber \\
\dot{\omega}_3 &=& 0 ,
\label{Eq-symtop}
\end{eqnarray}
where 
\begin{eqnarray}
\tilde\Omega \equiv \frac{I_3-I_1}{I_1} \omega_3 = \left(\frac{1}{I_1}-\frac{1}{I_3} \right)J_3 .
\end{eqnarray}
In quantum domain this corresponds to Hamiltonian (\ref{Ham123}) reduced to 
\begin{eqnarray}
\hat H_{\rm OAT} = \chi \hat J_3^2 
\label{Eq-OAT}
\end{eqnarray}
with
\begin{eqnarray}
\chi = \frac{1}{2I_1} - \frac{1}{2I_3}.
\label{Eq-chi-I12}
\end{eqnarray}
The subscript ``OAT'' in Eq. (\ref{Eq-OAT}) refers to the ``one-axis twisting'' scenario of spin squeezing  described below.

\subsection{Classical dynamics}
\label{SecClasPlate}
Textbooks show solutions of Eq. (\ref{Eq-symtop}) as regular precession of the top \cite{Symon,Goldstein},
\begin{eqnarray}
\omega_1 &=& A \cos \tilde\Omega t, \nonumber  \\
\omega_2 &=& A \sin \tilde \Omega t,
\label{Eq-topmotion}
\end{eqnarray}
where the amplitude is
$A = \sqrt{\omega_1^2+\omega_2^2} = \sqrt{\omega^2-\omega_3^2}$.
Thus,  in the frame fixed with the body, the axis of rotation circles with frequency $\tilde\Omega$ around the symmetry axis of the top. 
For $A\ll \omega_3$, i.e., for small angles between the rotation axis and the symmetry axis, an external observer
 sees the top wobbling with frequency $\omega_3+\tilde\Omega$. Two extreme cases of the mass distribution in the top correspond to a flat, plate-like top with $I_3 \to 2 I_1$, and a rod-like top with $I_3 \to 0$. The plate-like top has $\tilde\Omega \to \omega_3$ so that the wobbling frequency is $\approx 2\omega_3$, and the rod-like top has $\tilde\Omega\to -\omega_3$ so that the wobbling frequency tends to zero (one can see that when throwing up a spinning pencil).

Feynman in his book  ``Surely, You Are Joking, Mr. Feynman!'' \cite{Feynman-joking} tells a story:
{\it ``\dots I was in the [Cornell] cafeteria and some guy, fooling around, throws a plate in the air. As
the plate went up in the air I saw it wobble, and I noticed the red medallion of Cornell on the plate
going around. It was pretty obvious to me that the medallion went around faster than the wobbling.
I had nothing to do, so I start to figure out the motion of the rotating plate. I discover that when
the angle is very slight, the medallion rotates twice as fast as the wobble rate--two to one. It came out of a complicated equation!''} Feynman was surely joking when telling this story to R. Leighton who collected Feynman's memories, because the situation is just opposite: the wobbling is twice as fast as the rotation. This follows from the above arguments, and was clearly explained in a note  by B. F. Chao \cite{Chao-1989} four years after Feynman's book, as well as in a more detailed study in \cite{Tuleja}.

Consider now the situation when the top is  spun around an axis lying in the symmetry plane, as with coin tossing (see, e.g., \cite{Diaconis}). If  $\omega_3 =0$, then $\tilde\Omega=0$ and the rotational axis keeps its orientation. If the rotational axis is oriented slightly off the symmetry plane, it slowly precesses with a speed proportional to its deviation of the plane. For oblate tops ($I_1 < I_3$) the projection of the precession velocity to the symmetry axis has the same orientation as the projection of the rotational axis, $\tilde\Omega/\omega_3 >0$; for prolate tops  ($I_1 > I_3$) the situation is opposite, $\tilde\Omega/\omega_3 <0$.

\subsection{Quantum dynamics and spin squeezing}
The one-axis twisting (OAT) scenario of spin squeezing corresponding to Hamiltonian  (\ref{Eq-OAT}) was first
proposed theoretically by Kitagawa and Ueda \cite{Kitagawa}. Based on proposals specifying various mechanisms (e.g., \cite{Wineland1994,Duan-2001,Sorensen-Molmer-2002,Takeuchi-2005,Schleier-SmithPRA2010}) it was observed experimentally in  hyperfine states of individual atoms \cite{Chaudhury-2007,Fernholz-2008}, in collective spins of atomic samples interacting by spin-dependent collisions \cite{Orzel2001,Esteve2008,Gross2010,Riedel}, and by optically mediated dispersive interaction in near-resonant cavities \cite{Leroux-2010}. Other proposals for OAT realization include nuclear spins in quantum dots \cite{Rudner-2011,Korkmaz}, phonon-induced interactions of spins in diamond nanostructures \cite{Bennett-2013}, or cold paramagnetic molecules \cite{Bhattacharya-2015}.

For an intuitive picture of OAT spin squeezing, consider $N$ two-level atoms initially prepared in the same spin state: as a whole, the system is in spin coherent state. Collective spin states can be visualized on the Bloch sphere with coordinates  $J_{1,2,3}$ as in Fig. \ref{f-figTACT}(a). The initial spin coherent state is represented by a circle centered on the equator at $(J_1,J_2,J_3) = (J,0,0)$ with  radius $\sim \sqrt{N}/2$ corresponding to the fluctuations of $J_2$ and $J_3$. Points of the circle deviated by $\Delta J_3$ to the north or south from the equator move in the $J_2$ direction with velocity $\approx N\chi \Delta J_3$. Thus, Hamiltonian (\ref{Eq-OAT}) twists the Bloch sphere around axis $J_3$. This squeezes the circle into an ellipse, keeping its area constant: thus, noise in some quantum variable decreases while increasing in another. By a suitable rotation of the spin, one can arrange the decreased noise to occur in the variable used for measurements. 

As shown in  \cite{Opatrny2015}, maximum rate at which squeezing is generated by quadratic Hamiltonians is proportional to the difference between the maximum and minimum eigenvalues of the twisting tensor; for OAT with Hamiltonian (\ref{Eq-OAT}) it is just $\chi$. Note also that for general Hamiltonians a simple formula specifies the maximum rate of squeezing generation in Gaussian states, containing just the second derivatives of the {\em classical\/} Hamiltonian \cite{Opatrny-classqueez}. In a geometric interpretation, the squeezing rate can  be related to the difference of principal curvatures of the energy surface.

\subsection{``Spin squeezing'' in the classical dynamics}

A careful observer could see the OAT effect in the classical motion, as well. Throw up ensemble of plates rotating around axes confined into a narrow circular cone. Let the cone axis be fixed with respect to the plate, lying in the plane of the plate (say, in the direction from the center to the medallion of the Feynman plate). Following the dynamics described in Sec. \ref{SecClasPlate}, the rotational axis drifts with respect to the plate with a rate proportional to the rotational axis deviation from the plane of the plate. Consequently, the circular cone of the ensemble changes into an elliptical one. After some time, the directions of the plate rotation become squeezed from one side and stretched perpendicularly. 


\section{Free asymmetric top, tennis-racket instability, and two-axis countertwisting}
\label{Sec-asymmetrictop}
Assume a rigid body with the principal moments of inertia $I_1<I_3<I_2$, and  ${K}_k=0$. The equations of motion for the angular velocities are 
\begin{eqnarray}
\dot{\omega}_1 &=& \frac{I_2-I_3}{I_1}\omega_2 \omega_3   , \nonumber  \\
\dot{\omega}_2 &=& \frac{I_3-I_1}{I_2}\omega_3 \omega_1   , \nonumber \\
\dot{\omega}_3 &=& \frac{I_1-I_2}{I_3}\omega_1 \omega_2   .
\label{Eq-omegadot-noL}
\end{eqnarray}
In quantum domain the corresponding Hamiltonian can be cast into the form
\begin{eqnarray}
\hat H = \chi_+ \hat J_2^2 -\chi_- \hat J_1^2
\label{Ham-chi-pm}
\end{eqnarray}
with
\begin{eqnarray}
\chi_+ &=& \frac{1}{2I_3} - \frac{1}{2I_2} \nonumber  \\
\chi_- &=& \frac{1}{2I_1} - \frac{1}{2I_3}.
\end{eqnarray}
In the special case of 
\begin{eqnarray}
 I_3 = \frac{2I_1 I_2}{I_1 + I_2}
\end{eqnarray}
the Hamiltonian of (\ref{Ham-chi-pm}) takes the form
\begin{eqnarray}
\hat H_{\rm TACT} = \chi (\hat J_2^2 - \hat J_1^2)
\label{Ham-TACT}
\end{eqnarray}
with
\begin{eqnarray}
\chi  &=& \frac{I_2-I_1}{4I_1 I_2}.
\end{eqnarray}
Hamiltonian (\ref{Ham-TACT}) corresponds to the two-axis countertwisting (TACT) scenario of spin squeezing \cite{Kitagawa}.

\begin{figure}
\centerline{\epsfig{file=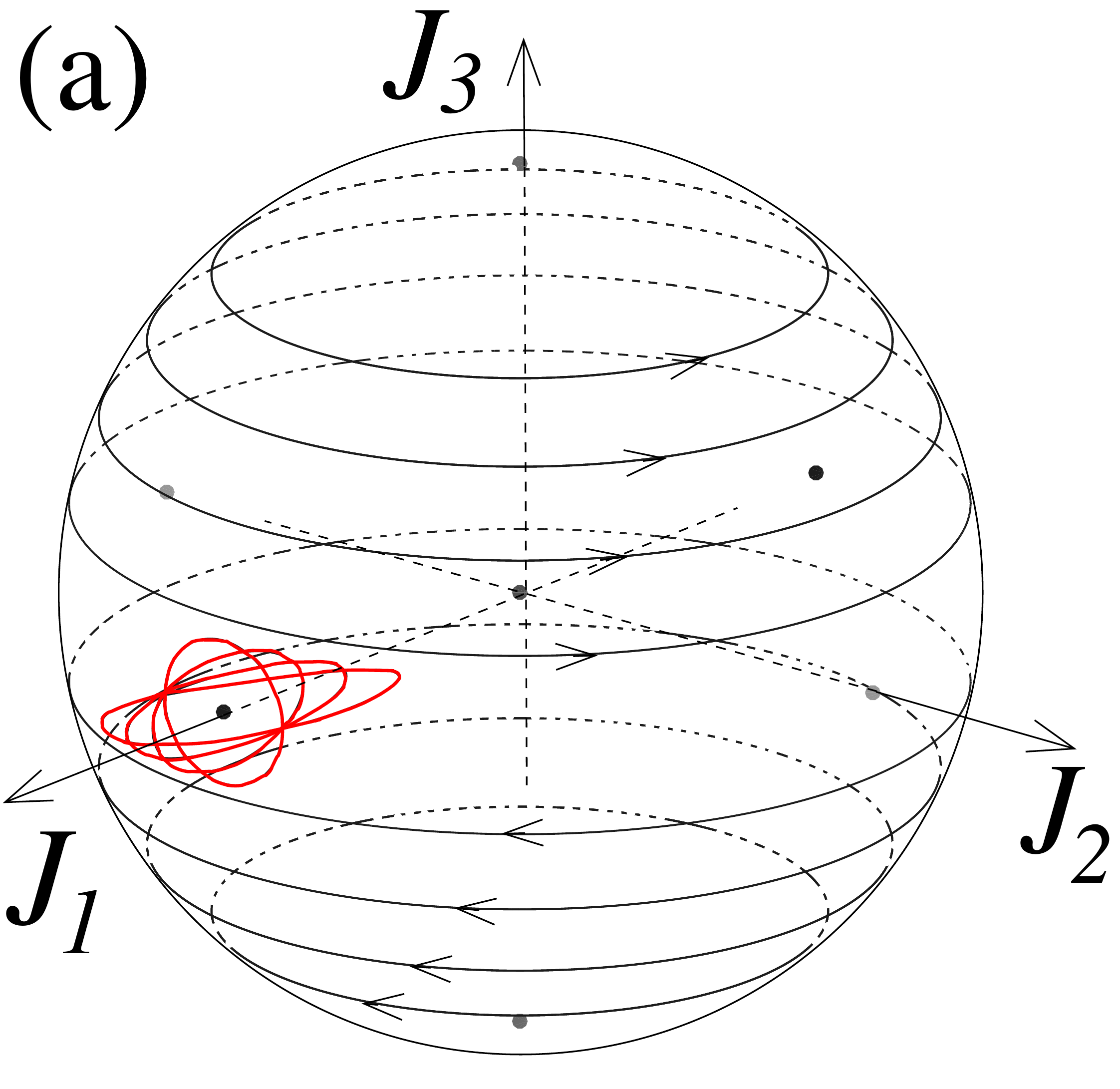,width=0.5\linewidth}
\epsfig{file=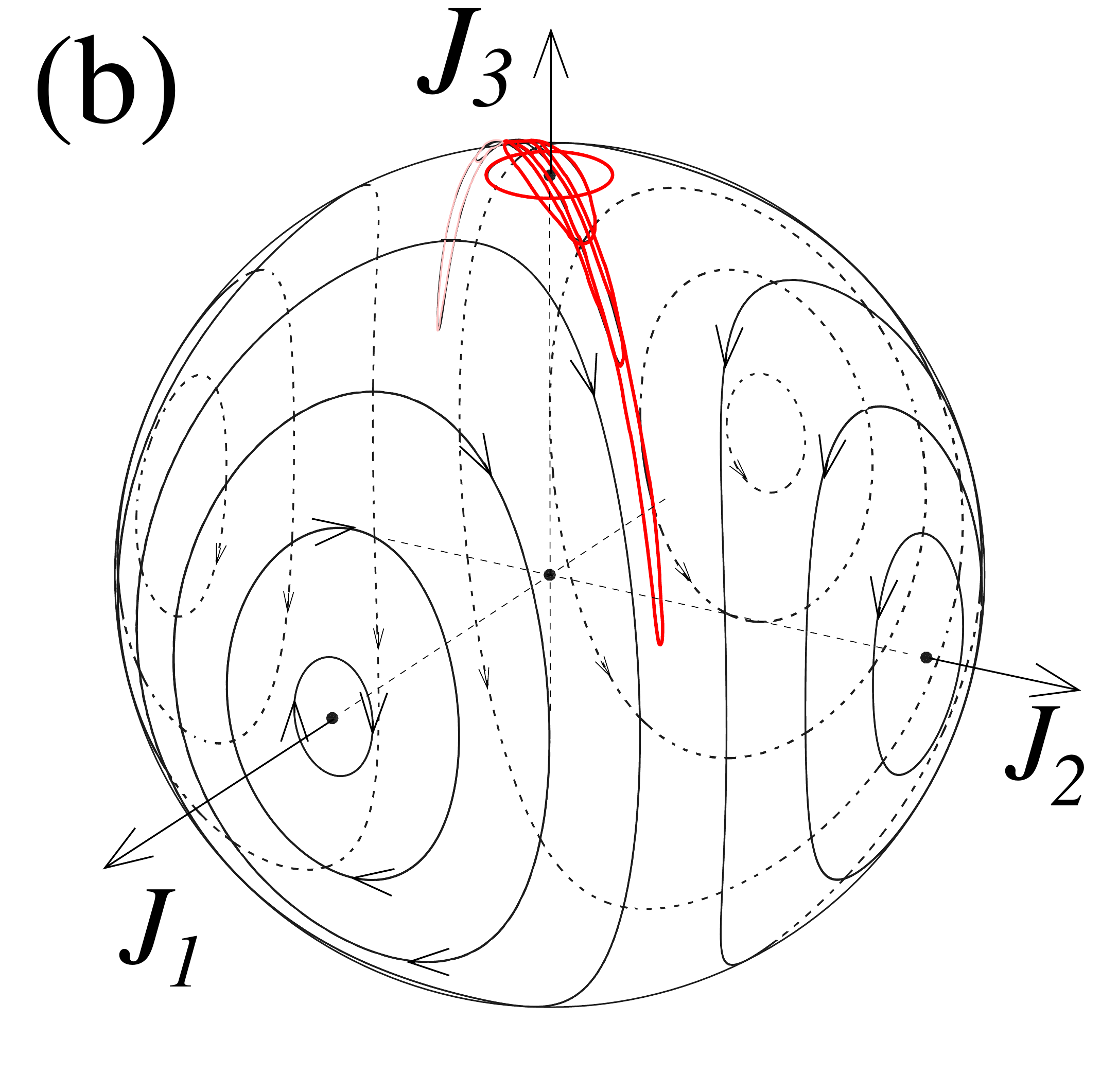,width=0.5\linewidth}}
\caption{\label{f-figTACT}
Evolution of the uncertainty region in spin squeezing scenarios, (a) OAT, (b) TACT.
}
\end{figure}

\subsection{Classical dynamics, Dzhanibekov effect}
As is well known from classical mechanics textbooks (see, e.g., \cite{Symon,Goldstein}), rotations around the two extreme principal axes 1 and 2 are stable, whereas rotation around the intermediate principal axis 3 is unstable. This can be seen by linearizing Eqs. (\ref{Eq-omegadot-noL}) with one  of $\omega_{1,2,3}$ much bigger than the remaining two. One can observe this when throwing up a spinning tennis racket: the rotations are stable if the axis of rotation is along the handle (smallest moment of inertia) or perpendicular to the plane of the head of the racket (biggest moment of inertia), and unstable if the axis of rotation is in the plane of the head of the racket, perpendicular to the handle (intermediate moment of inertia). If the initial angular velocity direction is slightly off the stable axis, the rotation axis precesses around it, but if it is slightly off the unstable axis, it diverges away (motion of vector $\vec{J}$ is shown in Fig. \ref{f-Bloch2}(a)). The dynamics of the tennis racket was studied in detail in \cite{Ashbaugh,Murakami,Rios-2016,VanDamme}. A simple geometric interpretation of  stability of the stationary points 
in terms of the intersecting energy ellipsoid and angular-momentum sphere is discussed in Sec. \ref{Sec-stability}.

Typically on Earth, one cannot observe the free spinning body for a long period. However, the effect is spectacular in zero gravity conditions, provided that the initial angular velocity direction is very close to the unstable axis. As the result, one can see the ``Dzhanibekov effect'' named after Russian cosmonaut Vladimir Dzhanibekov  who observed it while in space in 1985: 
a wing nut rotates while smoothly unscrewing from a screw. When leaving  the screw, the nut continues rotating along an axis that is very close to its unstable principal axis. After several turns, the nut suddenly changes its orientation and continues rotating. The orientation switches then continue in regular time intervals. The  Dzhanibekov effect was studied in detail in \cite{Petrov,Murakami,Rios-2016}. One can understand the motion by realizing than on trajectories that are close to the separatrix, the motion is very slow near the unstable points and relatively fast away from them.  


\subsection{Quantum dynamics and spin squeezing}


Hamiltonian (\ref{Ham-TACT}) was first proposed to produce spin squeezing  in the TACT scenario in \cite{Kitagawa}. The mechanism can be visualized on the Bloch sphere as in Fig. \ref{f-figTACT}(b):  the sphere is twisted in one sense around $J_1$ and in the opposite sense around  $J_2$. Spin states initially polarized along $J_3$  become squeezed as the uncertainty circle is stretched in one direction and compressed in the other. 

The TACT process can generate better squeezing properties than OAT, however, it is much more complicated to be performed with atomic spins than OAT. Therefore, various schemes for achieving effective TACT by applying the OAT twisting Hamiltonian (\ref{Eq-OAT}) and spin rotations have been proposed \cite{Liu2011,Shen-Duan-2013,Zhang2014,Huang-2015,Wu-2015}. Possible physical realizations of TACT were suggested for collective spins based on atomic interactions induced by coherent Raman processes through  molecular intermediate states \cite{Helmerson,Micheli2003}, for individual atomic spins by inducing nuclear-electronic spin interaction \cite{Fernholz-2008}, for Bose-Einstein condensate with spatially modulated nonlinearity \cite{OKD2014}, for optical fields in resonators with Kerr media \cite{Opatrny2015}, nuclear spins via electric quadrupole interaction \cite{Korkmaz},  dipolar spinor condensates \cite{Kajtoch}, or for  samples of multilevel atoms interacting with near-resonant cavities \cite{Yu-2016,Zhang-2017,Borregaard-2017}.

Note that studies of a quantum mechanical asymmetric top go back to the early days of quantum theory \cite{Witmer,Wang,King1943,King,Winter,Lukac,Pan1999,Manfredi}, however, their goal was finding the Hamiltonian spectrum rather than the squeezing dynamics. Even though recently exact diagonalization of the TACT Hamiltonian was studied \cite{Bhattacharya,Pan2017}, there was no discussion about the connection to the quantum  asymmetric top.
We also note that recently an analogy between the tennis racket motion and a driven two-level system was identified, relevant to spin control in nuclear magnetic resonance \cite{VanDamme-2017}.



\section{Symmetric top with a coaxial rotor, spin twisting with coaxial rotation}
\label{SecCoaxial}

\subsection{Quantum dynamics}
\label{CoaxialQantum}
Let us start this section with the quantum case.
Using Eq. (\ref{Ham-LMG2}) with $V=0$ is equivalent to using Hamiltonian (\ref{Ham-nlin}) in the form
\begin{eqnarray}
\hat H = \chi \hat J_3^2 + \Omega \hat J_3
\label{OATcoaxial}
\end{eqnarray}
with $\Omega = \epsilon$ and $\chi=-W$. This corresponds to twisting around axis $J_3$ and simultaneous rotation around the same axis.
Since the Hamiltonian is a function of $\hat J_3$, its eigenfunctions are those of $\hat J_3$, i.e., Dicke states with sharp values of $J_3$.

\begin{figure}
\centerline{\epsfig{file=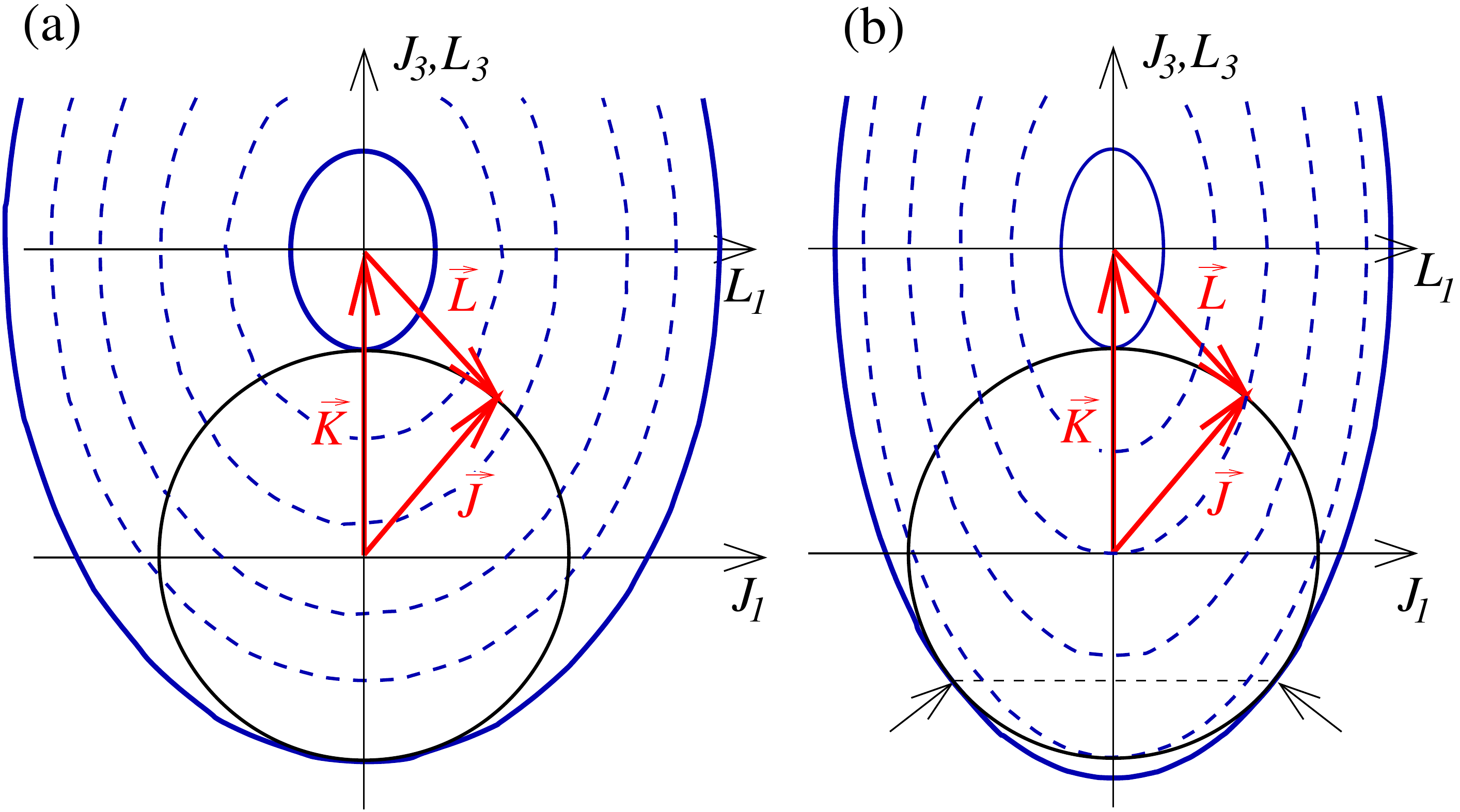,width=1\linewidth}}
\caption{\label{f-elipsy}
Constant angular momentum sphere (black) and constant energy ellipsoids (blue) corresponding to a symmetric top with a coaxial rotor.
The red vectors $\vec{J}$ and $\vec{L}$ refer to a generic point on the sphere. The maximum and minimum energy ellipsoids compatible with the given value of $J$ are plotted in full line, several other ellipsoids corresponding to intermediate energies are in dashed line. In case (a) the sphere and each of the extreme energy ellipsoids touch at a single point,  in case (b) the sphere and the maximum energy ellipsoid touch along a circle (indicated by a dashed line and short arrows).
}
\end{figure}

The  dynamics are split into two possible phases: (a) dominant rotation with  $|\chi| N <|\Omega|$, and  (b) dominant nonlinearity with $|\chi| N >|\Omega|$. In  case (a) the eigenstates corresponding to the extreme eigenvalues of $\hat H$ coincide with the eigenstates corresponding to the extreme eigenvalues of $\hat J_3$. In case (b) either the ground or the highest excited state of $\hat H$ is one of the intermediate eigenstates of $\hat J_3$. An exception in regime (b) occurs if $\Omega/\chi + N$ is an odd integer; then the ground (or the highest energy) state is degenerate, composed of two nearest Dicke states. 

In case (b), the eigenvalue of $J_3$ corresponding to the extreme energy state is $J_3 = {\rm round}[-\Omega/(2\chi)]$ for even $N$ and  $J_3 = {\rm round}[\frac{1}{2}-\Omega/(2\chi)]-\frac{1}{2}$ for odd $N$, where round$[\dots]$ means rounding to the nearest integer.  
This suggests a way for preparation of arbitrary Dicke states: initialize the system in a spin coherent state which is the ground state of some $\hat{\vec{J}}$, and then switch adiabatically the Hamiltonian from $\propto \hat{\vec{J}}$ to $\hat H$ of Eq. (\ref{OATcoaxial}) with a suitably chosen $\Omega$. If the change is sufficiently slow, the ground state follows the instantaneous Hamiltonian and the system ends up in the chosen Dicke state. The problem is that in the transition from  $\hat{\vec{J}}$ to $\hat H$ of (\ref{OATcoaxial}) some gaps in the energy spectrum close, so that the change would have to be infinitely slow to remain adiabatic. Methods of counterdiabatic driving to overcome this problems have been proposed in \cite{TO-Counterdiabatic}.


Spin coherent states localized on the Bloch sphere along the parallel $J_3  = -\Omega/(2\chi)$ in case (b) behave similarly as spin coherent states localized along the equator in  OAT: the center of the uncertainty area does not move, but the uncertainty circle is deformed into an ellipse and the state become squeezed (squeezing properties of Hamiltonian (\ref{OATcoaxial}) were  studied in \cite{Jin-2009}).

\subsection{Classical dynamics}
The classical model corresponds to a symmetric top, $I_1=I_2\neq I_3$, with a coaxial rotor, $K_1=K_2=0\neq K_3 \equiv K$, such as in Fig.  \ref{f-talir}(a). 
 The  equations of motion  and their solution have the same form as 
those of a free symmetric top, Eqs. (\ref{Eq-symtop}) and (\ref{Eq-topmotion}), but the precession frequency is changed to
\begin{eqnarray}
\tilde\Omega = \frac{(I_3-I_1)\omega_3 + {K}}{I_1} 
= \left(\frac{1}{I_1}-\frac{1}{I_3} \right) J_3 + \frac{K}{I_3} .
\label{Omegasymtop}
\end{eqnarray}

Similarly to the quantum case, the dynamics are split into two regimes with dominant $|{K}|/J > |1-I_3/I_1|$ (a), and  $|{K}|/J < |1-I_3/I_1|$ (b).
The critical parameter is the ratio between the angular momentum  ${K}$ of the rotor and the magnitude of the angular momentum $J$ of the combined system. This can be shown either using the results of 
the preceding subsection and the correspondence of quantum and classical models, or by the following geometric picture.
Consider contact points of the energy ellipsoid and angular momentum sphere as in Fig. \ref{f-elipsy}. In case (a), the lowest energy ellipsoid touches the total angular momentum sphere from outside and the highest energy ellipsoid is touched from inside, both extremal points of contact being on the $J_3$ axis. In this regime, the only direction of a rotational axis not moving with respect to the body is along $J_3$. On the other hand, in case (b) the energy ellipsoid and the total angular momentum sphere touch in a circle that corresponds either to the maximum (plate-like top) or minimum (rod-like top) energy with a given angular momentum. When spun around an axis oriented in that direction, the rotational axis does not move with respect to the body.

As an example, consider a flat symmetric top with $I_1=I_2=I_3/2$ as in Fig. \ref{f-talir}(a). 
(The reader can experiment  with a simple realization by gluing a fidget-spinner to a light plate.) 
In this case
Eq. (\ref{Omegasymtop})  yields $\tilde \Omega = \omega_3 + 2{K}/I_3$. Choosing ${K}=-\frac{1}{2}I_3 \omega_3$ leads to $\tilde \Omega = 0$, which for $|\omega_{1,2}|\ll |\omega_3|$ means the system is near the boundary between regimes (a) and (b). As a result, a rotational axis close to the symmetry axis of the body keeps its position with respect to the body, and the wobble frequency is equal to the rotation frequency, $\omega_3 + \tilde \Omega = \omega_3$. 

As another example choose   ${K}=-\frac{3}{4}I_3 \omega_3$. This leads to $\tilde \Omega = -\omega_3/2$ so that the wobble frequency of a plate is half the rotation frequency, $\omega_3 + \tilde \Omega = \omega_3/2$. 
Thus, with a little cheating of adding a properly spinning rotor, one can force a plate to behave exactly as described in Feynman's cafeteria story \cite{Feynman-joking}.

\begin{figure}
\centerline{\epsfig{file=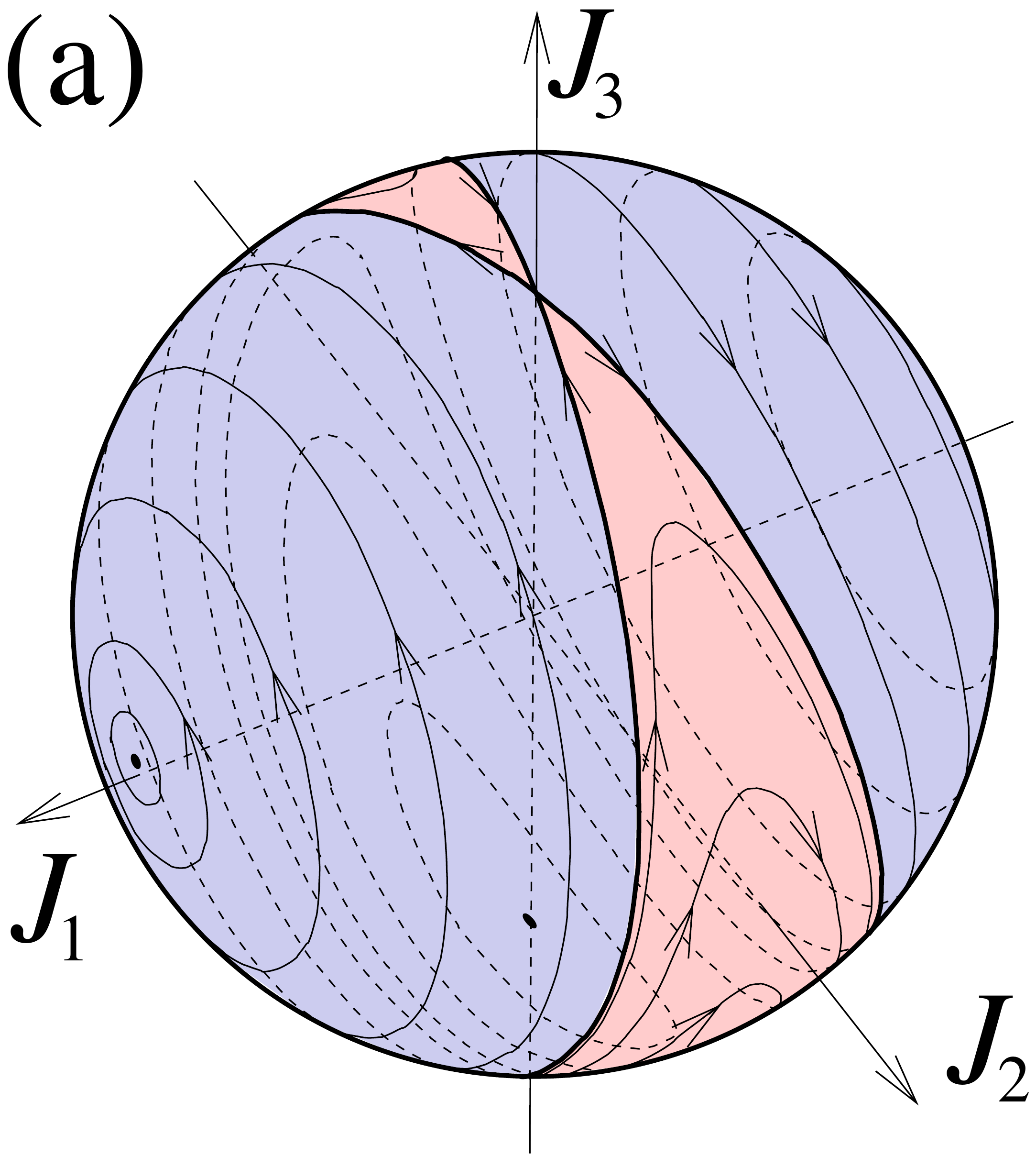,width=0.5\linewidth}
\epsfig{file=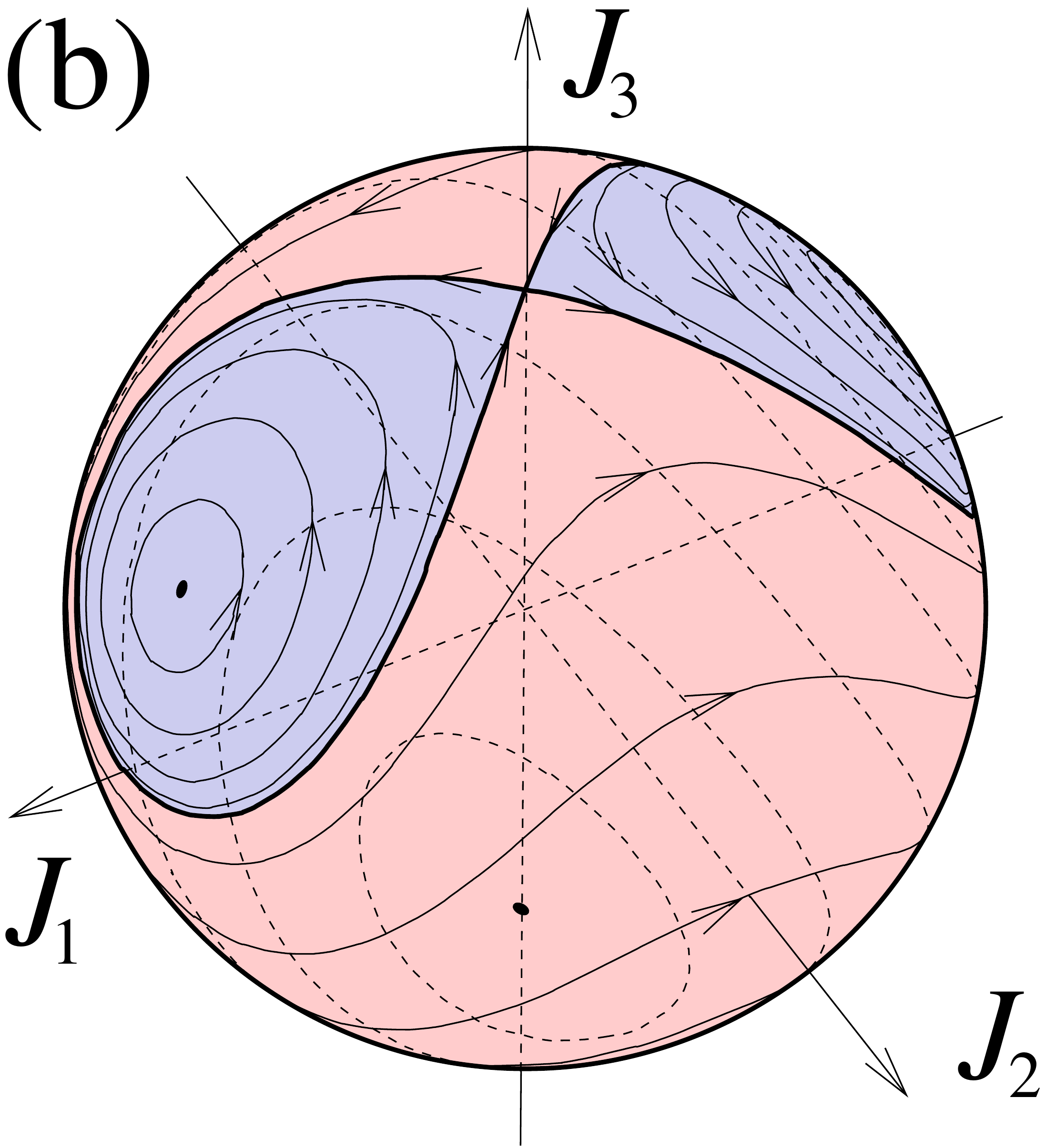,width=0.5\linewidth}}
\centerline{\epsfig{file=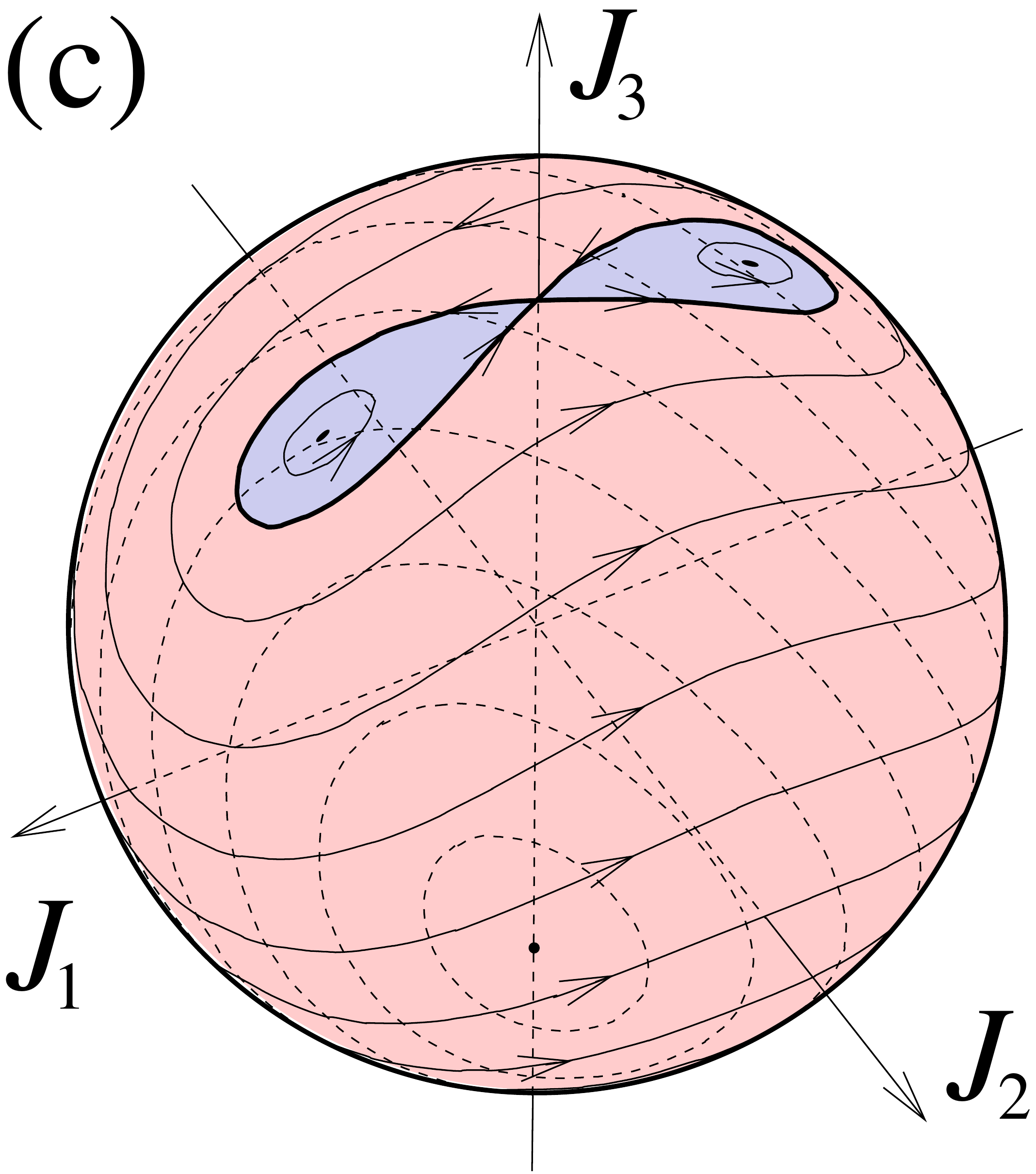,width=0.5\linewidth}
\epsfig{file=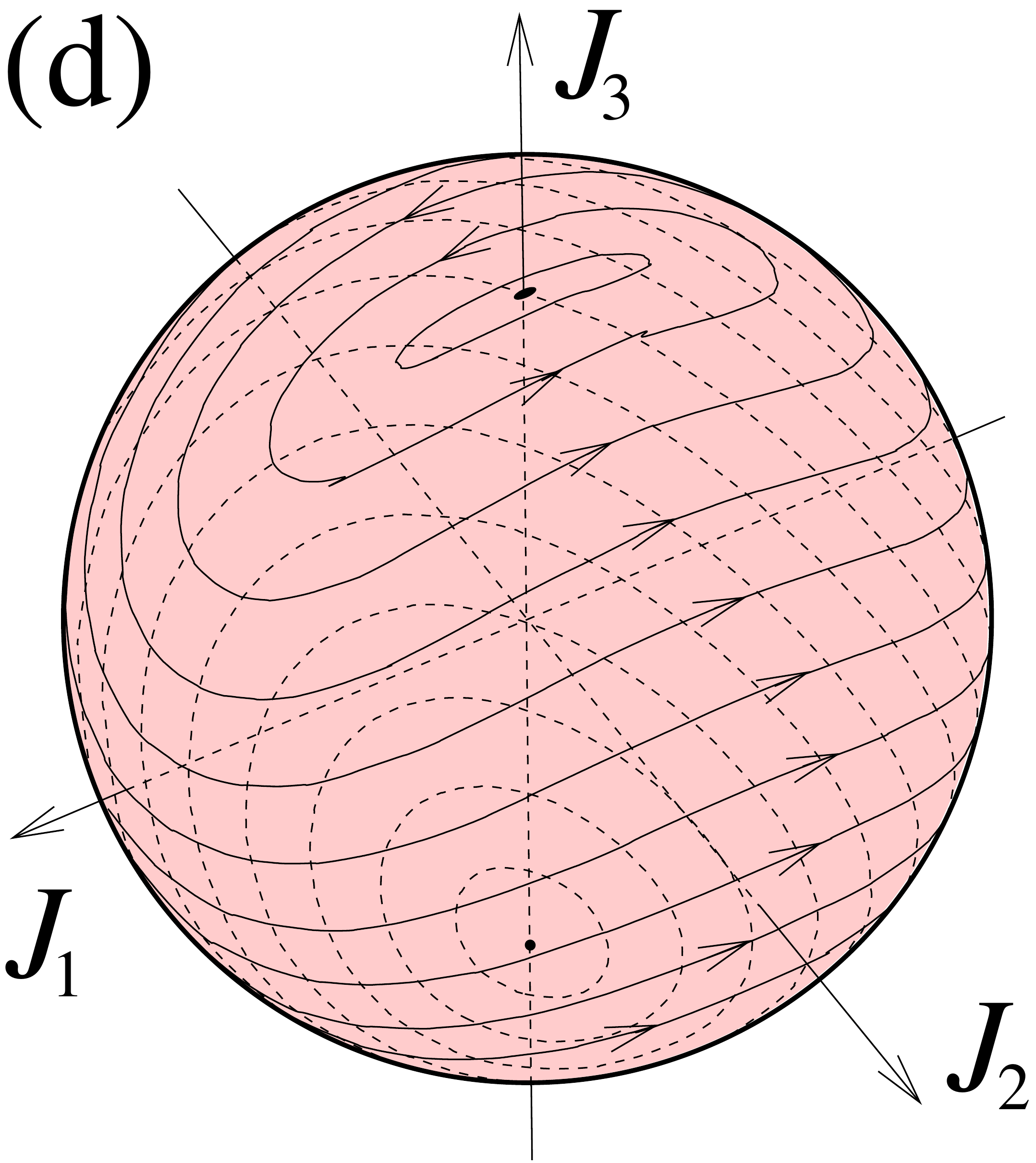,width=0.5\linewidth}}
\caption{\label{f-Bloch1}
Angular momentum trajectories for a symmetric top with a perpendicular axis rotor, or a twist-and-turn Hamiltonian (\ref{Ham-twistandturn}). The parameters are $\Omega/(\chi J)=0.2$ (a), 1 (b), 1.7 (c), and 2 (d). Panels (a)-(c) correspond to the Josephson regime with the blue area representing ``self-trapped'' states. Panel (d) corresponds to the boundary between the  Josephson and Rabi regimes where one unstable and two stable stationary points merge, leaving behind one stable stationary point for  $\Omega/(\chi J)>2$.
}
\end{figure}

\section{Symmetric top with a perpendicular axis rotor, twist-and-turn Hamiltonian}
\label{SecPerpendicularRotor}

\subsection{Quantum dynamics}

Let the  Hamiltonian have the form
\begin{eqnarray}
\hat H = \chi \hat J_1^2 + \Omega \hat J_3.
\label{Ham-twistandturn}
\end{eqnarray}
This is equivalent to the LMG Hamiltonian  (\ref{Ham-LMG2}) with $V=W$, $\Omega = \epsilon$ and $\chi=2V$. 
The corresponding evolution is twisting around  axis  $J_1$ and simultaneous rotation about the perpendicular axis $J_3$ (also called ``twist-and-turn'' dynamics \cite{Muessel}). 

Assume now for simplicity $N\gg 1$ so that $J \approx N/2$.
Similarly as for parallel rotation, there are two distinct regimes: that with dominant rotation $|\Omega| > 2J|\chi|$, and that with dominant twisting $|\Omega| < 2J|\chi|$ (see Fig. \ref{f-Bloch1}). 
This follows from a similar geometric consideration as in the preceding section, or from a general treatment given in Sec. \ref{SecPhase}. 
In the regime of dominant rotation, the  Hamiltonian is nondegenerate, with a single maximum and a single minimum on the Bloch sphere. For  $2J\chi=\pm \Omega$ a quantum phase transition occurs with the energy maximum or minimum being split into two, so that in the regime of dominant twisting a saddle point on the Bloch sphere occurs.

The dynamics due to Hamiltonian (\ref{Ham-twistandturn}) was studied in \cite{Smerzi} as  coherent atomic tunneling between two zero-temperature Bose-Einstein condensates confined in a double-well trap, and in \cite{Micheli2003} as evolution of a two-component condensate.
The linear term $\propto \Omega$ corresponds to tunneling of the atoms between the two wells \cite{Smerzi} or to Rabi oscillations between the internal states \cite{Micheli2003}, and the nonlinear term $\propto \chi$ refers to the mutual scattering of the atoms. Circling around a single minimum or maximum energy on the Bloch sphere correspond to the oscillation of the condensate between the two wells, whereas trajectories around one of the two local extrema correspond to {\em self-trapping\/} of the condensate in one of the wells. 
Another proposed realization of  Hamiltonian (\ref{Ham-twistandturn})  is a Bose-Einstein condensate in a ring trap with an optical lattice \cite{Kolar-2015}:  two counterpropagating modes are coupled by a periodic potential that changes the propagation direction of the atoms by Bragg reflection (linear term $\propto \Omega$). Sufficiently strong interaction of the atoms (nonlinear term $\propto \chi$) can keep them self-trapped in one of the rotational modes.

In two-state Bose-Einstein condensates, the limiting case of the linear regime with $|\Omega/\chi| \gg N$ has been  been dubbed ``Rabi regime'' whereas  the limiting case of the nonlinear regime with $|\Omega/\chi| \ll 1/N$ the ``Fock regime'', the transition regime with $1/N \ll |\Omega/\chi|\ll N$ being called ``Josephson regime'' \cite{Leggett-2001}. As shown in \cite{Pezze}, these three regimes correspond to different scaling rules for the dependence of the interferometric phase sensitivity  on the atomic number.
Experimental observation of transitions between the Josephson and Rabi dynamics in spins of a rubidium Bose-Einstein condensate was reported in \cite{Zibold2010}. 

Hamiltonian (\ref{Ham-twistandturn}) with suitably chosen ratio of the linear and nonlinear terms has been shown to be especially useful for producing spin squeezed states \cite{Muessel,Opatrny2015,Opatrny-classqueez}. Note that the nonlinear term  alone as in the OAT scheme tends to align the uncertainty area along the equator (see Fig. \ref{f-figTACT}(a)) so that the squeezing process becomes gradually less efficient. If a linear term with  $\Omega = \chi J$ is added, then the uncertainty ellipse is kept inclined by $\pi/4$ from the equator (see the trajectories in Fig. \ref{f-Bloch1}(b)) which is the optimum orientation for achieving maximum squeezing rate \cite{Opatrny2015}.  

\subsection{Classical dynamics}

The rigid body dynamics corresponds to a symmetrical top with a perpendicular rotor  such as in Fig. \ref{f-talir}(b). The  ``Rabi oscillations'' occur in the case with dominant rotor angular momentum, $|{K}_3|/J > |1-I_3/I_1|$: the rotational axis of the top circles around the axis of the rotor. If the axis of the body rotation is along the rotor axis, its direction is fixed and stable for both co-rotational and counter-rotational orientations.  

In the ``self trapping'' or ``Josephson'' regime, the angular momentum of the top is dominant, $|{K}_3|/J < |1-I_3/I_1|$. In this case,  the counter-rotation becomes unstable: the direction opposite to the rotation of the rotor becomes located on a separatrix dividing the $4\pi$ sphere of rotational axis orientations into three regions (see Fig. \ref{f-Bloch1} for visualization and Sec. \ref{Sec-stability} for more details on stability). In one region the rotational axis circles around the direction of the rotor, in the two other regions the rotational axis circles around a direction pointing between the rotor axis and the symmetry axis of the top.  Imagine the Feynman plate supplemented with a perpendicular axis rotor: if spun around a suitably chosen axis, the rotational axis becomes ``self-trapped'' with respect to the plate. The plate rotates stably around an axis that is at an angle relative to the plate axis---something that one would never see with common cafeteria plates. 

Note that there is no classical analogy of the Fock regime. When translating the condition  $|\Omega/\chi| \ll 1/N$ into the Euler-top language with correct dimensionality, one finds $|{K}_3| \ll \hbar |1-I_3/I_1|$.

\begin{figure}
\centerline{\epsfig{file=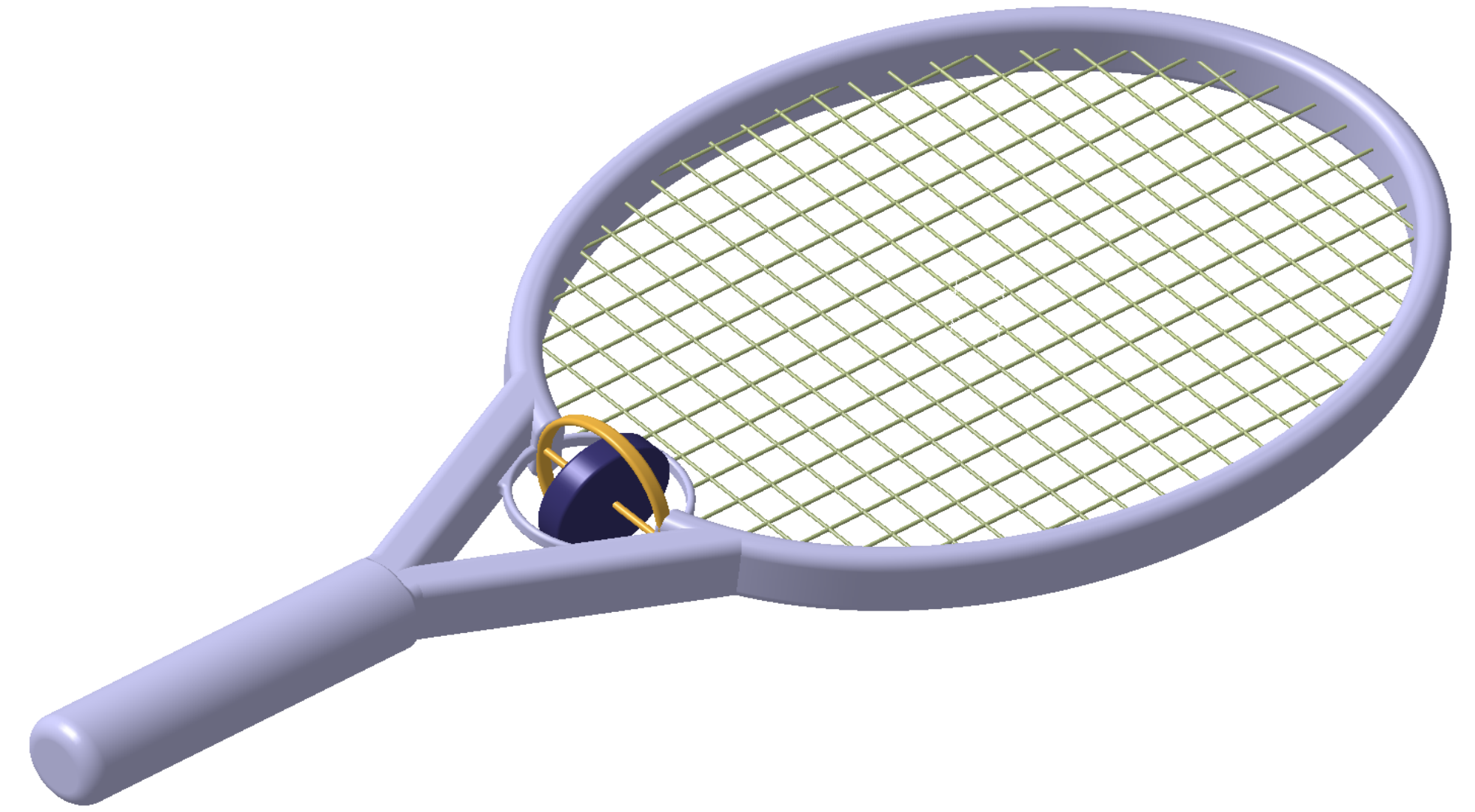,width=0.9\linewidth}}
\caption{\label{f-raketa1}
Tennis racket  with a rotor  along the middle principal axis.
}
\end{figure}

\begin{figure}
\centerline{\epsfig{file=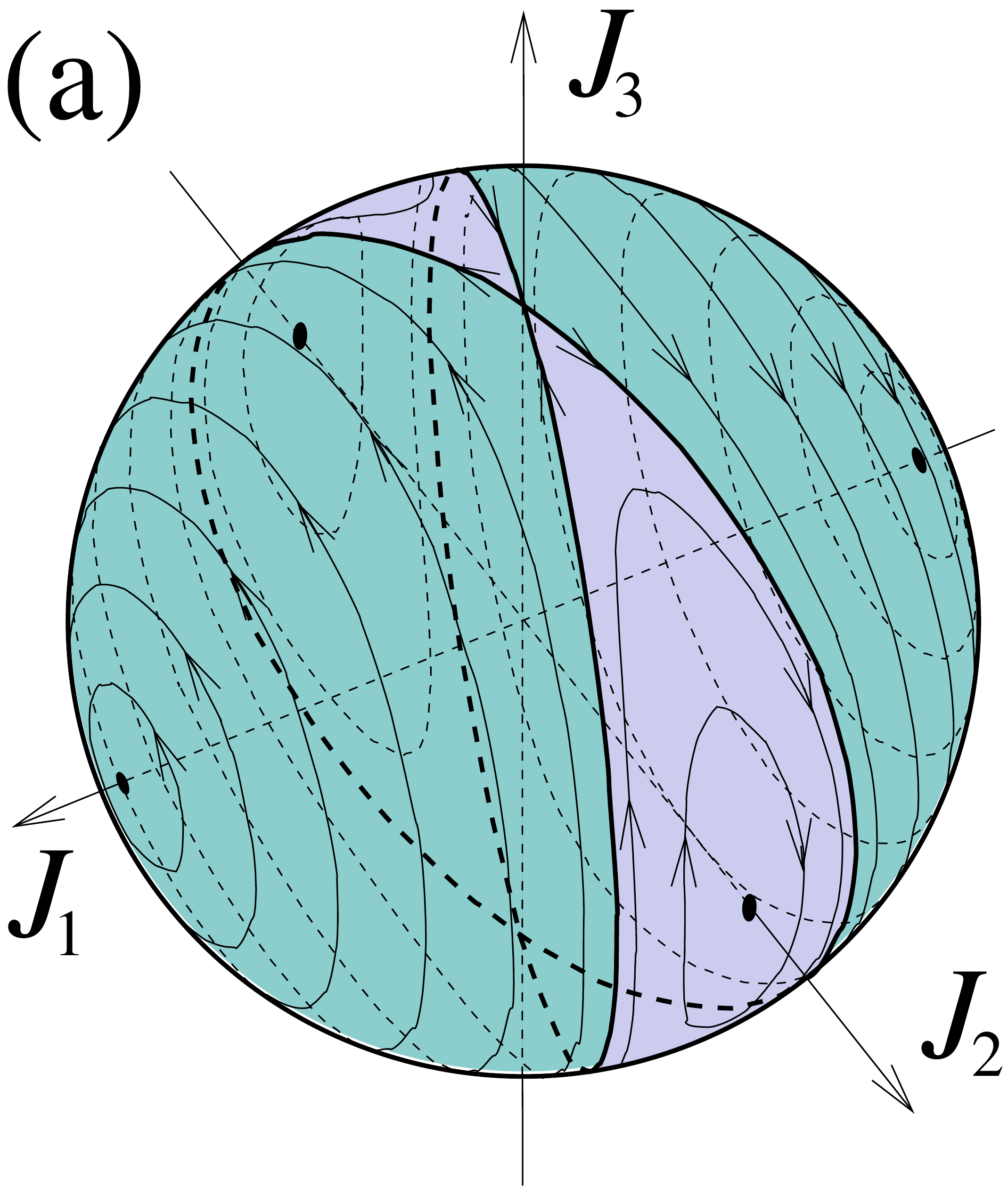,width=0.5\linewidth}
\epsfig{file=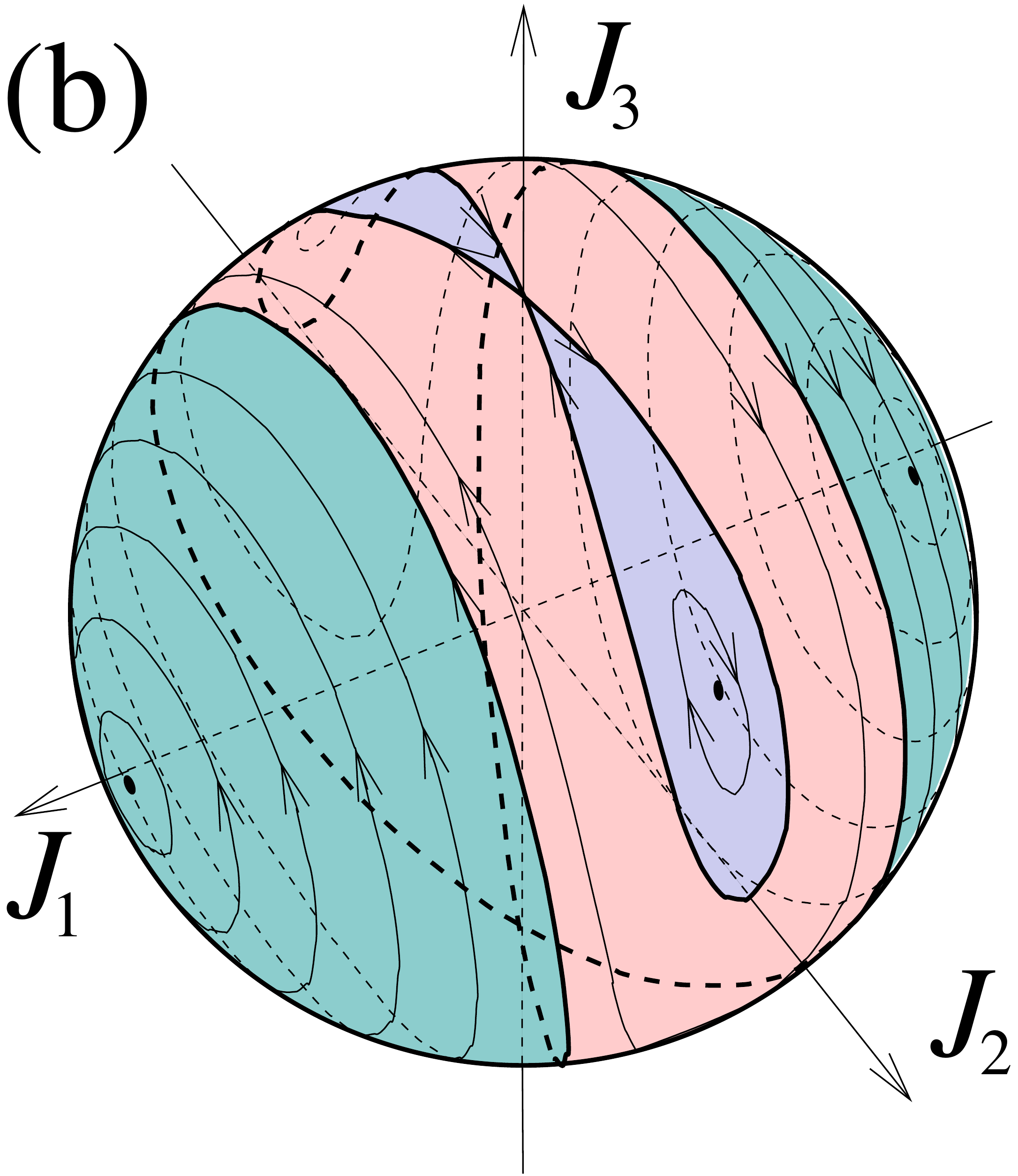,width=0.5\linewidth}}
\centerline{\epsfig{file=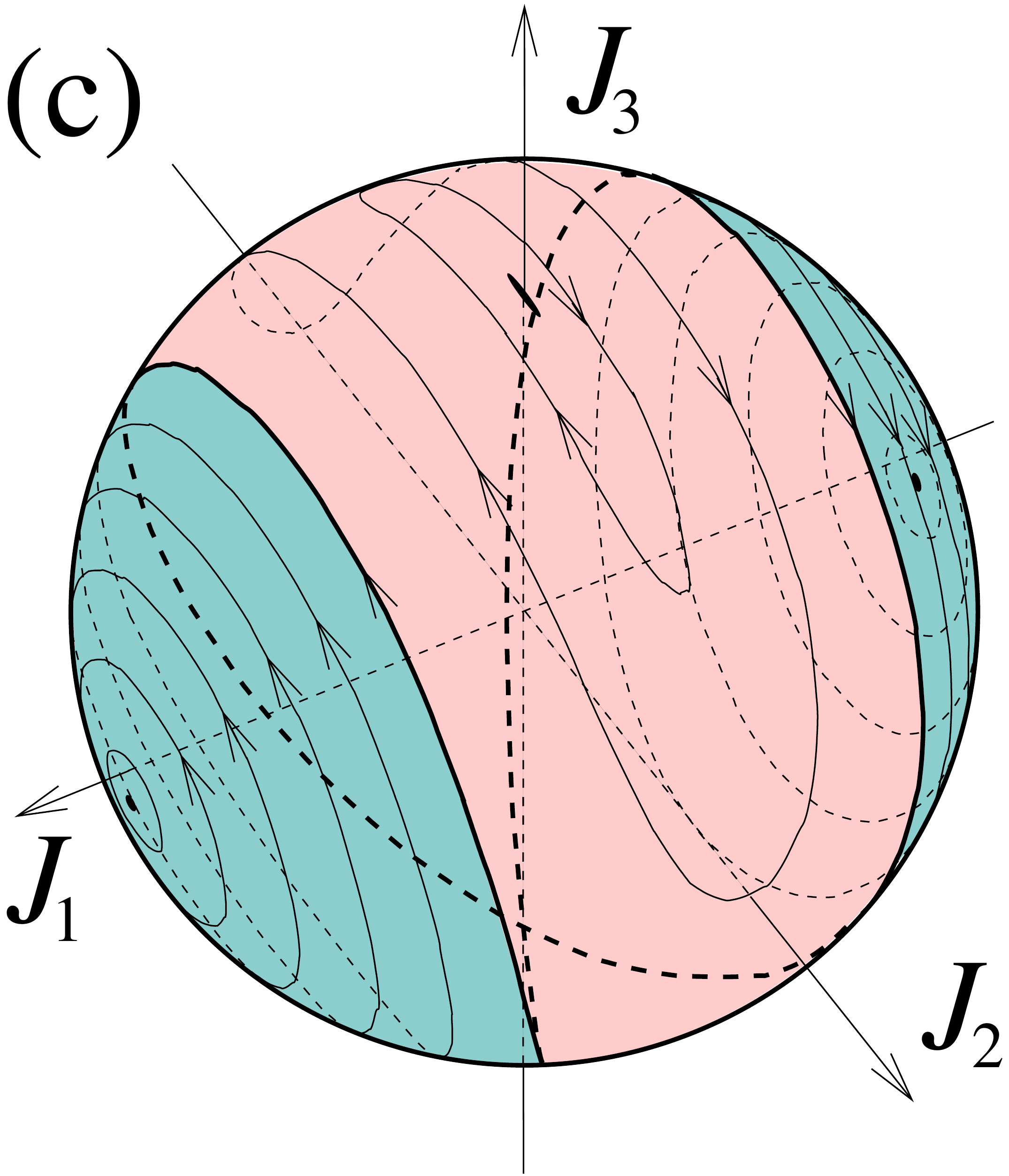,width=0.5\linewidth}
\epsfig{file=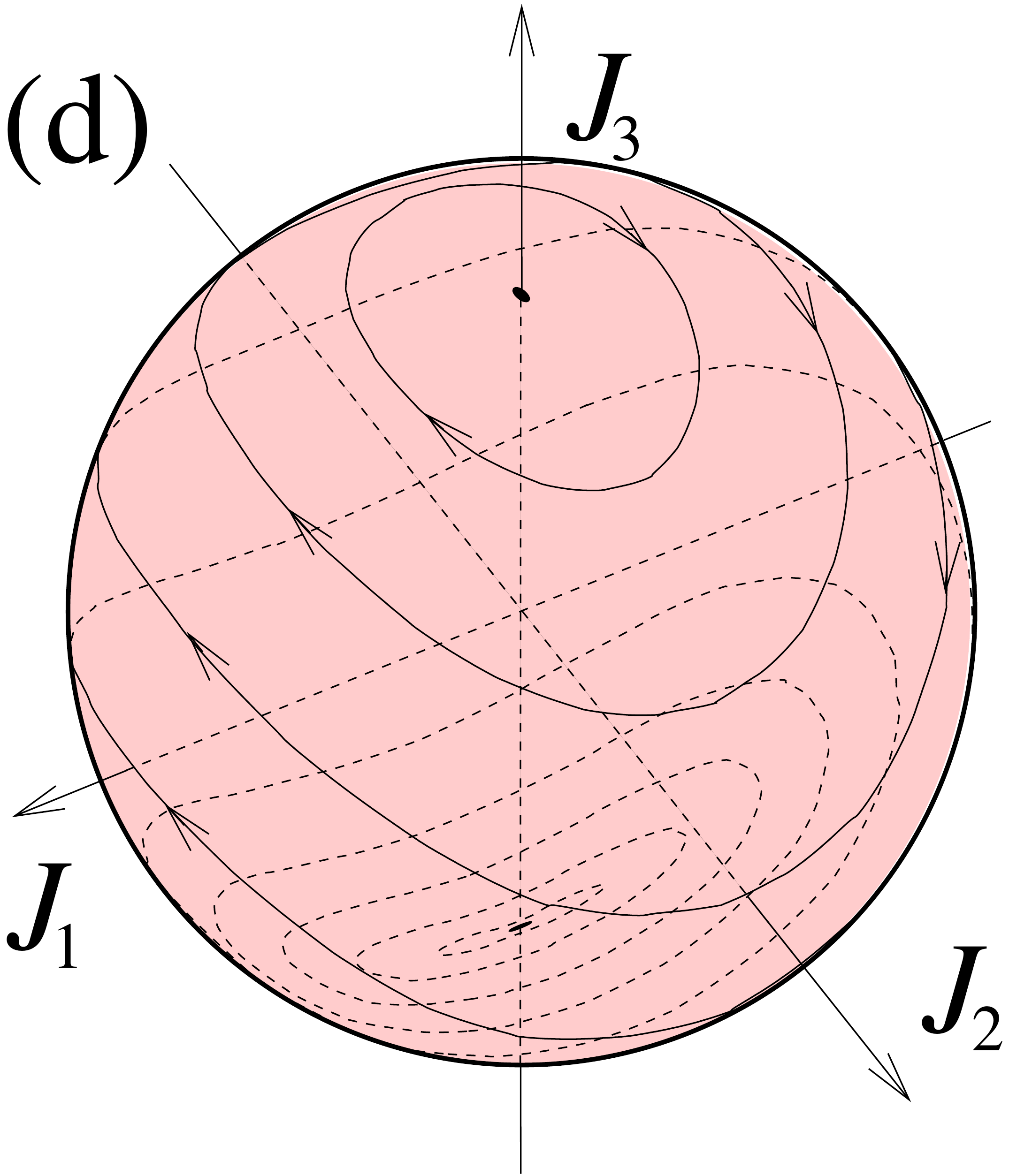,width=0.5\linewidth}}
\caption{\label{f-Bloch2}
Angular momentum trajectories for an  asymmetric top with a rotor along the middle axis. The twisting parameters of the corresponding quantum Hamiltonian are $\chi_3=0$, and $\chi_1 = -10 \chi_2$.
(a)  $\Omega_3=0$ (i.e., no rotor), (b) $|\Omega_3|=1.7 J|\chi_2|$, (c) $|\Omega_3|=2 J|\chi_2|$ (critical value for disappearing the saddle point along $+J_3$), and (d) $|\Omega_3|=2 J|\chi_1|$ (critical value for disappearing the saddle point along $-J_3$).
}
\end{figure}

\section{Asymmetric top with a principal axis rotor, general LMG and quantum phase transitions}
\label{AsymmetricPrincipal}

\subsection{Quantum dynamics}

Features of the general Hamiltonian (\ref{Ham-LMG2}) have been widely explored, especially with focus on quantum phase transitions and related critical phenomena \cite{Kwok-Lin,Zibold2010, Vidal2004,Castanos,Engelhardt,Gallemi,Campbell-2016,Leyvraz,Ribeiro2007,Ribeiro,Hooley}. 
The concept of
quantum phase transition typically refers to closing the gap between the ground and the first excited state by varying a system parameter \cite{Sachdev}. In contrast to thermal phase transitions where many states are involved and features of the system are suddenly changed by varying temperature, quantum phase transitions can happen at zero temperature. Rather than thermal,  the relevant fluctuations are of quantum nature. Recently the concept has been generalized to excited state quantum phase transitions (ESQPT) \cite{Caprio,Cejnar}. In ESQPT, the variation of parameters leads to sudden emergence of singularities in the energy spectrum: in a smooth density of energy levels a peak or a discontinuity occurs. These effects can be related to the Hamiltonian map on the Bloch sphere (see Fig. \ref{f-Bloch2}): a discontinuity in the energy spectrum corresponds to a local minimum or maximum of energy on the sphere, and a peak in the energy spectrum corresponds to a saddle point of energy. 


\subsection{Classical dynamics}

The classical model corresponds to an asymmetric top with a rotor whose axis is aligned with one of the principal axes of the top. 
Relevant problems include rotational stabilization of rigid bodies \cite{Krishnaprasad,Bloch-1992,Casu} with applications to attitude control of spacecraft by momentum wheels. Recently, such a model has been used to analyze motion of a diver exerting a twisted somersault \cite{Bharadwaj}: the body of the diver is modeled by an  asymmetric top  and the moving arms by a rotating disc.

As an example of the ESQPT analogy, let us consider stabilization of rotation of a tennis racket around the middle principal axis by a rotor as in Fig. \ref{f-raketa1}. The transition is visualized using the Bloch sphere in Fig.  \ref{f-Bloch2}.
With no rotor (Fig.  \ref{f-Bloch2}(a)), the sphere consists of two pairs of ``self-trapped'' regions where motion of the angular momenta encircle the stable directions $\pm J_1$ and $\pm J_2$. These regions are separated by a line called separatrix,  going through the unstable stationary angular momenta $\pm J_3$.

Adding the rotor with some small angular momentum $K_3$, the separatrix splits into two  (Fig.  \ref{f-Bloch2}(b)). A new region between the separatrices emerges as a stripe of trajectories encircling the sphere. With increasing $|K_3|$,  the stripe becomes wider and the stable fixed points move towards the unstable points. With a critical value of  $|K_3|$, one pair of  stable points  merge with one unstable point, resulting in a stable point (Fig.  \ref{f-Bloch2}(c)). This is a new phase in which the racket co-rotating with the rotor around the intermediate principal axis becomes stable, although counter-rotation is still unstable. 

With further increasing   $|K_3|$, the remaining pair of stable points approach the unstable point till they merge (Fig.  \ref{f-Bloch2}(d)). For  $|K_3|$ above this second critical value the system is in phase with only two stationary angular momenta, both stable.


\section{Stationary angular momenta and their stability in generalized LMG}
\label{SecPhase}


Even though the possibility to generalize LMG to arbitrary directions of the linear term was briefly mentioned in \cite{Vidal2006}, we are not aware of any systematic study of such a model. We consider here such a generalization given by Hamiltonian (\ref{Ham123}) and present a simple geometric approach to find the occurrence of stationary points in the angular momentum space and determine their stability.
The results are used to find new cases of ESQPT.

\subsection{Stationary values of the angular momentum}
In the angular momentum space, stationary values correspond to the points where the constant energy ellipsoid touches the constant total-angular-momentum sphere. In the classical model, this occurs where the gradient of energy (\ref{eqEnergie}) is colinear with the gradient of the squared total momentum (\ref{eqLsquare}), 
\begin{eqnarray}
{\rm grad}\ E_{\rm body} = \lambda \ {\rm grad} \ {J^2} 
\end{eqnarray}
for some $\lambda$.
This leads to the relation between the angular momentum components
\begin{eqnarray}
 \label{eqL1}
 J_1 &=& \frac{I_3 {K}_1 J_3}{(I_3-I_1)J_3+I_1 {K}_3}, \\
 J_2 &=& \frac{I_3 {K}_2 J_3}{(I_3-I_2)J_3+I_2 {K}_3},
 \label{eqL2}
\end{eqnarray}
which, when used in Eq.  (\ref{eqLsquare}), leads to the polynomial equation for $J_3$,
\begin{eqnarray}
 \sum_{n=0}^6 a_n J_3^n = 0,
 \label{eqn6}
\end{eqnarray}
where the coefficients $a_n$ are expressed in Appendix \ref{Sec-Coefficients}.

\begin{figure}
\centerline{\epsfig{file=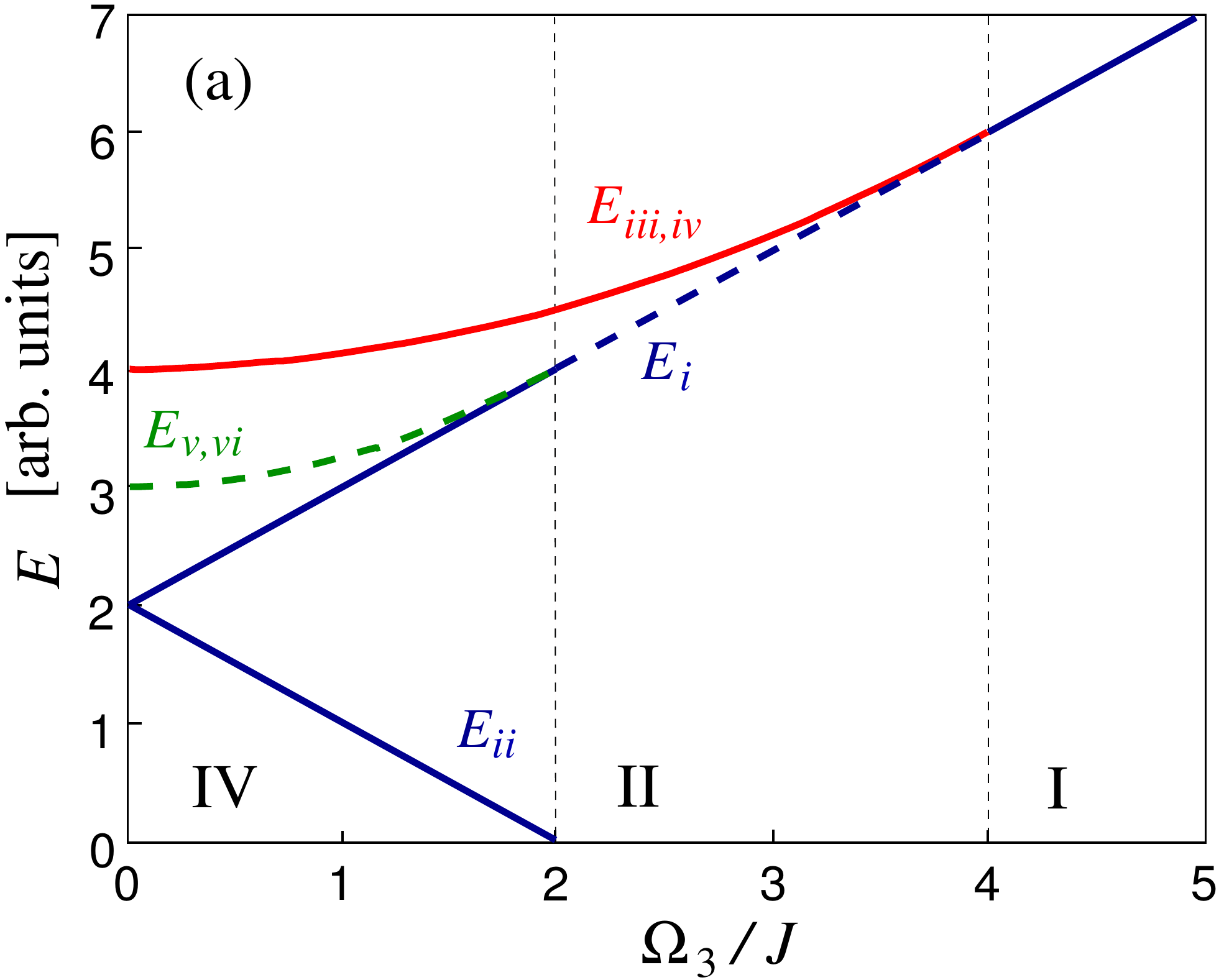,width=0.8\linewidth}}
\centerline{\epsfig{file=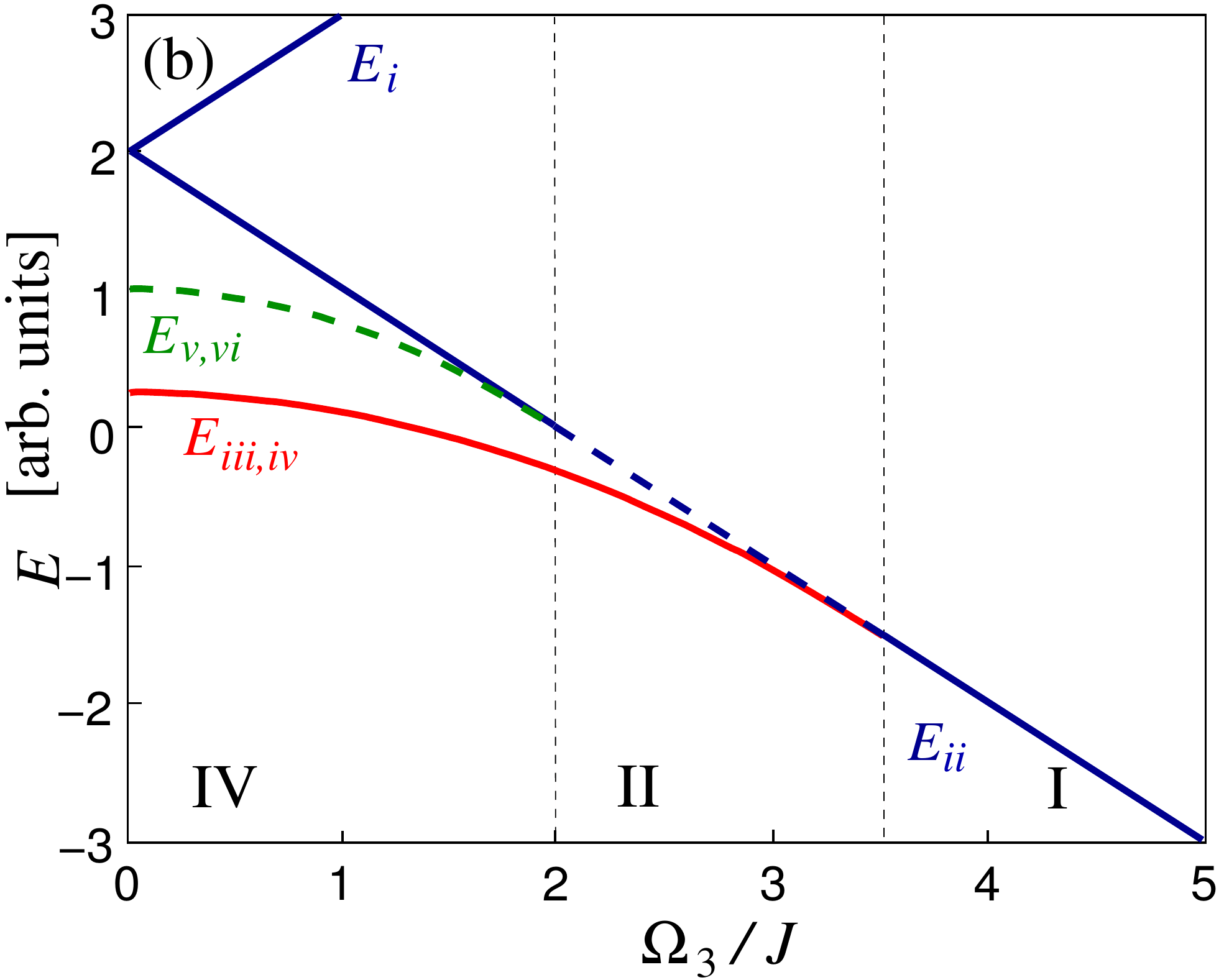,width=0.8\linewidth}}
\centerline{\epsfig{file=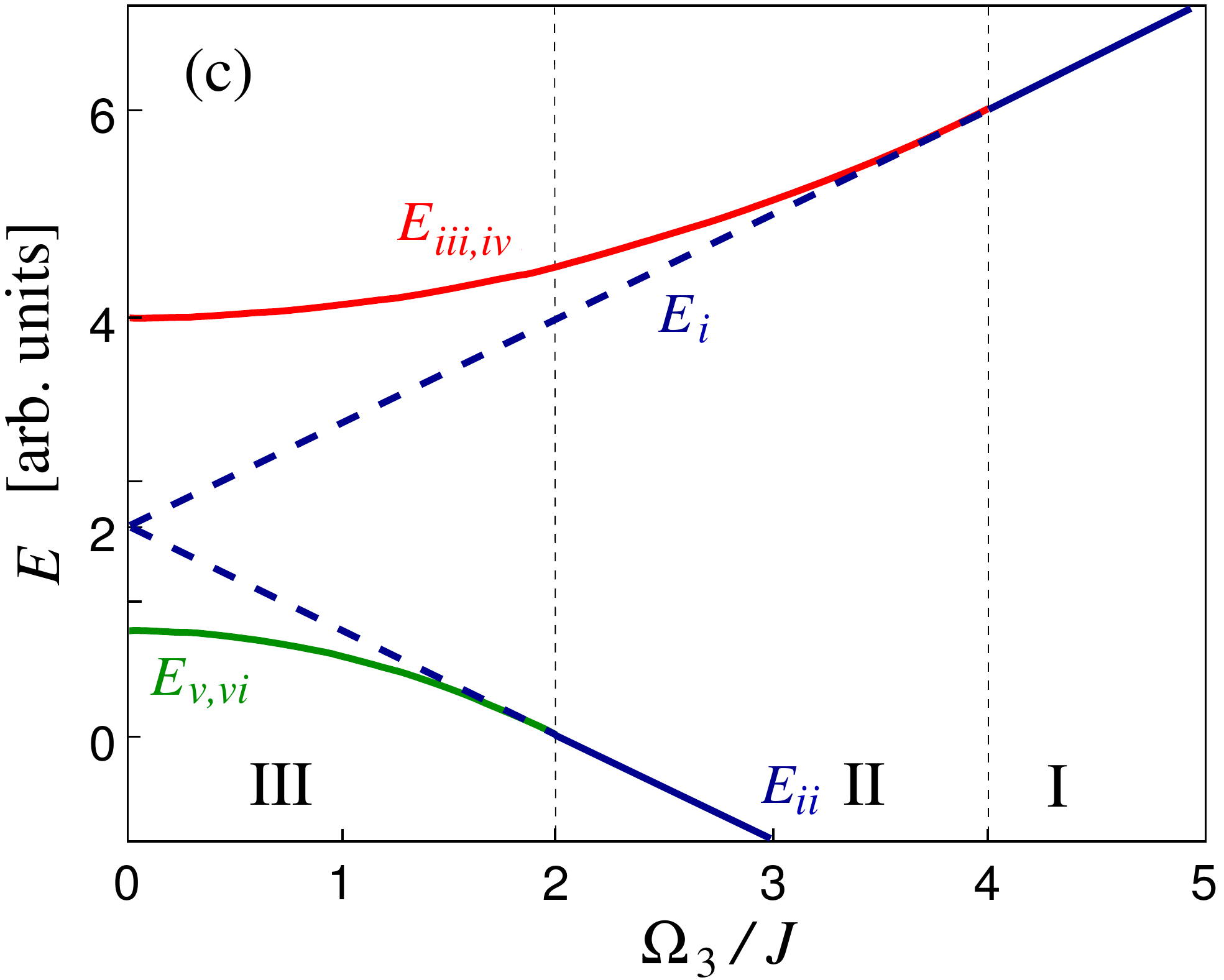,width=0.8\linewidth}}
\caption{\label{f-statpoints}
Energies of the stationary angular momenta, Eqs. (\ref{EnLMG1})---(\ref{EnLMG2}), of the original LMG. Full line corresponds to stable, dashed line to unstable values of $\vec{J}_{i-vi}$. Roman numbers I---IV refer to zones specified in Ref.  \cite{Ribeiro2007}. The twisting tensor eigenvalues are (in arbitrary units): (a) $\chi_1=4$, $\chi_2=3$, $\chi_3=2$, 
 (b) $\chi_1=0.25$, $\chi_2=1$, $\chi_3=2$, (c) $\chi_1=1$, $\chi_2=4$, $\chi_3=2$.
}
\end{figure}

\begin{figure}
\centerline{\epsfig{file=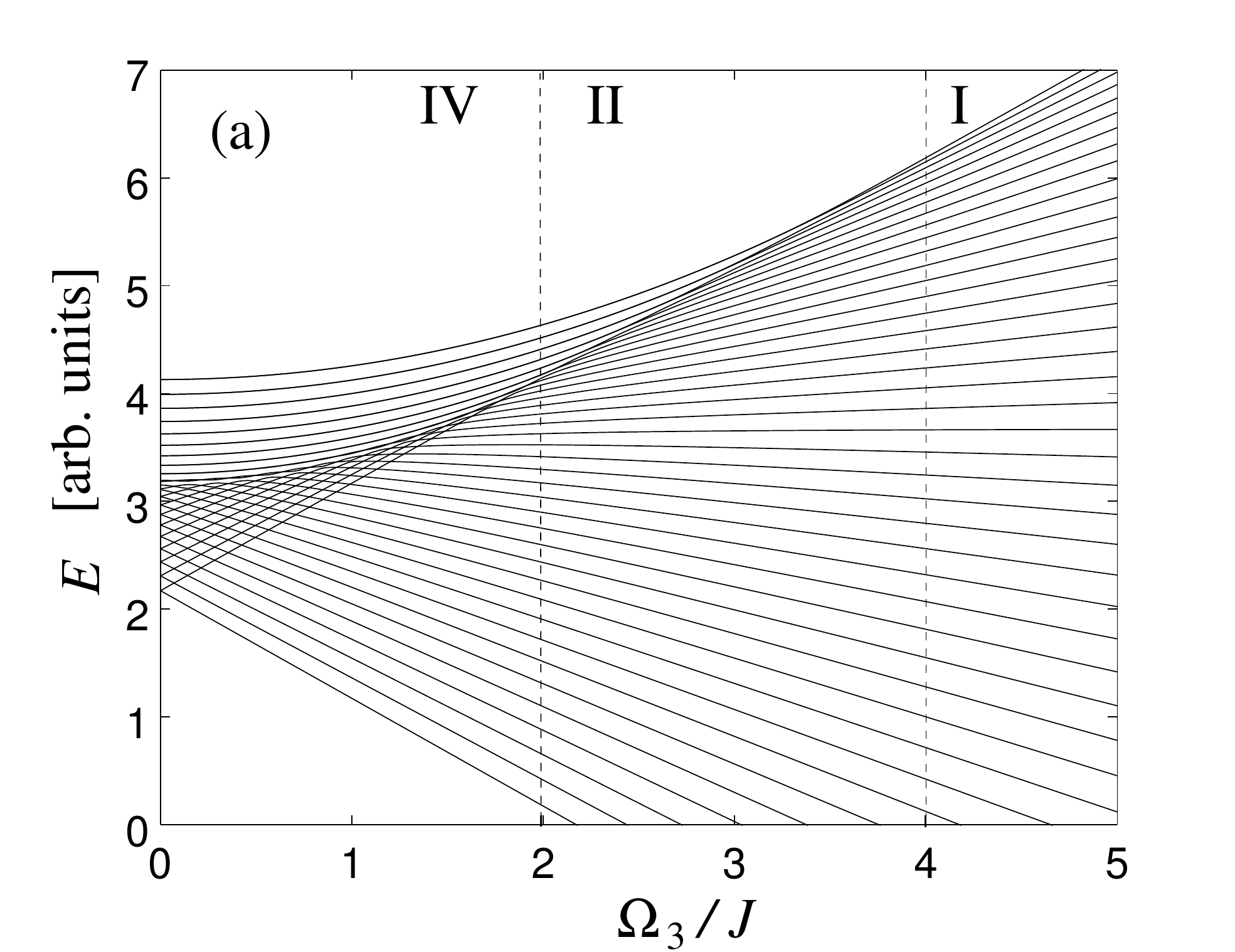,width=0.84\linewidth}}
\centerline{\epsfig{file=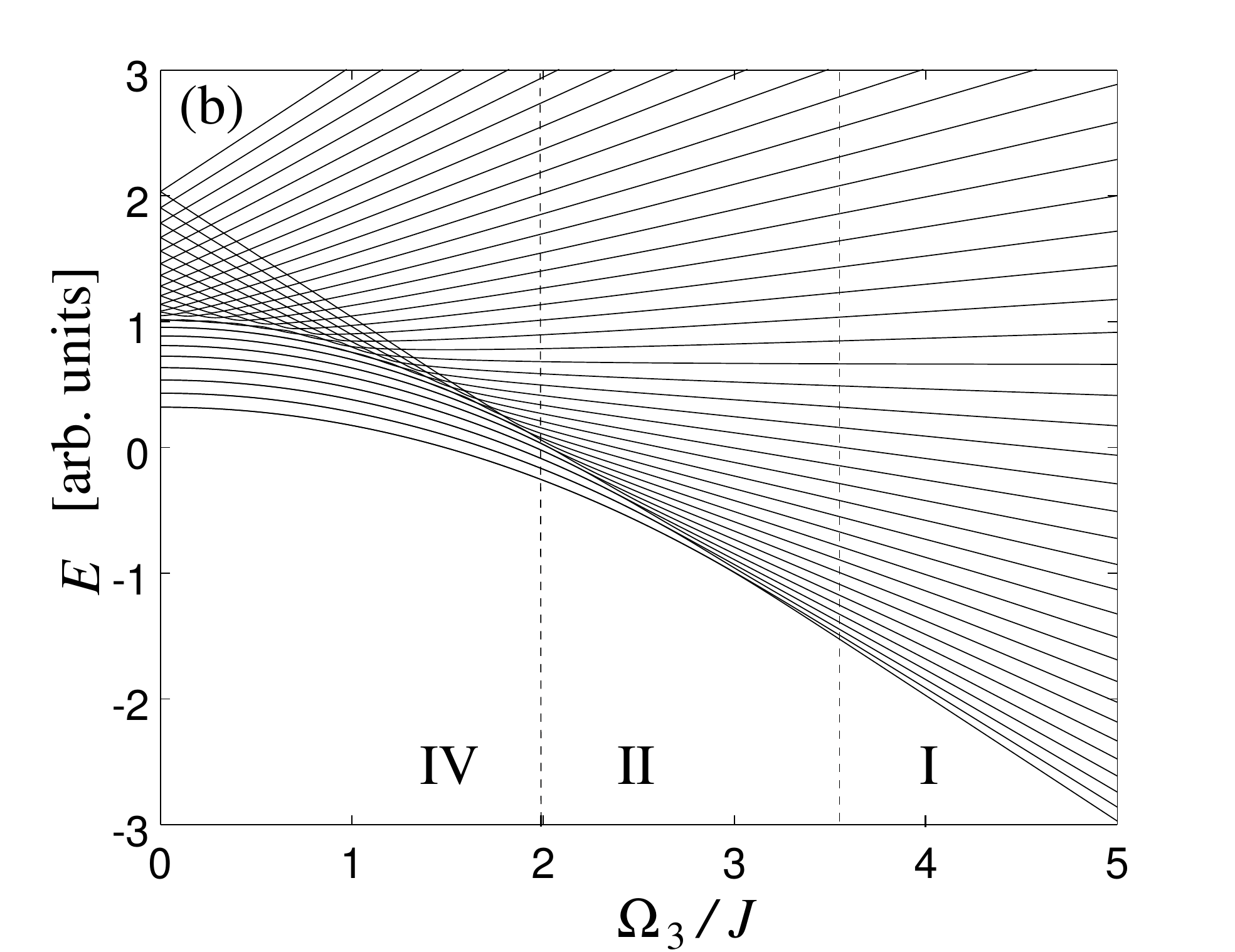,width=0.84\linewidth}}
\centerline{\epsfig{file=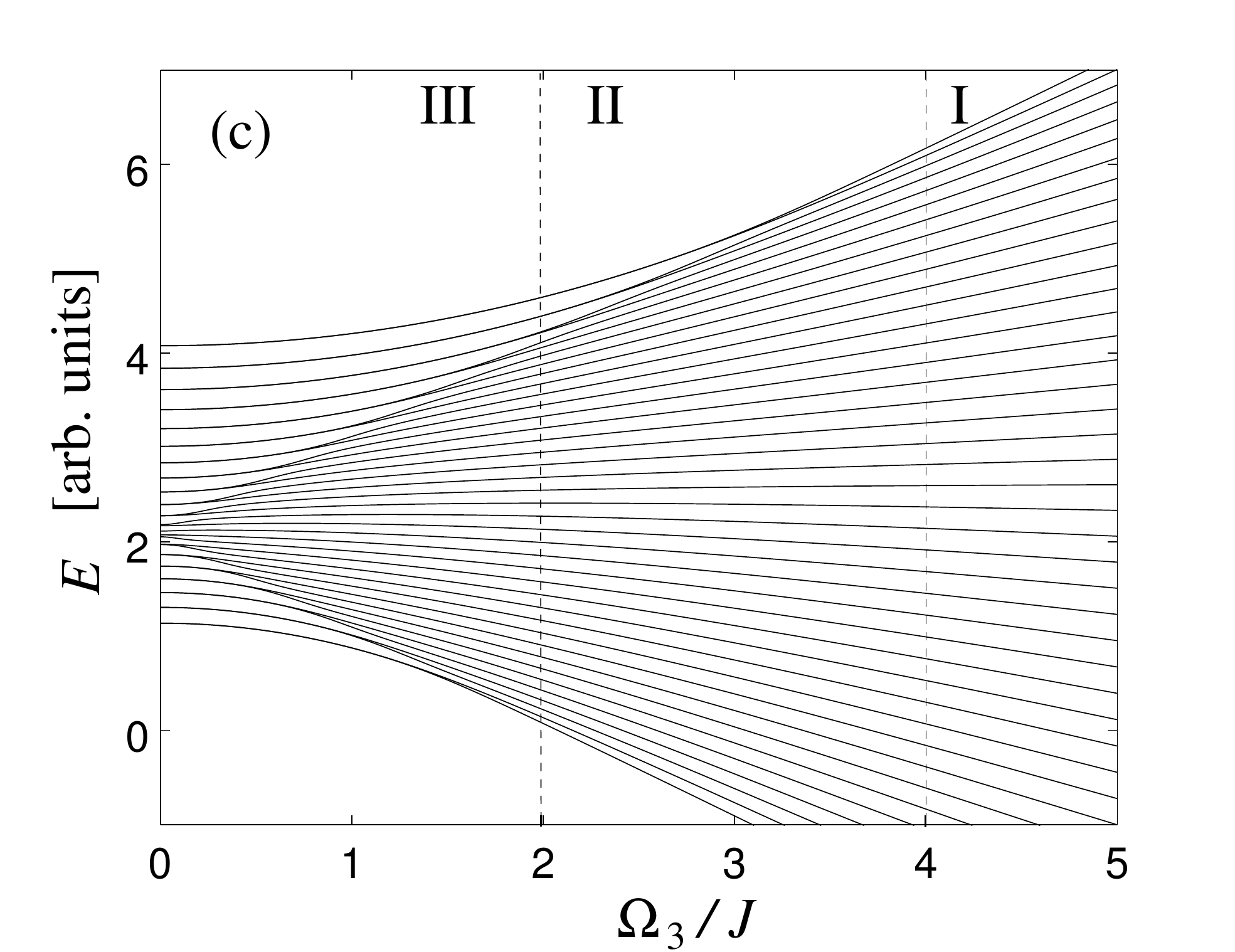,width=0.84\linewidth}}
\caption{\label{f-spectraLMG}
Eigenvalues of the Hamiltonian (\ref{Ham123}) with $\Omega_{1,2}=0$ and the values of $\chi_{1,2,3}$ equal to those of Fig. \ref{f-statpoints}. The number of particles is $N=40$ (corresponding to $J = 20$ and Hilbert space of 21 states).
}
\end{figure}

\begin{figure}
\centerline{\epsfig{file=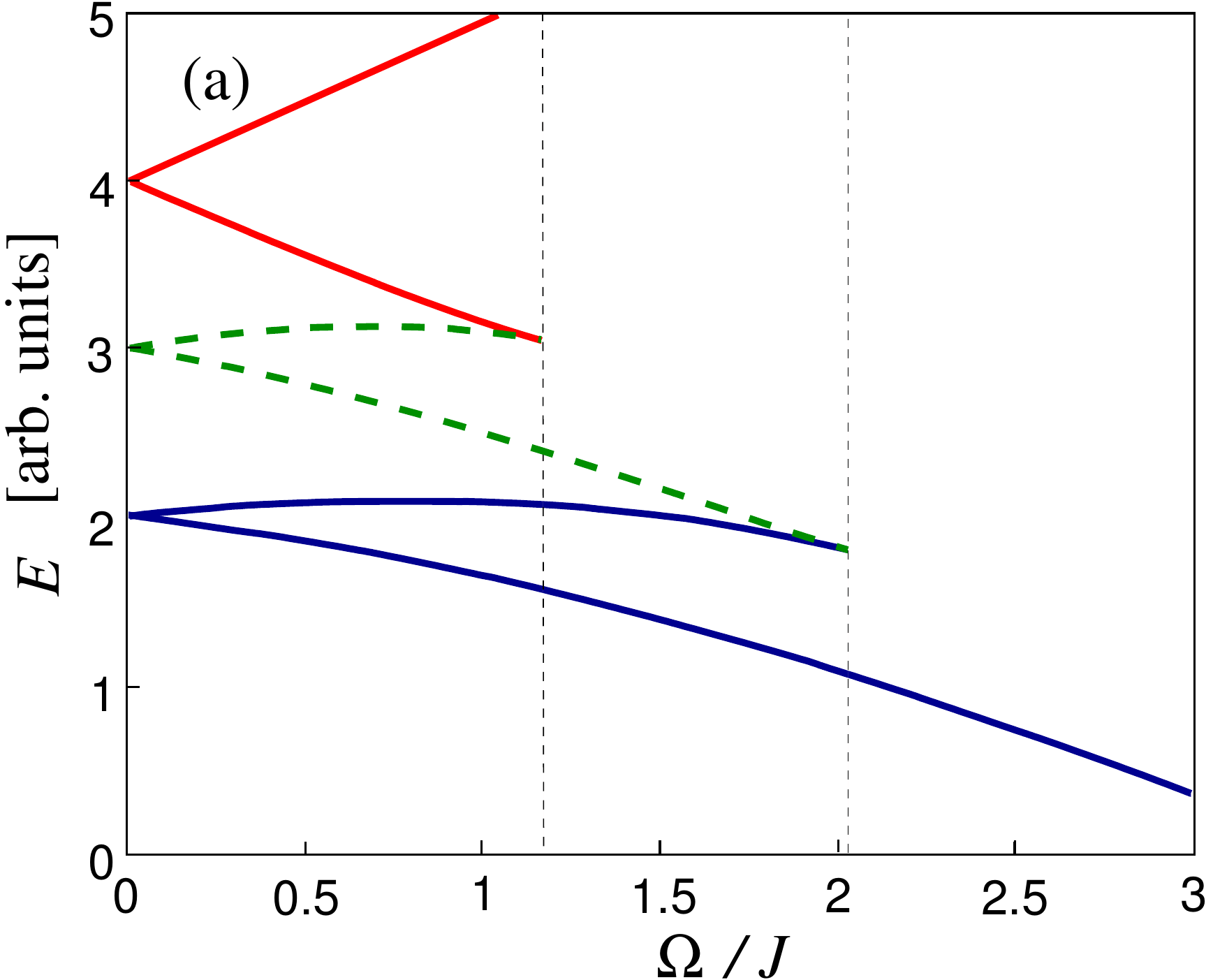,width=0.8\linewidth}}
\centerline{\epsfig{file=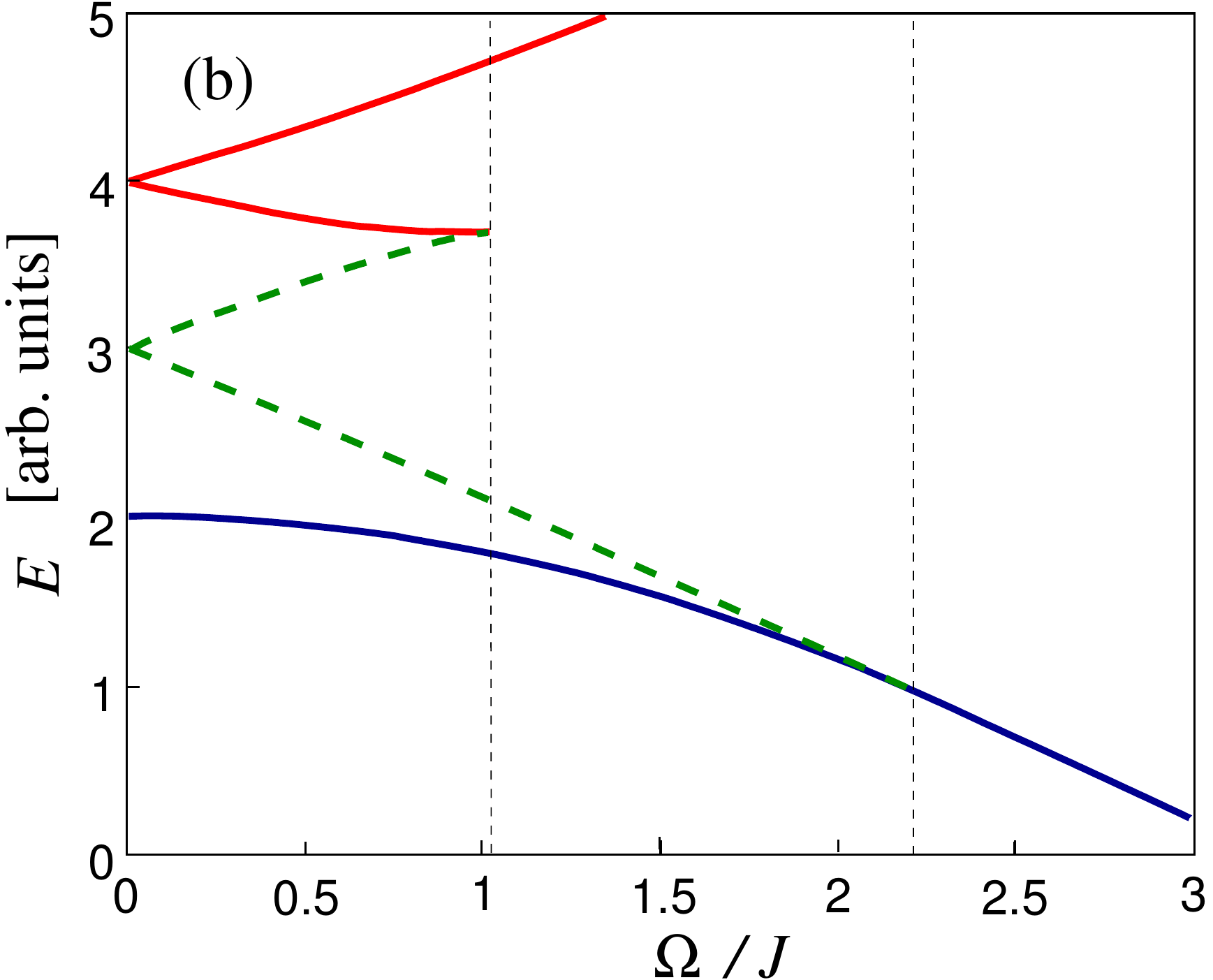,width=0.8\linewidth}}
\centerline{\epsfig{file=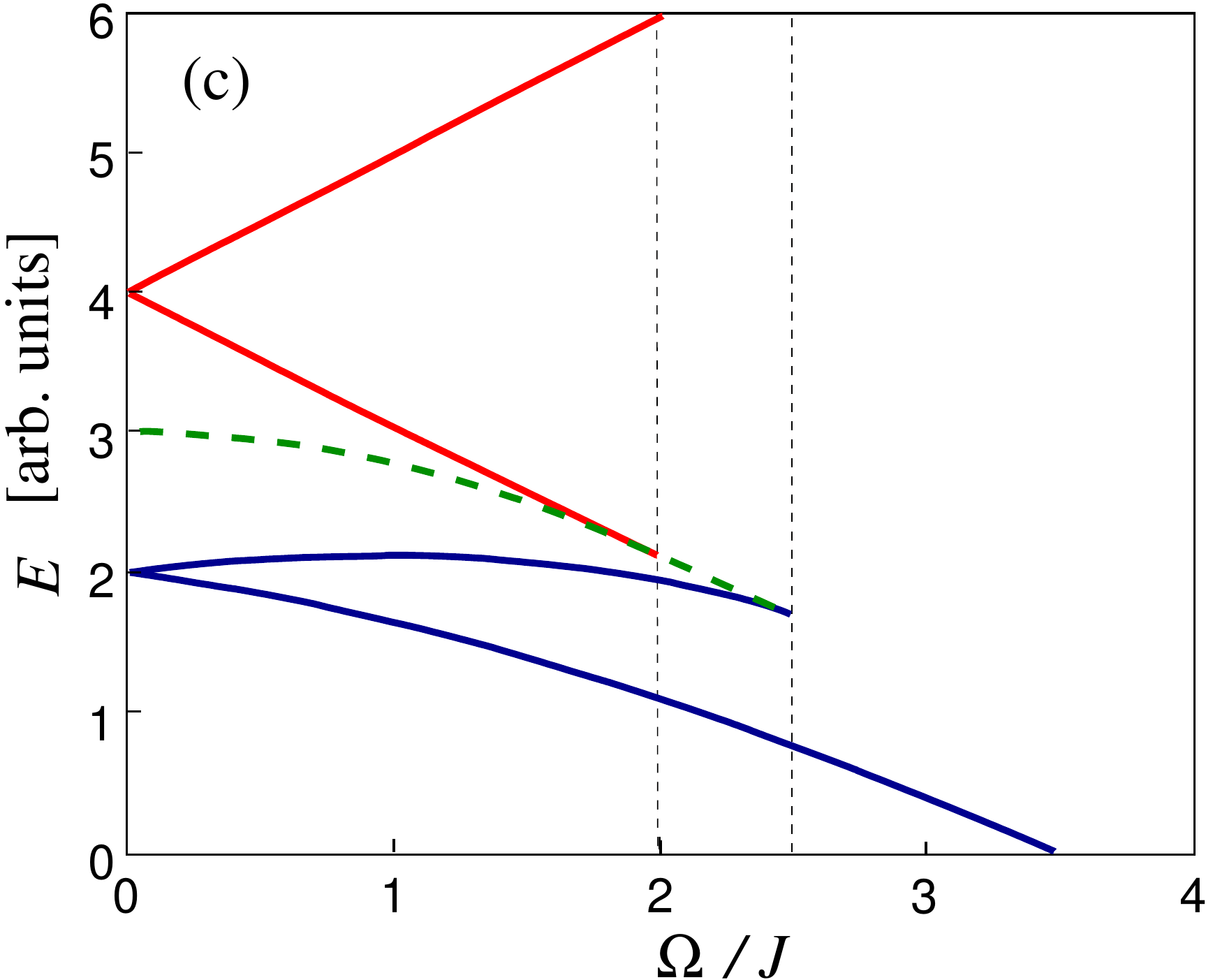,width=0.8\linewidth}}
\caption{\label{f-genLMG-statpoints}
Energies of the stationary angular momenta of the generalized LMG. Full line corresponds to stable, dashed line to unstable angular momenta.  The twisting tensor eigenvalues are (in arbitrary units) $\chi_1=4$, $\chi_2=3$, $\chi_3=2$, the ratio of components of vector $\vec{\Omega}$ are $\Omega_1:\Omega_2:\Omega_3$ as follows, (a) 2:1:1, (b) 1:2:0, (c) 2:0:1. 
 }
\end{figure}

\begin{figure}
\centerline{\epsfig{file=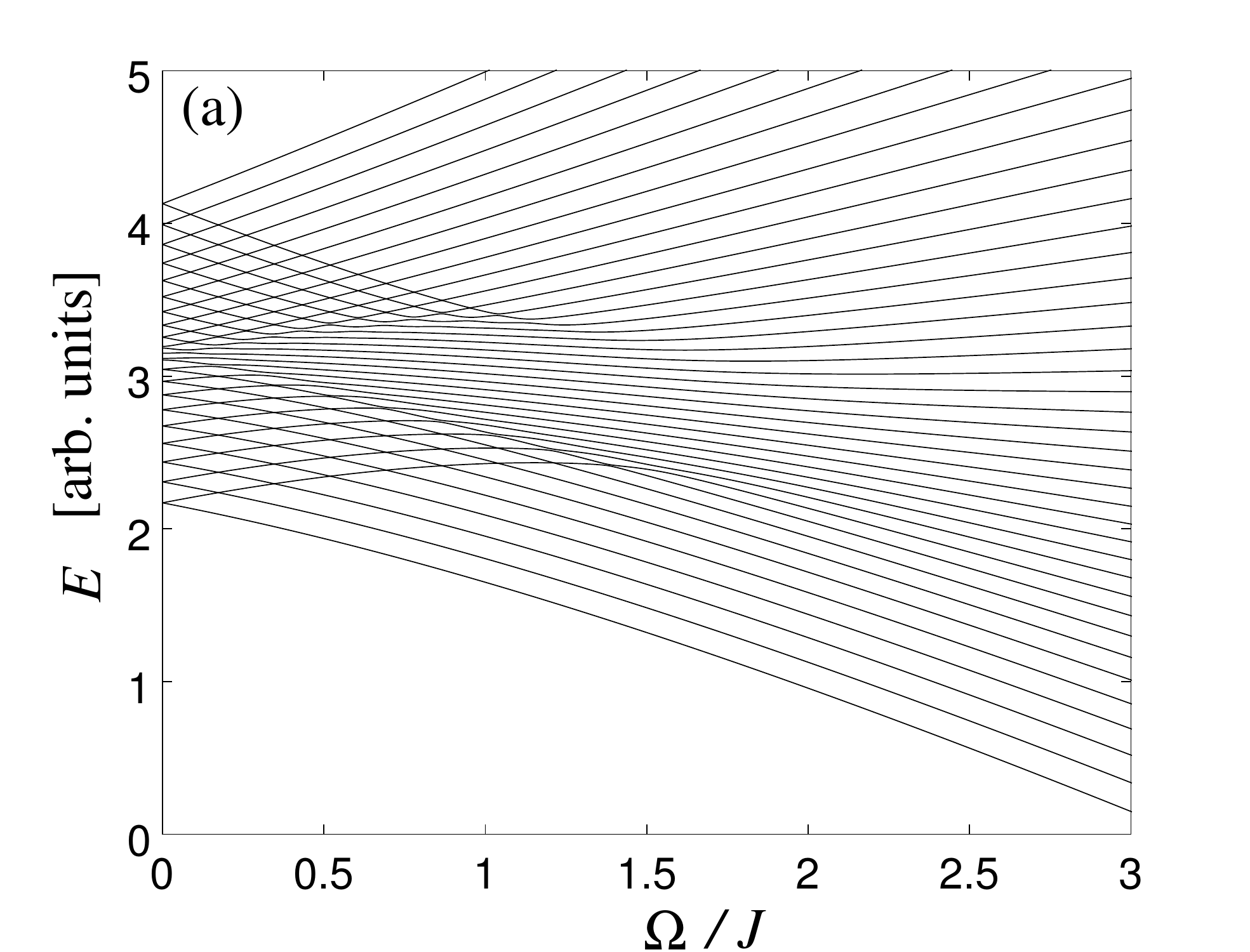,width=0.84\linewidth}}
\centerline{\epsfig{file=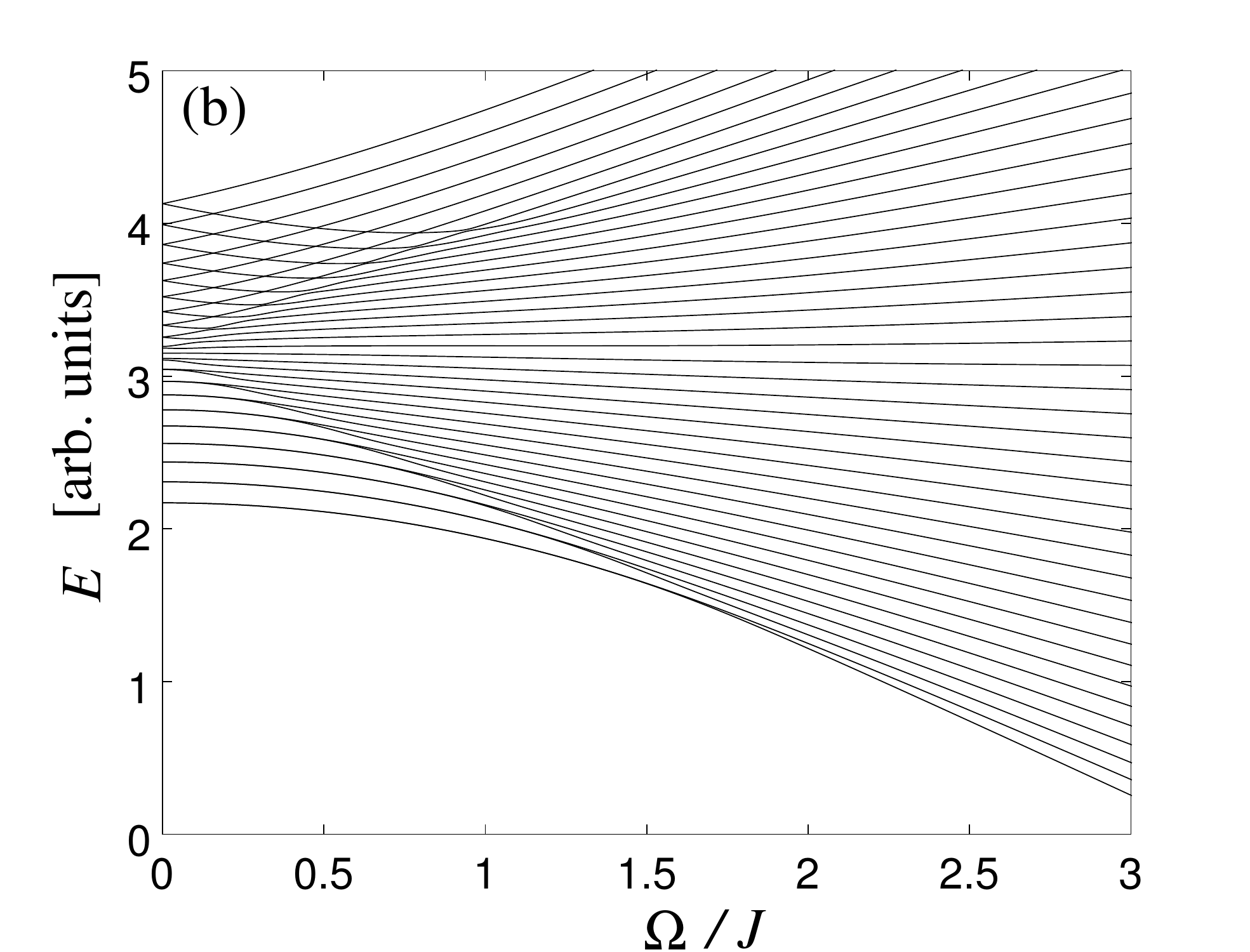,width=0.84\linewidth}}
\centerline{\epsfig{file=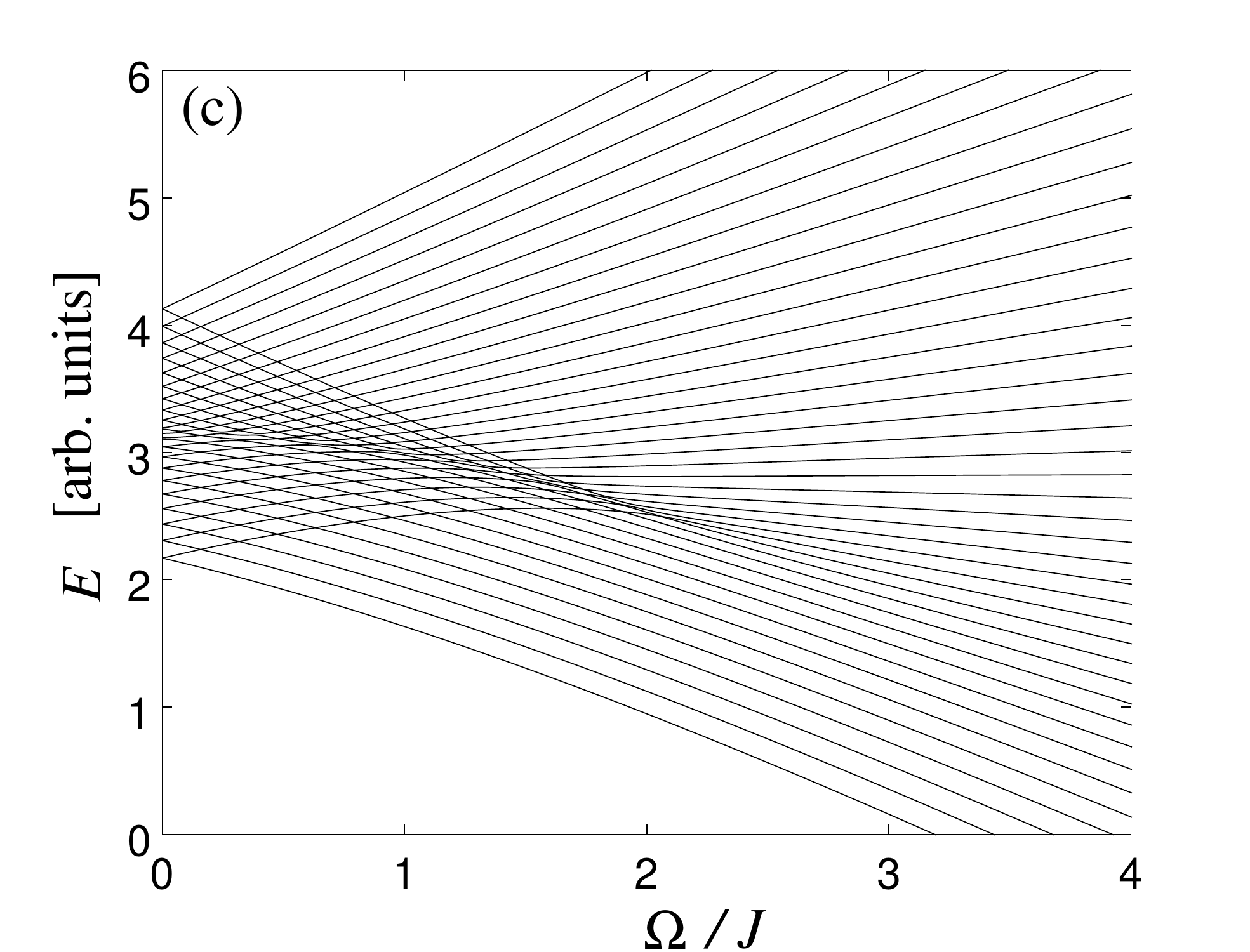,width=0.84\linewidth}}
\caption{\label{f-genLMG-spectr}
Eigenvalues of the Hamiltonian (\ref{Ham123}) of a generalized LMG with the same values of $\chi_{1,2,3}$ and $\Omega_{1,2,3}$ as in Fig. \ref{f-genLMG-statpoints}. The number of particles is $N=40$.
 }
\end{figure}

Equation (\ref{eqn6}) has up to 6 real roots which, together with Eqs. (\ref{eqL1}) and (\ref{eqL2}), specify the stationary values of $\vec{J}$.

\subsection{Stationary point stability}
\label{Sec-stability}
There is a simple geometrical picture allowing us to find the stability of a given stationary point. 
Assume first that the centers of the angular momentum sphere and of the energy ellipsoid are in the same half-space defined by the tangential plane of the contact point. 
At the point of contact, the energy ellipsoid has two principal radii of curvature, $R_1$ and $R_2$. Assume now that both radii are larger than the radius of the sphere, $R_{1,2}> J$. The ellipsoid then touches the sphere from outside. For slightly higher energy there is no contact between the sphere and the ellipsoid, and for slightly lower energy the ellipsoid and the sphere intersect in a closed curve. Thus, the contact point corresponds to a local maximum of energy, i.e., a stable stationary point encircled by states of slightly lower energy. Similarly for  both $R_{1,2}< J$ the ellipsoid touches the sphere from inside, and the contact point is stable  stationary point of the local energy minimum. On the other hand, if, say, $R_1 < J < R_2$, then there exist two directions along which the ellipsoid radius coincides with $J$. Along these directions the ellipsoid intersects the sphere. The contact point then corresponds to the energy saddle on the angular momentum sphere, with the intersection lines corresponding to trajectories approaching to or departing from the (unstable) stationary point. 

Assume now that the centers of the   sphere and of the   ellipsoid are in opposite half-spaces defined by the tangential plane of the contact point. Then the sphere and the ellipsoid touch each other from outside and the contact point corresponds to a stable stationary angular momentum.

In Appendix \ref{Curvature} we derive the principal curvatures at a general point of an ellipsoid. To analyze various phases then means finding stationary points by solving the algebraic equation (\ref{eqn6}) and deciding about their stability by finding the principal radii of the energy ellipsoid using Eq. (\ref{Rellipsoid}).

\subsection{Special case: phase transitions in the original LMG}
\label{PhasetransLMG}

Consider first the special situation with $\Omega_1=\Omega_2=0$ (or equivalently $K_1=K_2=0$). 
To simplify the analysis, assume that $\chi_{1,2,3}>0$ (one can always achieve this by a suitable additive constant), and suppose that $\chi_{1,2}\neq \chi_3$.
In this case Eq. (\ref{eqn6}) can be factorized as
\begin{eqnarray}
(J_3^2-{J}^2) \left[J_3-\frac{\Omega_3}{2(\chi_1-\chi_3)} \right]^2  \left[J_3-\frac{\Omega_3}{2(\chi_2-\chi_3)} \right]^2
=0 . \nonumber \\
\end{eqnarray}
One can see that the stationary angular momenta are always those with  $J_3 = \pm {J}$ (and thus $J_{1,2}=0$), and depending on the magnitude of $\Omega_3$  (or ${K}_3$), also the vectors with 
$J_{3} =\Omega_3/[2(\chi_{1,2}-\chi_3)]$;
the existence of the latter cases depends on whether the resulting $J_3$ fulfills the condition $|J_3|<J$.
Thus, the stationary angular momenta are
\begin{eqnarray}
\vec{J}_{i,ii} &=& \left( 
\begin{array}{c}
0 \\ 0 \\  \pm {J}
\end{array}
\right) ,  \\
\vec{J}_{iii,iv} &=&
\left( 
\begin{array}{c}
\pm \sqrt{J^2 - \frac{\Omega_3^2}{4(\chi_1-\chi_3)^2}} \\ 0 \\ \frac{\Omega_3}{2(\chi_1-\chi_3)}
\end{array}
\right) , \\
\vec{J}_{v,vi} &=& 
\left( 
\begin{array}{c}
0 \\ \pm  \sqrt{J^2 - \frac{\Omega_3^2}{4(\chi_2-\chi_3)^2}}  \\  \frac{\Omega_3}{2(\chi_2-\chi_3)}
\end{array}
\right) .
\end{eqnarray}
Stationary vectors $\vec{J}_{i,ii}$ occur always, whereas $\vec{J}_{iii,iv}$ occur when $|\Omega_3|< 2|\chi_1-\chi_3|{J}$, and 
 $\vec{J}_{v,vi}$ occur when $|\Omega_3|< 2|\chi_2-\chi_3|{J}$.

Energies of the stationary points are obtained from the  Hamiltonian, Eq. (\ref{Ham123}), by substituting values of $\vec{J}_{i-vi}$  for operators $\hat{\vec{J}}$. The results can be used as estimates of the singular points of the energy spectrum.
We get 
\begin{eqnarray}
\label{EnLMG1}
E_{i,ii} &=& \chi_3 J^2  \pm \Omega_3 J, \\
E_{iii,iv} &=& \chi_1 J^2 + \frac{\Omega_3^2}{4(\chi_1-\chi_3)}, \ \frac{|\Omega_3|}{J}< 2|\chi_1-\chi_3|, \\
E_{v,vi} &=&  \chi_2 J^2 + \frac{\Omega_3^2}{4(\chi_2-\chi_3)}, \ \frac{|\Omega_3|}{J}< 2|\chi_2-\chi_3|.
\label{EnLMG2}
\end{eqnarray}

To decide about the stability, we find the principal curvature radii of the energy ellipsoid at the stationary points  as follows,
\begin{eqnarray}
\vec{J}_{i}: & & R_1 = \frac{\chi_3}{\chi_1}\left| J+\frac{\Omega_3}{2\chi_3} \right|,  \\
& &  R_2 =   \frac{\chi_3}{\chi_2}\left| J+\frac{\Omega_3}{2\chi_3} \right| , \\
\vec{J}_{ii}: & & R_1 = \frac{\chi_3}{\chi_1}\left| J-\frac{\Omega_3}{2\chi_3} \right|,  \\
& &  R_2 =   \frac{\chi_3}{\chi_2}\left| J-\frac{\Omega_3}{2\chi_3} \right| , \\
\vec{J}_{iii,iv}: & & R_1 =  \frac{\chi_1 J}{\chi_3 \left( 1-\frac{\Omega_3^2}{4\chi_3 (\chi_3-\chi_1)J^2} \right)} ,  \\
& &  R_2 = \frac{\chi_1}{\chi_2}  , \\
\vec{J}_{v,vi}: & & R_1 =  \frac{\chi_2 J}{\chi_3 \left( 1-\frac{\Omega_3^2}{4\chi_3 (\chi_3-\chi_2)J^2} \right)} ,  \\
& &  R_2 =  \frac{\chi_2}{\chi_1}  .
\end{eqnarray}

Comparing the values $R_{1,2}$ with $J$ according to the criteria in Sec. \ref{Sec-stability}, we find the following different regimes (see Fig. \ref{f-statpoints}).
\begin{enumerate}
\item
Case $\chi_3<\chi_2<\chi_1$ (Fig. \ref{f-statpoints}(a)): $\vec{J}_i$ is unstable for $2J(\chi_2-\chi_3)<\Omega_3<2J(\chi_1-\chi_3)$ and stable outside this interval;  $\vec{J}_{ii}$ is unstable for $-2J(\chi_1-\chi_3)<\Omega_3<-2J(\chi_2-\chi_3)$ and stable otherwise;  $\vec{J}_{iii,iv}$ are stable in the whole interval of their existence $-2J(\chi_1-\chi_3)<\Omega_3<2J(\chi_1-\chi_3)$ and  $\vec{J}_{v,vi}$ are unstable in their whole interval $-2J(\chi_2-\chi_3)<\Omega_3<2J(\chi_2-\chi_3)$. 

Starting from $\Omega_3=0$, the system has two degenerate energy minima, two degenerate saddle points, and two degenerate maxima. Varying $\Omega_3$ in the interval $|\Omega_3|< 2J(\chi_2-\chi_3)$, the degeneracy of the energy minima is lifted, and apart from the degenerate saddles and maxima, the system has one global and one local energy minimum; this corresponds to zone IV defined in  \cite{Ribeiro2007,Ribeiro}.  At $|\Omega_3|= 2J(\chi_2-\chi_3)$ the two saddle points merge with the local energy minimum forming a single saddle point, so that in the intervals  $2J(\chi_2-\chi_3)<|\Omega_3|< 2J(\chi_1-\chi_3)$ the system has one global energy minimum, one saddle point, and two degenerate energy maxima. This interval corresponds to zone II of \cite{Ribeiro2007}.   At $|\Omega_3|= 2J(\chi_3-\chi_1)$ the two degenerate maxima and the saddle point merge to form a single global maximum. Then, for   
 $2J(\chi_3-\chi_1)<|\Omega_3|$ the system has just one global energy minimum and one global maximum, corresponding to zone I of \cite{Ribeiro2007}.

Note that case  $\chi_3<\chi_1<\chi_2$ has qualitatively the same properties, except that $\chi_1$ and $\chi_2$ change their roles.

\item 
Case $\chi_1<\chi_2<\chi_3$ (Fig. \ref{f-statpoints}(b)): the situation is the same as for  $\chi_3<\chi_2<\chi_1$, except that the terms $\chi_1-\chi_3$ and $\chi_2-\chi_3$ change signs, and maxima and minima switch their roles. Similarly for  $\chi_2<\chi_1<\chi_3$ where $\chi_1$ and $\chi_2$ change their roles.

\item
Case $\chi_1<\chi_3<\chi_2$ (Fig. \ref{f-statpoints}(c)): $\vec{J}_i$ is unstable for $-2J(\chi_3-\chi_1)<\Omega_3<2J(\chi_2-\chi_3)$ and stable outside this interval;  $\vec{J}_{ii}$ is unstable for $-2J(\chi_2-\chi_3)<\Omega_3<2J(\chi_3-\chi_1)$ and stable otherwise.  $\vec{J}_{iii-vi}$ are stable in the whole intervals of their existence. 

Assume now, to be specific, that $\chi_3-\chi_1<\chi_2-\chi_3$.
Starting from $\Omega_3=0$ up to $|\Omega_3|= 2J(\chi_3-\chi_1)$, the degeneracy of the two saddle points is lifted, while there are still two degenerate maxima and two degenerate minima of energy. This regime corresponds to zone III  of \cite{Ribeiro2007}.
At $|\Omega_3|= 2J(\chi_3-\chi_1)$ the two minima and the lower-energy saddle point merge to form a single global minimum. In the intervals $2J(\chi_3-\chi_1) <|\Omega_3| < 2J(\chi_2-\chi_3)$ the system has one global energy minimum, one saddle point and two global energy maxima, corresponding to zone II   of \cite{Ribeiro2007}. At $|\Omega_3| = 2J(\chi_2-\chi_3)$ the two maxima and the saddle point merge to form a single energy maximum; for  $|\Omega_3| > 2J(\chi_2-\chi_3)$ the system has one global energy maximum and one minimum, corresponding to zone I  of \cite{Ribeiro2007}. 

\end{enumerate}

These results based on a classical analogy can be compared with spectra of quantum Hamiltonians diagonalized numerically. As can be seen comparing Figs. \ref{f-statpoints} and \ref{f-spectraLMG}, energies of the stationary points calculated according to Eqs. (\ref{EnLMG1})--(\ref{EnLMG2}) clearly match  the singularities of the quantum spectra.

\begin{figure}
\centerline{\epsfig{file=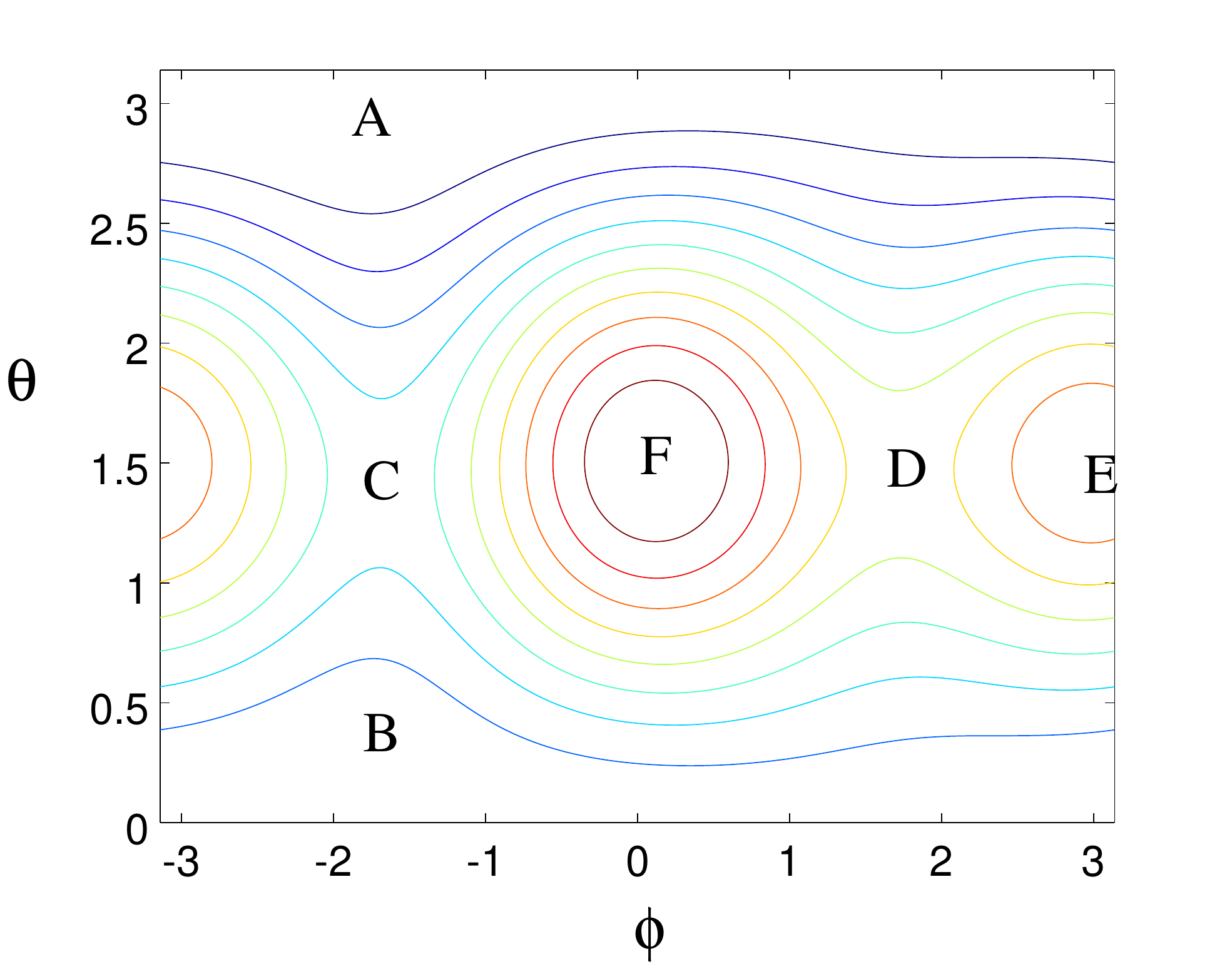,width=1.0\linewidth}}
\centerline{\epsfig{file=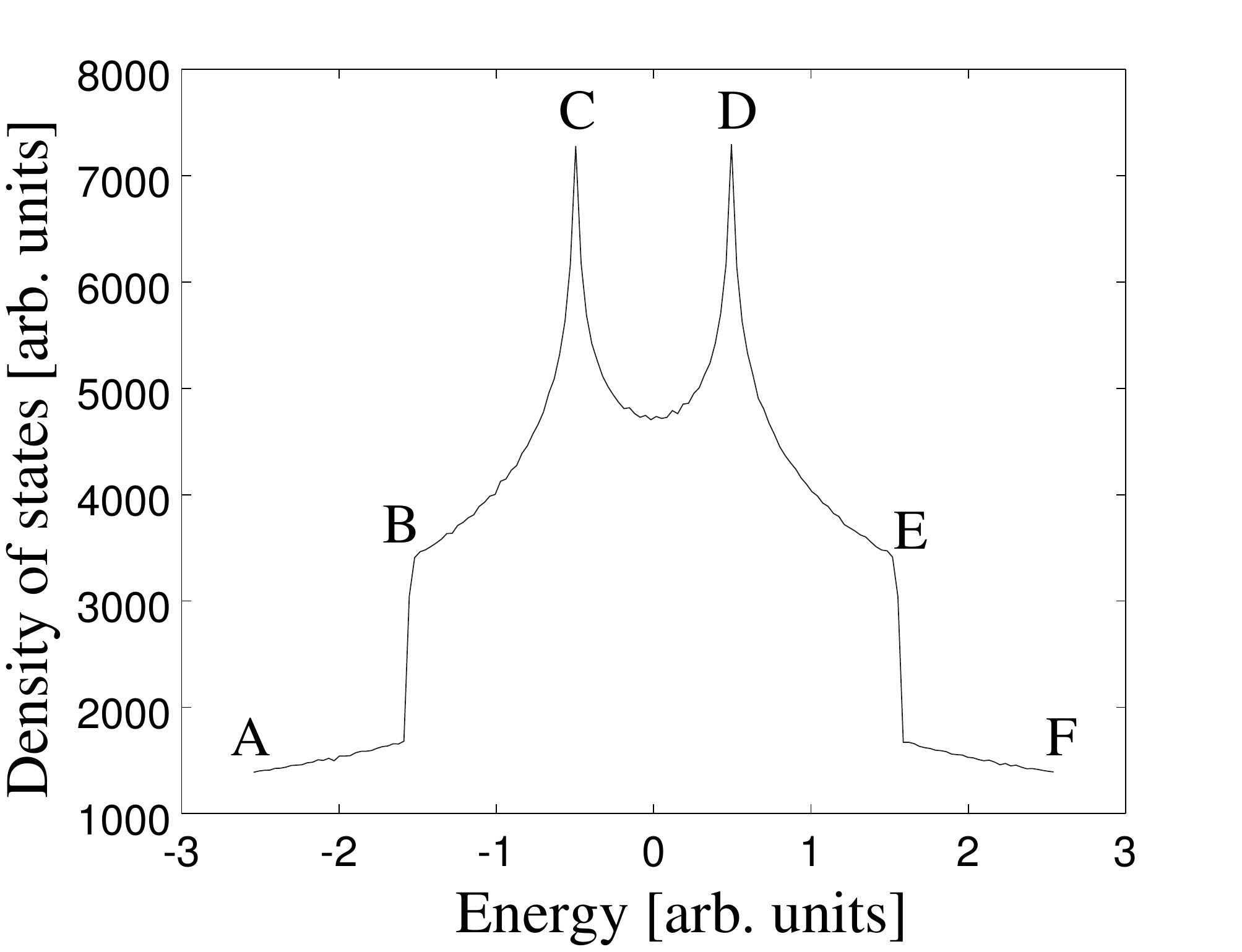,width=1.0\linewidth}}
\caption{\label{f-countours1}
Upper panel: Contours of the equal energy of the generalized LMG model with $(\chi_1,\chi_2,\chi_3) = (2,0,-2)$ and $(\Omega_1,\Omega_2,\Omega_3) = (0.5,0.5,0.5)$. Coordinates $\phi$ and $\theta$ refer to the direction of vector $\vec{J}$. Lower panel: energy spectrum corresponding to the same parameters. The beginning and end of the graph correspond to the global minimum and maximum, respectively. The two discontinuities correspond to the local minimum and local maximum, and the peaks correspond to the two saddle points.
}
\end{figure}

\subsection{Phase transitions in the generalized LMG}

For general values of $\vec{\Omega}$ one can factorize Eq. (\ref{eqn6}) numerically. Energies of the resulting values are shown in Fig. \ref{f-genLMG-statpoints}, and the corresponding Hamiltonian eigenvalues in Fig. \ref{f-genLMG-spectr}. The general features are as follows. Starting at $\vec{\Omega} = \vec{0}$, the system has three pairs of degenerate stationary angular momenta with energies $\chi_{1,2,3}J^2$. Ramping up $\vec{\Omega}$, the degeneracy is lifted for those stationary angular momenta in whose direction  $\vec{\Omega}$ has a nonzero component. In Figs. \ref{f-genLMG-statpoints}(a) and \ref{f-genLMG-spectr}(a) this is the case for all three components (Fig. \ref{f-countours1} shows typical energy contours and the corresponding density of states). In Figs. \ref{f-genLMG-statpoints}(b) and (c) one component of  $\vec{\Omega}$ vanishes and the degeneracy of the corresponding energy remains (note that in the original LMG model in Fig. \ref{f-statpoints} two components vanish so that only one degeneracy is lifted). 

One can see that for a general direction, two critical values of $\Omega$ occur: at each of them one of the local extrema of energy merges with one of the saddle points and these two stationary points disappear. Thus, it is natural to distinguish three generic phases of the generalized LMG system, according to the number of saddle points of energy on the angular momentum sphere: those with zero, one, and two saddle points. In case of various symmetries, more detailed classification may be relevant. In particular, considering the original LMG model in \cite{Ribeiro2007,Ribeiro}, two zones were identified within the phase with two saddle points: zone III in which the saddles have different energies, and zone IV with energy degenerate saddles and lifted degeneracy of either energy minima or maxima.  

Other special cases can be found in the generalized LMG model if $\vec{\Omega}$ is confined to a plane perpendicular to one of the principal directions of tensor $\chi$. In particular, in the phase with a single saddle point, one of the energy extrema may become degenerate---this is the case of Figs. \ref{f-genLMG-statpoints}(b) and \ref{f-genLMG-spectr}(b)  (note that in the original LMG, both energy extrema are degenerate in zone II in which a single saddle occurs). In the phase with two saddle points, one of the energy extrema may become degenerate (Figs. \ref{f-genLMG-statpoints}(b) and  \ref{f-genLMG-spectr}(b), 
or the saddle points can be degenerate  (Figs. \ref{f-genLMG-statpoints}(c) and \ref{f-genLMG-spectr}(c)).
These cases can be considered as new sub-phases in the generalized LMG model.

\section{Floquet time crystals}
\label{SecTimeCrystal}
The concept of time crystals was introduced by F. Wilczek \cite{Wilczek-2012,Shapere-2012}, referring to processes in which spontaneous breaking of time symmetry occurs, in analogy to  broken spatial  symmetry in usual crystals. Interesting phenomena were predicted for systems with periodic driving as so called ``Floquet time crystals'' \cite{Sacha-2015,Else-2016,Yao-2017}, whose observations have recently been reported in trapped ions \cite{Zhang-2017-crystal} and in nitrogen-vacancy centres in diamond \cite{Choi-2017}. 
In the Floquet time crystals, the external driving has period $\tau$ and thus the Hamiltonian has a discrete time symmetry. Yet, under certain conditions the system behavior breaks this time symmetry and periodic phenomena occur in times corresponding to a multiple of $\tau$, i.e., $n\tau$. 

\begin{figure}
\centerline{\epsfig{file=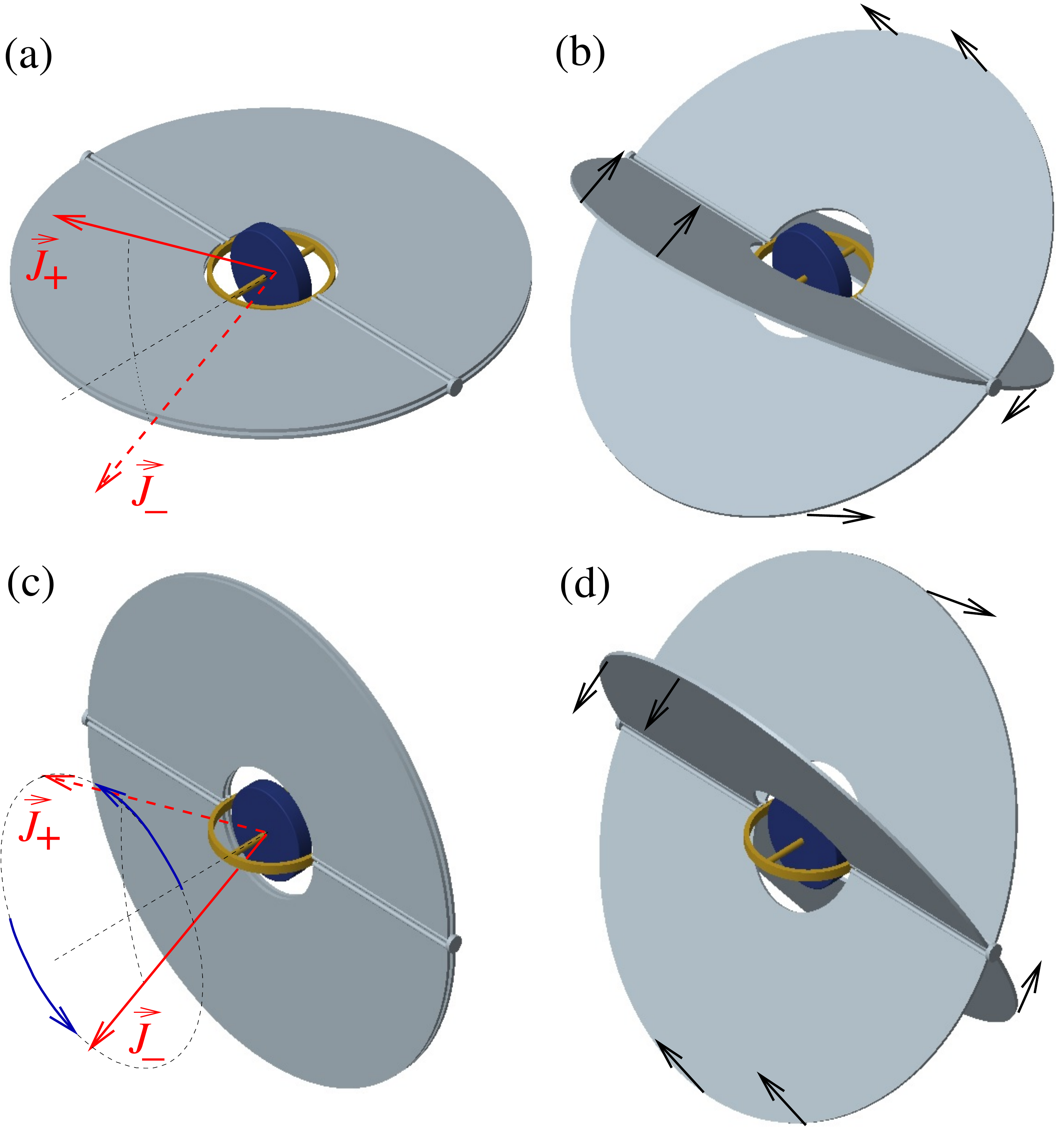,width=1.0\linewidth}}
\caption{\label{f-FloquetGyro}
Reshaping a body in the mechanical analogue of the LMG Floquet time crystal. The body starts as a symmetric top with a perpendicular rotor, having two degenerate stable rotational states with angular momenta $\vec{J}_{\pm}$ (a). The body is then reshaped (b) to take a form of a symmetric top with a coaxial rotor (c) so that the original angular momenta  $\vec{J}_{\pm}$ precess around the body axis. After swapping $\vec{J}_{\pm}$, the body is reshaped (d) back to the original form (a).
}
\end{figure}

Recently, Floquet time crystal in the LMG model has been proposed \cite{Russomanno-2017}. In their scheme, the system is initialized in one of the degenerate energy extremal state. Then, a kick rotates the system around the axis of the LMG linear term by $\pi$ (in Fig. \ref{f-Bloch2} that would be a $\pi$ rotation around $J_3$). As a result, the system swaps to the other degenerate state. If the kicks occur with period $\tau$ and the system is initially close to one of the local energy extrema, oscillations of some physical quantity may occur with  period $2\tau$.

\begin{figure}
\centerline{\epsfig{file=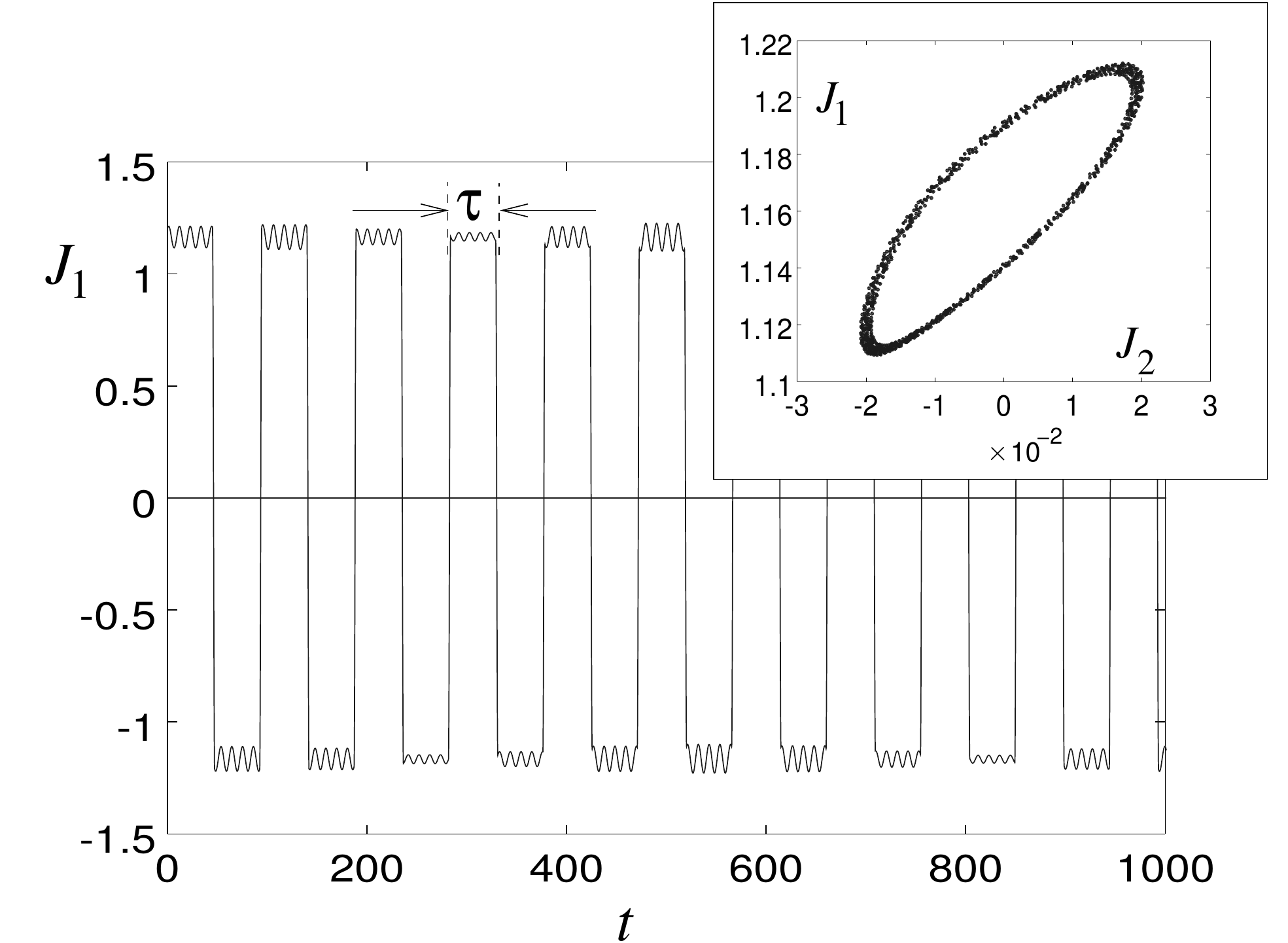,width=1.0\linewidth}}
\caption{\label{f-Floquet}
Evolution of the angular momentum component $J_1$ in the mechanical Floquet time crystal scenario. The time and angular momentum are dimensionless, their scales following from the choice $I_0=1$ and $K_3=1$. The initial values are $(J_1,J_2,J_3)=(1.2,0.02,1.98)$ and the time parameters are $\tau_0 = 45.20$ and $\tau_{\rm switch} = 2.09$.
Inset: stroboscopic values of $J_{1,2}$ for 1000 repetitions.
}
\end{figure}

It is almost straightforward to build a classical analogue of the LMG Floquet time crystal of \cite{Russomanno-2017}. Assume a plate-like symmetric top with $I_2=I_3\equiv I_0$, and $I_1=2I_0$, with a perpendicular rotor with angular momentum $K_{1,2}=0$, $K_3\neq 0$ as in Fig. \ref{f-FloquetGyro}(a). Following Sec. \ref{PhasetransLMG}, two stable stationary angular momenta occur at $\vec{J}_{\pm}$ with $J_1 = \pm \sqrt{J^2 -4K_3^2}$, $J_2=0$, and $J_3=2K_3$ for any $J>2|K_3|$. Assume that the system is prepared near one of these stationary points, say $\vec{J}_+$ with $J_1 = + \sqrt{J^2 -4K_3^2}$. We now need to swap the stationary states. Since a sudden kick instantaneously changing the rotational axis is unphysical, one can consider an alternate scenario. Assume that the body is reshaped, changing the moments of inertia to $I_1=I_2=I_0$ and $I_3 = 2I_0$ as in Fig \ref{f-FloquetGyro}(b) and (c), this transformation happening much faster than the precession. After reshaping, the body is a symmetric top with a coaxial rotor. Following Sec. \ref{SecCoaxial} and Eq. (\ref{Omegasymtop}), the rotational axis precesses with angular velocity 
\begin{eqnarray}
\tilde \Omega = \frac{K_3+J_3}{2I_0}.
\label{OmegaFloquet}
\end{eqnarray} 
If  $J_3=2K_3$, then  $\tilde \Omega = 3K_3/(2I_0)$ 
so that after time $\tau_{\rm swap} = 2\pi I_0/(3K_3)$ the angular momentum is changed to $J_1 = - \sqrt{J^2 -4K_3^2}$. Then, the body reshapes back (Fig. \ref{f-FloquetGyro}(d)) and continues motion with the rotational axis near the new stationary direction $\vec{J}_-$. Assume the body is left to evolve, changing periodically its shape from a symmetric top with perpendicular rotor for time $\tau_0$ to a symmetric top with a coaxial rotor for time $\tau_{\rm swap}$. The driving period is $\tau =\tau_0 +\tau_{\rm swap}$, however, the system returns to the initial stationary angular momentum with period $2\tau$, as in a Floquet time crystal.

The dynamics is not completely equivalent to the quantum model of \cite{Russomanno-2017} where the kick instantaneously rotates each state by the same angle (ideally $\pi$). Since $\tilde \Omega$ in Eq. (\ref{OmegaFloquet}) depends on $J_3$, angular momenta deviating from the stationary values $\vec{J}_{\pm}$ would, after $\tau_{\rm swap}$, be rotated by different angles. Depending on the system parameters and on the initial state, the deviations may accumulate over time, leading to chaotic dynamics. Nevertheless, one can find intervals of initial values of $J_{1,2,3}$, and of times $\tau_0$ and $\tau_{\rm swap}$, for which regular motion corresponding to a Floquet time crystal is observed.
An example of such a motion is in Fig. \ref{f-Floquet}.

\section{Conclusion}
\label{SecConclusion}

Analogies between the Euler-top dynamics and quantum evolution of collective spins allow us to have simple physical pictures of quantum phenomena such as spin squeezing by OAT or by TACT scenarios. When adding a rotor with axis along one of the principal axes of the top, the system corresponds to the LMG model.
We find remarkable the close links between the quantum phase transitions in the LMG model and the behavior of stationary angular momenta in classical rigid body dynamics. Allowing for arbitrary  orientation of the rotor axis in the classical domain leads to a generalized LMG model in the quantum domain, predicting new scenarios of quantum phase transitions. These could be observed once a full TACT scheme is implemented (e.g., using the recent  proposals \cite{Zhang-2017,Borregaard-2017}) with additional suitable linear terms. Vice versa, the LMG Floquet time crystal proposed in the quantum domain \cite{Russomanno-2017} finds its classical counterpart in a periodically reshaped Euler top.

Some questions that remain beyond the present study might be worth further investigation. Is there, within the proposed models, any measurable analogue of the classical body orientation? Taking into account the possibility of multimode spin squeezing \cite{Kruse-2016,Opatrny-2017}, is there any relevant scenario of rigid body dynamics to which it would correspond?

Feynman concludes his wobbling-plate story with enthusiasm \cite{Feynman-joking}: {\it ``I went on to work out equations of wobbles. Then I thought about how electron orbits start to
move in relativity. Then there's the Dirac Equation in electrodynamics. And then quantum
electrodynamics. 
[\dots]
It was effortless. It was easy to play with these things. It was like uncorking a bottle: Everything
flowed out effortlessly. I almost tried to resist it! There was no importance to what I was doing, but
ultimately there was. The diagrams and the whole business that I got the Nobel Prize for came from
that piddling around with the wobbling plate.''} 

We believe that enthusiasm for physics of wobbling tops is worth sharing. Tossing a coin or throwing up a tennis racket, an attentive observer could see in their motion how spin squeezing works. If some of the readers happen to jump twisted somersaults,  they  might experience the LMG-type phase transitions themselves. Or, if you are astronaut having some free time on orbit, you might observe self trapping of Bose-Einsten condensates  and other critical phenomena in a variant of the Dzhanibekov effect when gluing a fidget spinner to a box you leave rotating.

\acknowledgments
T.O. and L.R. dedicate this paper to J. Tillich who taught us classical mechanics many years ago and raised our interest in the tennis racket instability.
T.O. is grateful to P. Cejnar and J. Vidal for stimulating discussions. This work was supported by the  Czech Science Foundation, grant No. 17-20479S.

\appendix

\begin{widetext}
\section{Coefficients of the stationary angular momenta polynomials}
\label{Sec-Coefficients}
Using Eqs. (\ref{eqL1}) and  (\ref{eqL2}) in Eq.  (\ref{eqLsquare}) leads to the polynomial equation (\ref{eqn6}) with coefficients
\begin{eqnarray}
a_0 &=& - {J^2}{K}_3^4 \frac{I_1^2 I_2^2}{(I_3-I_1)^2 (I_3-I_2)^2} , \\
a_1 &=&  - 2 {J^2}{K}_3^3 \frac{ I_1 I_2 (I_1 I_3 - 2I_1 I_2 + I_2 I_3)}{(I_3-I_1)^2 (I_3-I_2)^2}, \\
a_2 &=& \frac{I_1^2 I_2^2 {K}_3^4 +  I_3^2 {K}_3^2 \left( I_1^2 {K}_2^2+ I_2^2 {K}_1^2 \right) 
 -  {J^2}{K}_3^2 \left[ (I_1 I_3 - 2I_1 I_2 + I_2 I_3)^2 + 2 I_1 I_2 (I_3-I_1)(I_3-I_2) \right]}{(I_3-I_1)^2 (I_3-I_2)^2} , \\
a_3 &=& \frac{ 2 {K}_3^3 I_1 I_2(I_1 I_3 - 2I_1 I_2 + I_2 I_3)
 + 2  {K}_3 I_3^2 \left[ I_2(I_3-I_2) {K}_1^2 +  I_1(I_3-I_1) {K}_2^2 \right]}{(I_3-I_1)^2 (I_3-I_2)^2}
\nonumber \\ 
&& -  \frac{2 {J^2}{K}_3 (I_1 I_3 - 2I_1 I_2 + I_2 I_3)}{(I_3-I_1) (I_3-I_2)}, \\
a_4 &=&{K}_3^2  \frac{  (I_1 I_3 - 2I_1 I_2 + I_2 I_3)^2 + 2 I_1 I_2 (I_3-I_1)(I_3-I_2)   }{(I_3-I_1)^2 (I_3-I_2)^2}
+ I_3^2 \left[\frac{ {K}_1^2}{(I_3-I_1)^2} + \frac{ {K}_2^2}{(I_3-I_2)^2} \right]
- {J^2} \\
a_5 &=& 2 {K}_3\frac{ I_1 I_3 - 2I_1 I_2 + I_2 I_3}{(I_3-I_1) (I_3-I_2)}, \\
a_6 &=& 1. 
\end{eqnarray}
For the quantum mechanical variables the coefficients are 
\begin{eqnarray}
a_0 &=& -\frac{J^2 \Omega_3^4}{16(\chi_1-\chi_3)^2 (\chi_2-\chi_3)^2}, \\
a_1 &=& \frac{J^2 \Omega_3^3 (\chi_1+\chi_2-2\chi_3)}{4(\chi_1-\chi_3)^2 (\chi_2-\chi_3)^2}  , \\
a_2 &=& \Omega_3^2\frac{\Omega_1^2+\Omega_2^2+\Omega_3^2 - 4 J^2\left[ (\chi_1+\chi_2-2\chi_3)^2 + 2 (\chi_1-\chi_3)(\chi_2-\chi_3) \right]}{16(\chi_1-\chi_3)^2 (\chi_2-\chi_3)^2} ,  \\
a_3 &=& \Omega_3\frac{\Omega_1^2 (\chi_3-\chi_2) +  \Omega_2^2(\chi_3-\chi_1) -\Omega_3^2(\chi_1+\chi_2-2\chi_3)}{4(\chi_1-\chi_3)^2 (\chi_2-\chi_3)^2} + J^2 \Omega_3 \frac{\chi_1+\chi_2-2\chi_3}{(\chi_1-\chi_3)(\chi_2-\chi_3)}, \\
a_4 &=& \frac{\Omega_3^2\left[ (\chi_1 +\chi_2 - 2\chi_3)^2 + 2(\chi_1-\chi_3)(\chi_2-\chi_3) \right]}{4(\chi_1-\chi_3)^2 (\chi_2-\chi_3)^2}
 + \frac{\Omega_1^2}{4(\chi_1-\chi_3)^2} + \frac{\Omega_2^2}{4(\chi_2-\chi_3)^2} - J^2 , \\
 a_5 &=& -\frac{\Omega_3 (\chi_1+\chi_2-2\chi_3)}{(\chi_1-\chi_3)(\chi_2-\chi_3)} ,\\
a_6 &=& 1.
\end{eqnarray}

\section{Principal radii of curvature of an ellipsoid}
\label{Curvature}

Consider an ellipsoid
\begin{eqnarray}
\left(\frac{x}{a} \right)^2 + \left(\frac{y}{b} \right)^2 
+ \left(\frac{z}{c} \right)^2 = 1
\label{xyzellipsoid}
\end{eqnarray}
which can be parametrized as
\begin{eqnarray}
x &=& a \sin \theta \cos \phi, \\
y &=& b \sin \theta \sin \phi, \\
z &=& c \cos \theta .
\end{eqnarray}
In general, one can calculate the principal curvatures $\kappa_{1,2}$ from the mean curvature $H=(\kappa_1+\kappa_2)/2$ and Gauss curvature $G=\kappa_1 \kappa_2$ as (see, e.g., \cite{DifGeom})
\begin{eqnarray}
\kappa_{1,2} = H\pm \sqrt{H^2-G},
\label{kappas}
\end{eqnarray}
where
\begin{eqnarray}
H &=&\frac{g_{11}h_{22}-2g_{12}h_{12}+g_{22}h_{11}}{2(g_{11}g_{22}-g_{12}^2)}, \\
G &=&\frac{h_{11}h_{22}-h_{12}^2}{g_{11}g_{22}-g_{12}^2}, 
\end{eqnarray}
and 
\begin{eqnarray}
g_{ij} &=& \vec{x}_i\cdot \vec{x}_j, \\
\vec{x}_i &=& \frac{\partial \vec{x}}{\partial u_i}, \\
h_{ij} &=& \vec{n}\cdot \vec{x}_{ij}, \\
\vec{x}_{ij} &=& \frac{\partial^2 \vec{x}}{\partial u_i \partial u_j}, \\
\vec{x} &=& [x,y,z]^{T},
\end{eqnarray}
and $\vec{n}$ is a unit normal vector to the surface.
For the ellipsoid parametrization we use $u_1=\phi$ and $u_2=\theta$.
We thus find
\begin{eqnarray}
\vec{x}_1 = \left(\begin{array}{c} 
-a \sin \theta \sin \phi \\ b \sin \theta \cos \phi \\ 0 
\end{array} \right), \qquad
\vec{x}_2 = \left(\begin{array}{c} 
a \cos \theta \cos \phi \\ b \cos \theta \sin \phi \\ -c \sin  \theta
\end{array} \right),
\end{eqnarray}
so that
\begin{eqnarray}
g_{11} &=& (a^2 \sin^2 \phi + b^2 \cos^2 \phi) \sin^2 \theta, \\
g_{12} &=& (b^2-a^2) \sin \theta \cos \theta \sin \phi \cos \phi, \\
g_{22} &=&  (a^2 \cos^2 \phi + b^2 \sin^2 \phi) \cos^2 \theta + c^2 \sin^2 \theta. 
\end{eqnarray}
The normal vector is found as
\begin{eqnarray}
\vec{n} = \frac{\vec{x}_1 \times \vec{x}_2}{|\vec{x}_1 \times \vec{x}_2|} =
 -\frac{1}{Q}\left(\begin{array}{c} 
bc \sin^2 \theta \cos \phi \\ ac \sin^2 \theta \sin \phi \\ ab \sin  \theta \cos \theta
\end{array} \right), 
\end{eqnarray}
where
\begin{eqnarray}
Q &=& \left( b^2 c^2 \sin^4 \theta \cos^2 \phi + a^2 c^2 \sin^4 \theta \sin^2 \phi 
+ a^2 b^2 \sin^2 \theta \cos^2 \theta \right) ^{1/2}.
\end{eqnarray}
The other important vectors are
\begin{eqnarray}
\vec{x}_{11} = \left(\begin{array}{c} 
-a \sin \theta \cos \phi \\ -b \sin \theta \sin \phi \\ 0 
\end{array} \right), \qquad
\vec{x}_{12} = \left(\begin{array}{c} 
-a \cos \theta \sin \phi \\ b \cos \theta \cos \phi \\ 0 
\end{array} \right), \qquad
\vec{x}_{22} = \left(\begin{array}{c} 
-a \sin \theta \cos \phi \\ -b \sin \theta \sin \phi \\ -c \cos \theta 
\end{array} \right),
\end{eqnarray}
out of which we can calculate
\begin{eqnarray}
h_{11} &=& \frac{abc \sin^3 \theta}{Q} ,\\
h_{12} &=& 0, \\
h_{22} &=&  \frac{abc \sin \theta}{Q} .
\end{eqnarray}
We thus find
\begin{eqnarray}
H&=& \frac{abc}{2}\cdot \frac{a^2 (\sin^2 \phi + \cos^2 \phi \cos^2 \theta) + b^2 (\cos^2 \phi + \sin^2 \phi \cos^2 \theta) + c^2 \sin^2 \theta}{(a^2 b^2 \cos^2 \theta + a^2 c^2 \sin^2 \phi \sin^2 \theta + b^2 c^2 \cos^2 \phi \sin^2 \theta)^{3/2}}
\end{eqnarray}
and 
\begin{eqnarray}
G&=&  \frac{a^2 b^2 c^2}{(a^2 b^2 \cos^2 \theta + a^2 c^2 \sin^2 \phi \sin^2 \theta + b^2 c^2 \cos^2 \phi \sin^2 \theta)^2}
\end{eqnarray}
which allow us to find $\kappa_{1,2}$ according to (\ref{kappas}).
Expressing then in $\kappa_{1,2}$  angular variables $\theta , \phi$ in terms of the cartesian ones fulfilling Eq. (\ref{xyzellipsoid}), and taking the reciprocal value of $\kappa_{1,2}$, one finds the principal radii of the ellipsoid in the form 
\begin{eqnarray}
R_{1,2} = \frac{2 a^2 b^2 c^2  \left(
\frac{x^2}{a^4} +\frac{y^2}{b^4} + \frac{z^2}{c^4}\right)^{3/2}}{a^2+b^2+c^2 - x^2-y^2-z^2 \pm \sqrt{(a^2+b^2+c^2 - x^2-y^2-z^2)^2 - 4a^2b^2c^2 \left(
\frac{x^2}{a^4} +\frac{y^2}{b^4} + \frac{z^2}{c^4}\right)}} .
\label{Rellipsoid}
\end{eqnarray}
Note that for a sphere $a=b=c=R$ one finds $R_{1,2}=R$. At the vertex of an ellipsoid, $x=a, y=0, z=0$, Eq. (\ref{Rellipsoid}) yields $R_1=c^2/a$, and $R_2=b^2/a$. 
Along the equator $z=0$  Eq. (\ref{Rellipsoid}) yields 
\begin{eqnarray}
\label{princR1}
R_1 =  \frac{1}{ab} \left( \frac{a^2 y^2}{b^2} + \frac{b^2 x^2}{a^2} \right)^{3/2}, \qquad
R_2 = \frac{c^2}{ab} \left( \frac{a^2 y^2}{b^2} + \frac{b^2 x^2}{a^2} \right)^{1/2}. 
\label{princR2}
\end{eqnarray}
\end{widetext}


\end{document}